\documentclass[prd,aps,superscriptaddress,nofootinbib,showpacs,twocolumn,
preprintnumbers]{revtex4}

\usepackage[intlimits]{amsmath}
\usepackage{amsfonts}
\usepackage{amssymb,amscd}
\usepackage{bbold}
\usepackage{graphics}
\usepackage[dvips]{epsfig}    
\usepackage[usenames,dvipsnames,svgnames,table]{xcolor}
\usepackage{placeins}

\newcommand{\beq}{\begin{equation}}
\newcommand{\eeq}{\end{equation}}
\newcommand{\beqa}{\begin{eqnarray}}
\newcommand{\eeqa}{\end{eqnarray}}
\newcommand{\be}{\begin{equation}}
\newcommand{\ee}{\end{equation}}

\def\qq{\text{\boldmath$q$\unboldmath}}

\def\pp{\text{\boldmath$p$\unboldmath}}

\def\kk{\text{\boldmath$k$\unboldmath}}

\def\gbold{\text{\boldmath$\gamma$\unboldmath}}

\begin{document}

\title{Bayesian analysis of quark spectral properties from the Dyson-Schwinger equation}

\author{Christian~S.~Fischer}
\affiliation{Institut f\"ur Theoretische Physik, Justus-Liebig Universit\"at Gie{\ss}en, 35392 Gie{\ss}en, Germany}
\affiliation{HICforFAIR-Gie{\ss}en}
\author{Jan~M.~Pawlowski}
\affiliation{Institut f\"{u}r Theoretische Physik, Universit\"{a}t Heidelberg, Philosophenweg 16, 69120 Heidelberg, Germany}
\affiliation{ExtreMe Matter Institute EMMI, GSI Helmholtzzentrum f\"{u}r Schwerionenforschung mbH, 64291 Darmstadt, Germany} \author{Alexander~Rothkopf}
\affiliation{Institut f\"{u}r Theoretische Physik, Universit\"{a}t Heidelberg, Philosophenweg 12, 69120 Heidelberg, Germany}
\author{Christian~A.~Welzbacher}
\affiliation{Institut f\"ur Theoretische Physik, Justus-Liebig Universit\"at Gie{\ss}en, 35392 Gie{\ss}en, Germany}

\date{\today}
\begin{abstract}
We report results on the quark spectral function in the Landau gauge at finite temperature
determined from its Dyson-Schwinger equation. Compared to earlier quenched results \cite{Mueller:2010ah} 
this study encompasses unquenched $N_f=2+1$ fermion flavors in the medium. 
For the reconstruction of real-time spectra we deploy a recent Bayesian approach (BR method) \cite{Burnier:2013nla} 
and develop a new prior in order to better assess the inherent systematic uncertainties.  
We identify the quark quasi-particle spectrum and analyze the (non-)appearance of zero modes at or around the 
pseudo-critical temperature. In both, the fully unquenched system and a simpler truncation using a model 
for the gluon propagator we observe a characteristic three-peak structure at zero three-momentum. The 
temperature dependence of these structures in case of the gluon propagator model is different than observed 
in previous studies. For the back-coupled and unquenched case we find interesting modifications at and around
the pseudo-critical transition temperature. 
\end{abstract}

\maketitle

\section{Introduction}

The wealth of data produced in heavy ion collision experiments at the
Relativistic Heavy-Ion Collider (RHIC) and the Large Hadron Collider
(LHC) has lead to interesting insights about the nature of the
quark-gluon plasma (QGP) in various temperature regimes (see e.g.
Refs.~\cite{Muller:2006ee,Shuryak:2008eq,BraunMunzinger:2008tz,Andronic:2014zha,Foka:2016zdb}
and references therein). Thermal and transport properties of the QGP
are encoded in the correlation functions of QCD. In particular they
can be assessed from real time properties of QCD's most basic
correlation functions, the quark
\cite{Nickel:2006mm,Harada:2007gg,Harada:2009zq,Karsch:2007wc,Karsch:2009tp,Mueller:2010ah,Qin:2010pc,Qin:2013ufa,Gao:2014rqa}
and gluon propagators
\cite{Strauss:2012dg,Haas:2013hpa,Christiansen:2014ypa,Dudal:2013yva,Ilgenfritz:2017kkp}. A
prominent example is the dilepton production in a heavy ion
collision. It can been related to the spectral properties of
thermalized quasi-particles and specifically to the dispersion
relation of
quarks~\cite{Braaten:1990wp,Peshier:1999dt,Arnold:2002ja,Kim:2015poa}. Another
important example are QCD transport coefficients, that have been
expressed in terms of single particle spectral functions of the
fundamental fields, the quarks and
gluons~\cite{Haas:2013hpa,Christiansen:2014ypa}.  In summary, a
detailed understanding of a potential quasiparticle spectrum in the
QGP, in particular close to the chiral phase transition, is highly 
desirable.

In this work we focus on the quark spectral function encoding the
quark dispersion relation and decay width in the medium. At large
temperatures reliable results have been obtained in the hard-thermal
loop (HTL) expansion
\cite{Braaten:1989mz,Baym:1992eu,Blaizot:1993bb}. In this regime, the
quark spectral function shows two excitations in the dispersion
relation, the ordinary quark with a positive ratio of chirality to
helicity, and a collective 'plasmino' mode with a corresponding
negative ratio. Both have thermal masses of the order $gT$ and decay
widths of the order $g^2T$, where $g$ is the coupling constant and $T$ the
temperature. The two excitations are accompanied by a continuum
contribution from a branch cut in the quark propagator due to Landau
damping, i.e.\ the absorption of a space-like quark by a hard gluon or
hard antiquark.

Beyond systematic weak-coupling expansions, there is no
straightforward approach for the determination of the spectral
function. Model calculations offer qualitative insights
\cite{Schaefer:1998wd,Kitazawa:2005mp,Kitazawa:2006zi,Harada:2007gg}
which, however, need to be corroborated in more fundamental
approaches. Models for quark spectral functions constructed along the
lines of the HTL results have been fitted to data from quenched
lattice QCD \cite{Karsch:2007wc,Karsch:2009tp} and quenched
Dyson-Schwinger calculations \cite{Mueller:2010ah}. Again, such an
approach offers qualitative insights but suffers from potential biases
involved in the model building. This holds in particular for
temperatures around the (pseudo-)critical one, where HTL is not
expected to be reliable. 

In principle, functional approaches like Dyson-Schwinger equations
(DSE) and the functional renormalization group (FRG) offer the
possibility to determine the two-point correlators in the complex
momentum plane thus allowing for a direct extraction of the
corresponding spectral function. This has been performed successfully
for the gluon propagator at zero temperature in
\cite{Strauss:2012dg}. At finite temperatures direct computations in
the complex freqeuncy plane have been carried out in matter systems,
see e.g.\
\cite{Kamikado:2013sia,Tripolt:2014wra,Pawlowski:2015mia,Strodthoff:2016pxx}.
However, the additional conceptual and numerical challenges have delayed similar direct
analyses in QCD so far.

There are, however, approaches that allow us to extract spectral
functions from numerical data in the spatial Euclidean momentum
region. These approaches utilise the fact that the spectrum is related
to the Euclidean correlator via an integral transform, which needs to
be inverted. This is a classic ill-posed inverse problem and Bayesian
inference can be used to give meaning to it, by systematically
incorporating additional prior information available. Among the
different implementations of the Bayesian reconstruction strategy is
the popular Maximum Entropy Method
(MEM)~\cite{Jarrell:1996,nr:1997,Asakawa:2000tr}, which originates in
two-dimensional image reconstruction. It has been deployed for the
study of quarks in cold and dense matter \cite{Nickel:2006mm} and
quarks at and around the (pseudo-)critical temperature
\cite{Qin:2010pc,Qin:2013ufa,Gao:2014rqa}. A method similar in spirit
as the MEM but instead using a quadratic regulator has been applied to
study gluon spectra in \cite{Dudal:2013yva}.

As has been discussed e.g.\ in \cite{Burnier:2013nla} the MEM based
approaches have to deal with several issues. The main difficulty is that of
flat directions in the regulator functional in case of positive
definite spectra. In practice this leads to very slow convergence in
case that a large number of data points is supplied. The second point
is related to the weighting of data and prior information, which
conventionally is implemented via computing the so called evidence
probability distribution. This step relies on a Gaussian
approximation, which in practice is difficult to justify, see also
App.\ref{sec:MEMcheck}.

In order to overcome these and further difficulties, a novel
implementation of the Bayesian strategy has been recently developed
\cite{Burnier:2013nla}. It is specifically designed for the solution
of one-dimensional inverse problems. Its generalization to arbitrary
spectra \cite{Rothkopf:2016luz} has been applied for extracting
spectral properties of gluons at finite temperature
\cite{Ilgenfritz:2017kkp} in lattice QCD. In this study we both deploy
the original BR method and develop in addition a new "low-ringing"
BR-type prior functional, which allows us to unambiguously distinguish
between peaked structures present in the underlying correlator data
and numerical ringing artifacts (see e.g.\ discussion in
\cite{Kim:2014iga}) common to inverse problems (c.f. Gibbs
phenomenon).

We apply our new method to temperature dependent quark propagators
obtained from two different truncation schemes for the quark and gluon
DSEs. On the one hand we re-analyse a model truncation for the gluon
propagator and compare to previous results
\cite{Qin:2010pc,Qin:2013ufa,Gao:2014rqa}. In these works a zero mode
in the spectral function at zero momentum has been identified in
addition to the two conventional symmetric peaks at finite
frequency. Since such a zero mode does not appear in the HTL-studies,
it has been attributed to the strong interaction physics governing the
transition around the (pseudo-)critical temperature and signalling the
formation of the quark-gluon plasma. While the appearance of this
structure has been found to be robust under variations within a class
of truncation schemes using models for the gluon \cite{Gao:2014rqa},
it remained to be seen whether this is also true for the fully
unquenched system. In this article we therefore study in addition a
truncation based on \cite{Mueller:2010ah,Fischer:2012vc} to include
results for the unquenched quark propagator with $N_f=2+1$.  This
truncation offers direct control over the Yang-Mills sector by
explicitly taking quark loop effects in the gluon propagator into
account. The resulting prediction for the unquenched gluon propagator
at finite temperature \cite{Fischer:2012vc} has been shown to agree
with corresponding lattice results of
\cite{Aouane:2012bk}. Furthermore, the temperature dependence of the
chiral condensate evaluated on the lattice \cite{Borsanyi:2010bp} has
been reproduced. We therefore may expect realistic and quantitative
results for the spectral functions as well.

The article is organised as follows. In the next section we summarize
the framework to determine the quark propagator at finite temperature
and chemical potential. Since all technical details have been given
elsewhere, see Refs.~\cite{Qin:2010pc,Qin:2013ufa,Gao:2014rqa} for the
model gluon and
Refs.~\cite{Fischer:2012vc,Fischer:2013eca,Fischer:2014ata,Eichmann:2015kfa}
for the unquenched system, we remain very brief.  The section
\ref{sec:Bayes} is devoted to the Bayesian BR method reconstruction
and the specific improvements we use in this work.  Our results for
the quark spectral functions are presented and discussed in section
\ref{sec:results}. We conclude in section \ref{sec:conclusion}.

\section{The quark propagator at finite temperature and chemical potential} \label{sec:DSE}

\subsection{Quark Dyson-Schwinger equation}\label{sec:truncation}

In order to determine the quark propagator at finite temperature and
chemical potential we work with the Euclidean metric version of QCD
and use the Matsubara formalism.  The renormalized quark
Dyson-Schwinger equation is then given by
\begin{align}
 S^{-1}(i\omega_p,\pp)
&=
Z_2\,S_0^{-1}(i\omega_p,\pp)+\Sigma(i\omega_p,\pp)
\,.
\label{quark:dse}
\end{align}
Here the inverse full quark propagator is denoted by
$S^{-1}(i\omega_p,\pp)$ and its inverse bare counterpart by
$S^{-1}_{0}(i\omega_p,\pp)$. We follow the conventions of
\cite{Mueller:2010ah} and explicitly use imaginary arguments for the
energy in all functions.  The dependence of all functions on the
renormalisation point is left implicit and $Z_2$ denotes the wave
function renormalization constant of the quark.  The quark Matsubara
frequencies are given by $\omega_p=(2n_p+1)\pi T$ with temperature
$T$.  The Dirac structure of the inverse propagators at finite T and
$\mu=0$ can be decomposed via
\begin{align} 
\nonumber S^{-1}_0(i\omega_p,\pp)
&=
i\gamma_4\,\omega_p
+
i\text{\boldmath{$\gamma$}\unboldmath}\cdot\pp
+
Z_m\,m(\mu)
\,,\\[2ex]
S^{-1}(i\omega_p,\pp)
&=
i\gamma_4\omega_p C(i \omega_p,\vert\pp\vert)
+
i\text{\boldmath{$\gamma$}\unboldmath}\cdot\pp\,A(i \omega_p,\vert\pp\vert)
\nonumber\\[2ex]
&\phantom{=}
+
B(i \omega_p,\vert\pp\vert)
\,.
\label{Dirac:decomp}
\end{align}
The dressing functions $A(i \omega_p,\vert\pp\vert)$,
$B(i\omega_p,\vert\pp\vert)$ and $C(i \omega_p,\vert\pp\vert)$ carry
all non-trivial information on the quarks.  The quark mass
renormalization constant $Z_m(\mu^2)$, the renormalized mass
$m(\mu^2)$ and $Z_2(\mu^2)$ at the renormalization scale $\mu$ are
determined within a MOM-renormalization scheme. The explicit form of
the quark self-energy can be written as
\begin{align}\nonumber
  \Sigma(i\omega_p,\pp)
  &=
    \frac{16\pi}{3}\dfrac{Z_2}{\tilde{Z}_3} \alpha(\mu)
    T\sum_{n_q}\int\frac{d^3q}{(2\pi)^3}\;\Big[\gamma_{\mu}S(i\omega_{q},\qq)
  \\[2ex]
  &\hspace{-.5cm}
    \times\Gamma_{\nu}(i\omega_q,\qq,i\omega_p,\pp)D_{\mu\nu}(i\omega_p-i\omega_q,\pp-\qq)\Big]
    \,,
\label{quark:self-e}
\end{align} 
with (Landau gauge) gluon propagator $D_{\mu\nu}$, quark-gluon vertex
$\Gamma_{\nu}$, gauge coupling $\alpha$ and ghost renormalization
constant $\tilde{Z}_3$.

\begin{figure}[t]
\includegraphics[width=20pc]{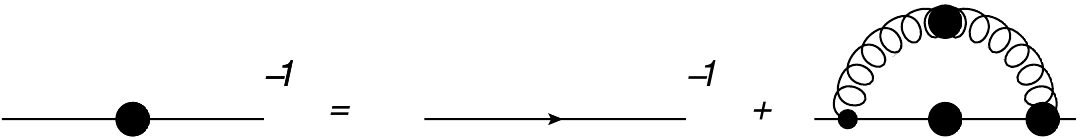}
\caption{\label{fig:qdse} The quark Dyson-Schwinger equation. Dressed
  propagators and vertices are denoted by large filled dots.}
\vspace{2pc}
\includegraphics[width=20pc]{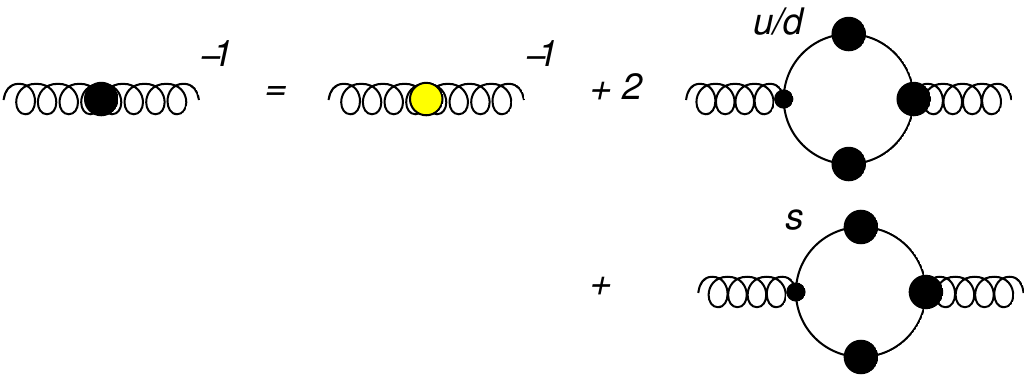}
\caption{\label{fig:gdse}The truncated gluon Dyson-Schwinger
  equation. The yellow dot represents input from quenched lattice QCD,
  whereas the dressed quark propagators in the quark-loop are
  calculated self-consistently from the quark-DSE.}
\end{figure}

In order to determine the quark propagator self-consistently from its
DSE, we also need to specify the dressed gluon propagator and the
dressed quark-gluon vertex. The truncation used in this work has been
developed over the past years
\cite{Fischer:2009wc,Fischer:2010fx,Fischer:2012vc,Fischer:2013eca,Fischer:2014ata}
and is guided by two main ideas. First, lattice results for the
temperature dependent quenched gluon propagator are used as external
input \cite{Fischer:2010fx,Maas:2011ez} and unquenched via adding an
explicit quark-loop for each of the $N_f=2+1$ quark flavors used in
this work. The resulting DSE for the gluon propagator is depicted in
Fig.~\ref{fig:gdse}.

This approach ensures that the unquenched gluon propagator inherits
the leading order temperature and chemical potential effects via the
quark loops, giving a contribution to the thermal mass. Second order
unquenching effects in the Yang-Mills diagrams are neglected. The
second element of our truncation is an approximation for the full
quark-gluon vertex, which combines information from the well-known
perturbative behavior at large momenta and an approximate form of the
Slavnov-Taylor identity at small momenta studied long ago by Ball and
Chiu \cite{Ball:1980ay}. The explicit form of this approximate
expression for the vertex is discussed in
Refs.\cite{Fischer:2010fx,Fischer:2012vc,Fischer:2014ata} and shall
not be repeated here for brevity.

A much simpler system in terms of numerical effort is obtained by
substituting the dressed gluon propagator in the quark DSE by a model
together with a bare quark-gluon vertex.  Taking into account a
Debey-like mass in the longitudinal gluon this reads $(k=q-p)$
\begin{eqnarray}
&&g^2 D_{\mu\nu}(i\omega_k,\kk)\Gamma_\nu(i\omega_q,\qq,i\omega_p,\pp) \notag\\[2ex]
&&= [P_{\mu\nu}^{T} D_{T}(\omega_k^2,\kk^2) 
+ P_{\mu\nu}^{L} D_{L}(\omega_k^2,\kk^2)]\gamma_\nu \,,
\end{eqnarray}
with $P_{\mu\nu}^{T,L}$ the transverse and longitudinal projection
operators with respect to the heat bath. The dressing functions are
given by
\begin{eqnarray}
D_{T} &=&\mathcal{D}(\kk^2 + \omega_k^2)\, , \quad
D_{L} =\mathcal{D}(\kk^2 + \omega_k^2 + m^2_g)\, .
\end{eqnarray}
where the functions 
\begin{eqnarray}
\mathcal{D}(s) & = &
\frac{8{\pi^{2}}D}{\sigma^4}e^{-s/\sigma^2} 
+ \frac{8{\pi^{2}} {\gamma_{m}}}{{\ln}[ \tau \! + \! (1 \! + \!
s/{\Lambda_{\text{QCD}}^{2}} ) ^{2} ] } \,{\cal F}(s)\,,\quad
\end{eqnarray}
are defined with ${\cal F}(s) = [1-\exp(-s/4 m_t^2)]/s$, $\tau=e^2-1$,
$m_t=0.5\,$GeV, $\gamma_m=12/25$, and
$\Lambda_{\text{QCD}}=0.234$GeV. With $\sigma D = (0.8 \mbox{GeV})^3$
we choose $\sigma=0.5$ as a representative value for the model
parameters. This is one (the simplest) example of a class of
truncation schemes that have been studied in Ref.~\cite{Gao:2014rqa}
and found to agree qualitatively with each other in the resulting
spectral functions for the quarks. As explained in the introduction we
use this model as a numerically easily accessible reference, which
already displays the interesting three-peak structure in the resulting
spectral functions.

\subsection{Quark spectral functions and representation}\label{sec:spectral_rep}

The spectral representation of the quark propagator is given by
\begin{align}\label{spectral}
  S(i\omega_p,\pp)
  &=
    \int_{-\infty}^\infty
    \!\frac{d\omega'}{2\pi}\,
    \frac{\rho(\omega',\pp)}{i\omega_p-\omega'}
    \,,\end{align}
  with spectral function $\rho(\omega,\pp)$ parameterized as
\begin{align}
\rho(\omega,\pp)
&=
2\pi\Big(
\rho_{4}(\omega,\vert\pp\vert)\gamma_4
+
\rho_{\rm v}(\omega,\vert\pp\vert)\left( i\gbold\cdot\pp\right)/\vert\pp\vert
\nonumber\\[2ex] 
&\phantom{=2\pi\big(}
-
\rho_{\rm s}(\omega,\vert\pp\vert) \label{eq:rhostruc}
\Big)
\,.
\end{align}
Similar to \cite{Mueller:2010ah} we choose conventions such that the scalar dressing functions 
themselves agree with those introduced by corresponding lattice studies \cite{Karsch:2009tp}.
With a positive definite metric, which is not the case for gauge-fixed QCD, the components  
of the spectral function would furthermore obey the inequality
\begin{align}
\rho_{4}(\omega,\vert\pp\vert)
&\geq
\sqrt{
\rho_{\rm v}(\omega,\vert\pp\vert)^2+\rho_{\rm s}(\omega,\vert\pp\vert)^2
}
\geq
0
\label{eq:RhoInequality}
\end{align}
as well as the standard sum rule
\begin{subequations} \label{eq:sumrules}
\begin{align}
Z_2\int_{-\infty}^\infty\!d\omega\, \rho_4(\omega,\vert\pp\vert)=1
\,,\label{eq:sum_rule4}
\end{align}
with wave function renormalization constant $Z_2$. The vector and
scalar spectral functions, $\rho_{\rm v}, \rho_{\rm s}$, sum up to
zero and hence are necessarily negative for some momentum regime,
\begin{align}
\int_{-\infty}^\infty\!d\omega\, \rho_{\rm v}(\omega,\vert\pp\vert)=0\,,\qquad 
\int_{-\infty}^\infty\!d\omega\, \rho_{\rm s}(\omega,\vert\pp\vert)=0\,.
\label{eq:sumrulevs}
\end{align}
\end{subequations}
Using specific projection operators for the chirally symmetric and/or
static ($\vert\pp\vert=0$) case one can define positive semi-definite
combinations of $\rho_4, \rho_v$ and $\rho_s$ using
Eq.~\eqref{eq:RhoInequality}, see e.g.\ \cite{Mueller:2010ah} for
details.  In this first study here we restrict ourselves to the
reconstruction of the component $\rho_4$ only. The corresponding left
hand side of Eq.~\eqref{spectral} is then also given by the $\gamma_4$
component of the quark propagator, i.e.
\begin{align}\nonumber 
  &S_4(i\omega_p,\vert\pp\vert)\\[2ex]
    =&\, \frac{i \omega_p\, C(\omega_p,\vert\pp\vert)}{\omega_p^2\, 
       C(\omega_p,\vert\pp\vert)^2+ \pp^2\, A(\omega_p,\vert\pp\vert)^2+B(\omega_p,\vert\pp\vert)^2}\,.
\end{align} 
This component has the advantage of being antisymmetric in $\omega_p$
and the corresponding part of the spectral function being symmetric in
$\omega$, which means that we may restrict ourselves in
Eq.~\eqref{spectral} to positive frequencies and arrive, using
Eq.~\eqref{eq:rhostruc}, at a purely imaginary kernel
\begin{align}\label{eq:spectralsym}
  S_4(i\omega_p,\pp)
  &=
    i \int_{0}^\infty
    \!d\omega'\,
    \frac{2\omega_p \rho_4(\omega',\pp)}{\omega_p^2+\omega'^2}
    \,,
\end{align}
which is easily implemented numerically for the spectral
reconstruction. I.e. we will directly formulate the inverse problem in
imaginary frequency space, since the rational kernel in
Eq.~\eqref{eq:spectralsym} suppresses spectral information much less
than its Euclidean counterpart and thus is ideally suited for use in
spectral reconstructions.

\section{Bayesian Spectral Reconstruction} \label{sec:Bayes}

\subsection{Bayesian inference}

In this study we use the concept of Bayesian inference to invert the
relation of Eq.~\eqref{eq:spectralsym} numerically. The quark
two-point function $S_4$ is evaluated along discretized imaginary
frequencies $\omega_p$ with $p\in[1\ldots N_{\rm data}]$, while the
spectral function is resolved along $N_\omega$ bins in real-time
frequencies $\omega^\prime_l$
\begin{align}
  -iS_4(i\omega_p) = \sum_{l=1}^{N_\omega} 
  \frac{2\omega_p}{\omega_p^2+\omega^{\prime 2}_l}\rho_l\,.\label{eq:numconv}
\end{align}
Since the data are obtained from a functional QCD computation it can
be provided at an arbitrary number of points $N_{\rm data}$ but
contains a finite numerical error $\Delta S_4$ due to e.g.\ the
evaluation of intermediate integrals in e.g.\
Eq.~\eqref{quark:self-e}. Therefore for any finite $N_{\rm data}$ and
$\Delta S_4$ a naive $\chi^2$ fit of the $N_\omega$ parameters
$\rho_l$ would lead to an infinite number of degenerate solutions all
reproducing the data within their uncertainty. Bayesian inference in
the form of Bayes theorem
\begin{align}
P[\rho|S_4,I]=\frac{ P[S_4 | \rho,I] P[\rho| I]}{P[\rho |I]}\,,\label{eq:BT}
\end{align}
provides a systematic prescription of how to regularize the otherwise
under-determined $\chi^2$ fit. It does so by incorporating additional
prior information $I$ on the spectrum, in addition to the correlator
data $S_4$. The posterior $P[\rho|S_4,I]$ for a test function $\rho$
denotes its probability to be the correct spectrum, given the computed
data and prior information. It is proportional to the likelihood
probability $P[S_4 | \rho,I]$ and the prior probability $P[\rho|I]$.

The former encodes how the correlator has been obtained and in case of
stochastically independent sampled data is related to the standard
$\chi^2$ fitting functional
\begin{align}
  P[S_4 | \rho,I]=&\, {\rm exp}[-L]\,,
\end{align}
with likelihood 
\begin{align}\nonumber 
  L=&\,\frac{1}{2}\sum_{p,q=1}^{N_{\rm data}}\big(S_4(i\omega_p)-
             S^\rho_4(i\omega_p)\big)\\[2ex] 
&\times C_{pq}^{-1}\big(S_4(i\omega_q)-S^\rho_4(i\omega_q)\big)\,.
\end{align}
Here $S^\rho_4(i\omega_p)$ denotes the correlator according to the
current choice of test function via Eq.~\eqref{eq:numconv} and
$C_{pq}$ the covariance matrix of the actual correlator data
$S_4$. Note that we have formulated the inverse problem in the
imaginary frequency domain and not based on Euclidean data. The reason
is that the exponentially damped integral kernel and the information
loss associated with the latter leads to much worse reconstruction
results than the rational K\"allen-Lehmann kernel used here (for an
explicit demonstration see
\cite{Rothkopf:2016luz,Pawlowski:2016eck}). Note also that in the case
of the Euclidean kernel it was found that in order to compute the
propagator from a test spectral function $\rho$ with high accuracy,
one needs to deploy a logarithmic frequency grid \cite{Welzbacher16}.

Since furthermore in our case the correlator is computed and not
sampled we will assign an estimated diagonal covariance matrix to the
discrete $S_4$ with constant relative errors
$\Delta S_4/S_4={\rm const.}$. $L[\rho]$ possesses many flat
directions, which translate into the non-uniqueness of the maximum
likelihood $\chi^2$ approach.

The second and decisive term in the numerator of Eq.~\eqref{eq:BT} is
the prior probability, which acts as a regulator to the likelihood
probability, lifting the flat directions of $L$. It tells us how
compatible the test function $\rho$ is with available prior
information. Conventionally it is expressed as
\begin{align}
P[\rho |I]={\rm exp}[-\alpha S[\rho,m]]\,,
\end{align}
where $\alpha$ represents a so called hyper-parameter, which weighs
the influence of the data versus the prior. Prior information is
encoded in $P[\rho |I]$ in two ways, on the one hand via the
functional form of S itself and via the so called default model
$m(\omega)$, which by definition is the most probable spectrum in the
absence of data, i.e. it represents the unique extremum of
$S[\rho,m]$.

Different implementations of the Bayesian strategy differ not only in
the regulator $S$ which they employ, but also in how the
hyperparameter $\alpha$ is treated, as well as how the most probable
spectrum
\begin{align}
\left.\frac{\delta P[\rho|S_4,I]}{\delta \rho_l}\right|_{\rho=\rho^{\rm Bayes}}=0\,,
\label{eq:BayesStat}
\end{align}
is determined numerically. Note that different Bayesian prescriptions
can yield different results, as long as $N_{\rm data}$ and
$\Delta S_4$ are finite. Only in the "Bayesian continuum limit" all
methods, if implemented correctly, must converge to the same
result. It is therefore important to test how reconstructions change
towards this limit using e.g.\ mock data tests.

In this study we will work with quarks at high temperature, where the
spectral functions are assumed to be positive definite, a property
which in turn will enter as prior information.

The popular Maximum Entropy Method for positive definite spectra
proposes to use the Shannon Jaynes Entropy as regulator, a choice
justified by arguments from two-dimensional image reconstruction
\begin{align}
  S_{\rm SJ}=\int d\omega \big(\rho-m-\rho {\rm log}
\big[\frac{\rho}{m}\big]\big)\,.
\end{align}
One carries out the reconstruction multiple times with different
values of $\alpha$ and then averages these intermediate results
$\rho_\alpha$ self consistently over a probability distribution for
$\alpha$, $P[\alpha | S_4,I]$. In the standard implementation by Bryan
the functional space from which the test function $\rho$ is chosen is
manually limited to a dimensionality of $N_{\rm data}$. It has been
shown that the extremum of $P[\rho |S_4,I]$ in general is not
contained in this choice of search space, which may result in slow
convergence. In additions the determination of $P[\alpha | S_4,I]$
relies on a Gaussian approximation.

In this study we instead use a more recent Bayesian approach to
spectral function reconstruction that has been developed with the
one-dimensional inverse problem of Eq.~\eqref{eq:numconv} in mind and
its regulator functional reads
\begin{align}
  S_{\rm BR}=\int d\omega \big( 1- \frac{\rho(\omega)}{m(\omega)} +
  {\rm log}\big[ \frac{\rho(\omega)}{m(\omega)} \big]\big).
\end{align}
The hyperparamter $\alpha$ is treated in a Bayesian fashion, in that
we assume full ignorance of its values $P[\alpha]=1$ and integrate it
out a priori
\begin{align}
 P[\rho|S_4,I(m)]=P[S_4|I]\int_0^\infty d\alpha P[\rho| I(\alpha,m)]\,.
  \label{eq:margin}
\end{align}
In addition we also require that $L=N_{\rm data}$ as the correct
spectrum, on average, would lead to such a value. The most probable
spectrum is then obtained from carrying out a numerical optimization
on $P[\rho|S_4,I(m)]$ according to Eq.~\eqref{eq:BayesStat}. In
contrast to the MEM we do not restrict the search space and use a
pseudo-Newton method (limited Broyden-Fletcher-Goldfarb-Shanno) to
find the global extremum in the full $N_\omega$ dimensional search
space.

The regulator $S_{\rm BR}$ was derived with the goal to limit its
influence on the outcome of the reconstruction to a minimum. I.e. it
shall influence the reconstruction as weakly as possible and "let the
data speak". Comparing $S_{\rm BR}$ to $S_{\rm SJ}$ or the quadratic
prior one finds that far away from the extremum $\rho=m$ $S_{\rm BR}$
shows the weakest curvature. While this makes it easy for structures
encoded in the data to manifest themselves in the reconstruction it
also means that common artefacts associated with inverse problems,
such as Gibbs ringing, are not efficiently suppressed.

\begin{figure}[t]
\includegraphics[scale=0.35]{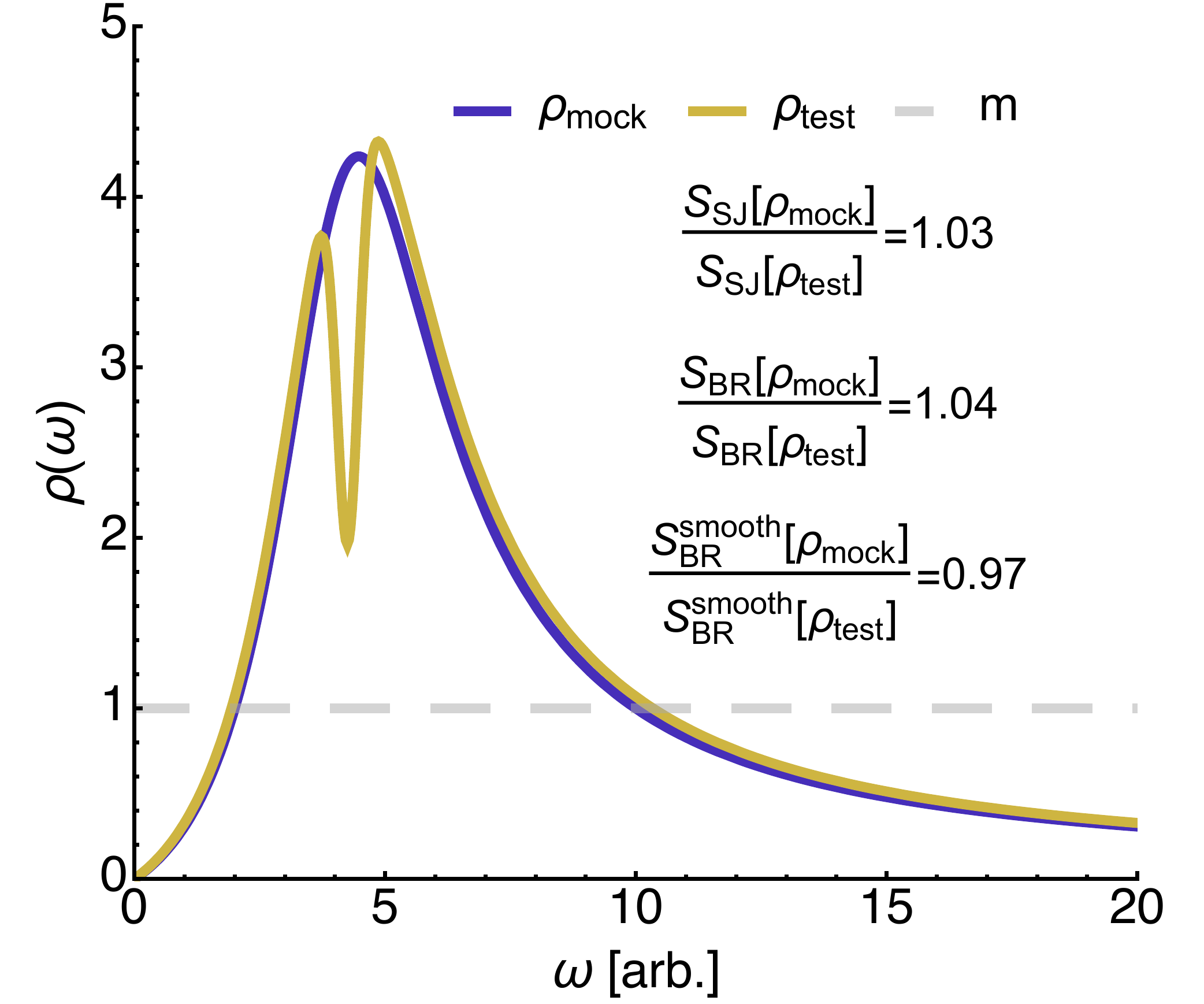}
\caption{Illustration of the susceptibility of the regulators
  $S_{\rm BR}$ and $S_{\rm SJ}$ on ringing artefacts. If a mock
  spectrum $\rho_{\rm mock}$ (dark blue) contains a broad feature then
  introducing a small wiggle leads actually to a smaller
  penalty.}\label{Fig:WiggleSpec}
\end{figure}

In turn, if the structures of physical interest are spectral peaks, as
will be the case in the following, we have to make sure that we
unambiguously distinguish ringing from genuine features.  Due to its
restricted search space the MEM usually produces smooth features and
it is considered safe from ringing. This impression is unfortunately
incorrect on the level of the regulator as illustrated in
Fig.~\ref{Fig:WiggleSpec}. For a flat default model $m=1$ we compare
the penalty assigned to a function $\rho_{\rm mock}$ with a single
broad feature as well as after introducing a wiggle close to its peak
$\rho_{\rm test}$. The area under the spectra has been kept the
same. What one finds is that both $S_{\rm BR}$ and $S_{\rm SJ}$ assign
a lower penalty to the distorted curve even though the former was
explicitly derived with a smoothness axiom. What distinguishes the two
curves is of course their arc-length, which increases for every
additional wiggly structure.

Hence we introduce a new BR-type regulator, which penalizes arc-length
$\ell=\int d\omega\sqrt{1+(d\rho/d\omega)^2}$ explicitly. In order to
leave as much of the original form of $S_{\rm BR}$ intact we add a
$\ell^2$ term and subtract unity. The result is
\begin{align}\nonumber 
  &S_{\rm BR}^{\rm smooth}\\[2ex]
  =&\int d\omega \left( - \frac{1}{2}\left(\frac{d\rho(\omega)}{d
             \omega}\right)^2  + 1- \frac{\rho(\omega)}{m(\omega)} + 
             {\rm log}\big[ \frac{\rho(\omega)}{m(\omega)} \big]\right).
\end{align}
In the presence of the derivative term, we are not any more able to
analytically compute the $\alpha$ dependent normalization of the prior
probability and thus cannot marginalize $\alpha$ as in done in
Eq.~\eqref{eq:margin}. Thus we revert to handling $\alpha$ in the same
way as in the "historic MEM" prescription, in which one adjusts
$\alpha$ in order that $(L-N_{\rm data})<10^{-1}$.

As we will see also explicitly in the mock data test in the following
section, this new regulator succeeds in efficiently suppressing
ringing artefacts. On the other hand peaks encoded in the correlator
become more washed out and it requires more data points and smaller
errors to resolve these peaks to the same accuracy as with the
original BR method. Hence our strategy is the following. Using the
smoothening prior we will identify in which part of the spectrum
actual peak structures reside and then extract their features using
the standard BR approach.

The reliability of the reconstruction can be estimated in three
different ways. Since in Bayes theorem in Eq.~\eqref{eq:BT} data and
prior information enters, we need to understand the dependence of the
reconstruction result on that input. For the former we can vary the
number of provided data points and carry out a Bootstrap Jackknife
resampling analysis of the correlator. Since here we do not use
sampled data we will instead successively lower the assigned error on
the data and observe convergence toward the Bayesian continuum
limit. The dependence on the prior can be estimated by repeating the
reconstructions with different choices of the default model, for which
we choose a flat function $m(\omega)=m_0$ and vary
$m_0=\{0.1,0.5,1.0,5.0,10\}$.

Bayesian methods also provide another measure for the robustness of
the reconstruction
\begin{align}
  \langle \delta \rho^2 \rangle_I \approx - 
  \int_I d\omega d\omega^\prime (\delta^2Q/\delta\rho^2)^{-1}/
  \int_I d\omega d\omega^\prime,\label{eq:curvatureerr}
\end{align} 
based on the curvature of the optimization functional
$Q[\rho,S_4,I]={\rm Log}[ P[\rho|S_4,I] ]$.  In previous works it has
been found that the actual dependence on the variation of data and
prior information is often larger than indicated by this quantity. One
possible reason for this discrepancy is the need for two assumptions
in the derivation of Eq.~\eqref{eq:curvatureerr}, namely that the
posterior is both highly peaked and may be approximated by a
Gaussian. In the following we will show error bands for the BR method
that are obtained from the variation of the default model and the
curvature measure, where the former dominates.

\subsection{Mock data tests}
\label{sec:mocktest}

Previous studies \cite{Qin:2013ufa} deploying a MEM-type approach
hinted at the presence of a three peak structure of the quark spectral
function in the Landau gauge at finite temperature. On the other hand it
is was not excluded that further peaked features may be
present. Therefore we must ascertain how well our reconstruction
method will be able to resolve different structures given a certain
quality of input data, for which we turn to a mock data analysis.

\begin{figure}[t]
\includegraphics[scale=0.5,trim= 0 0 0 0.5cm, clip=true]{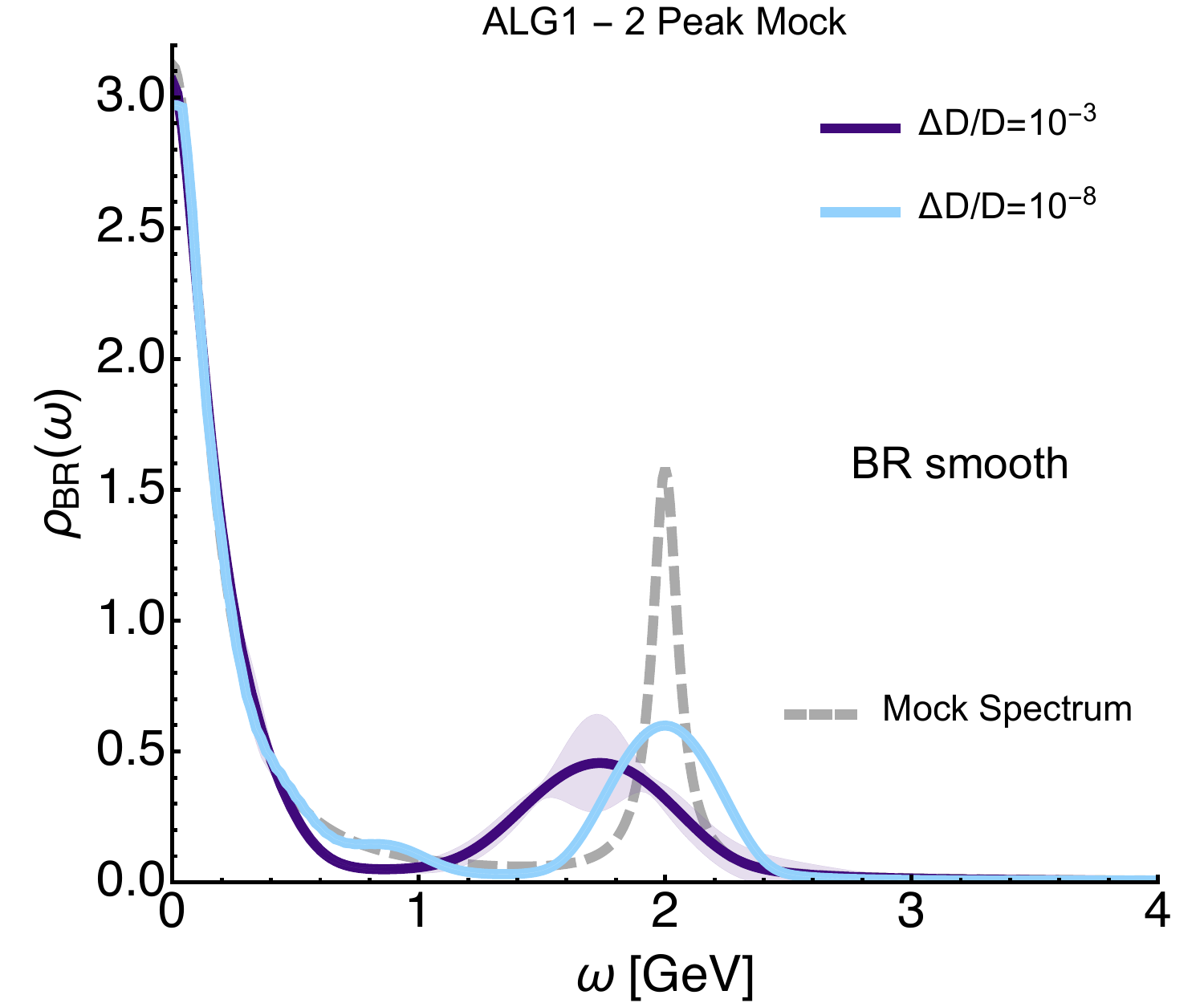}
\includegraphics[scale=0.5,trim= 0 0 0 0.5cm, clip=true]{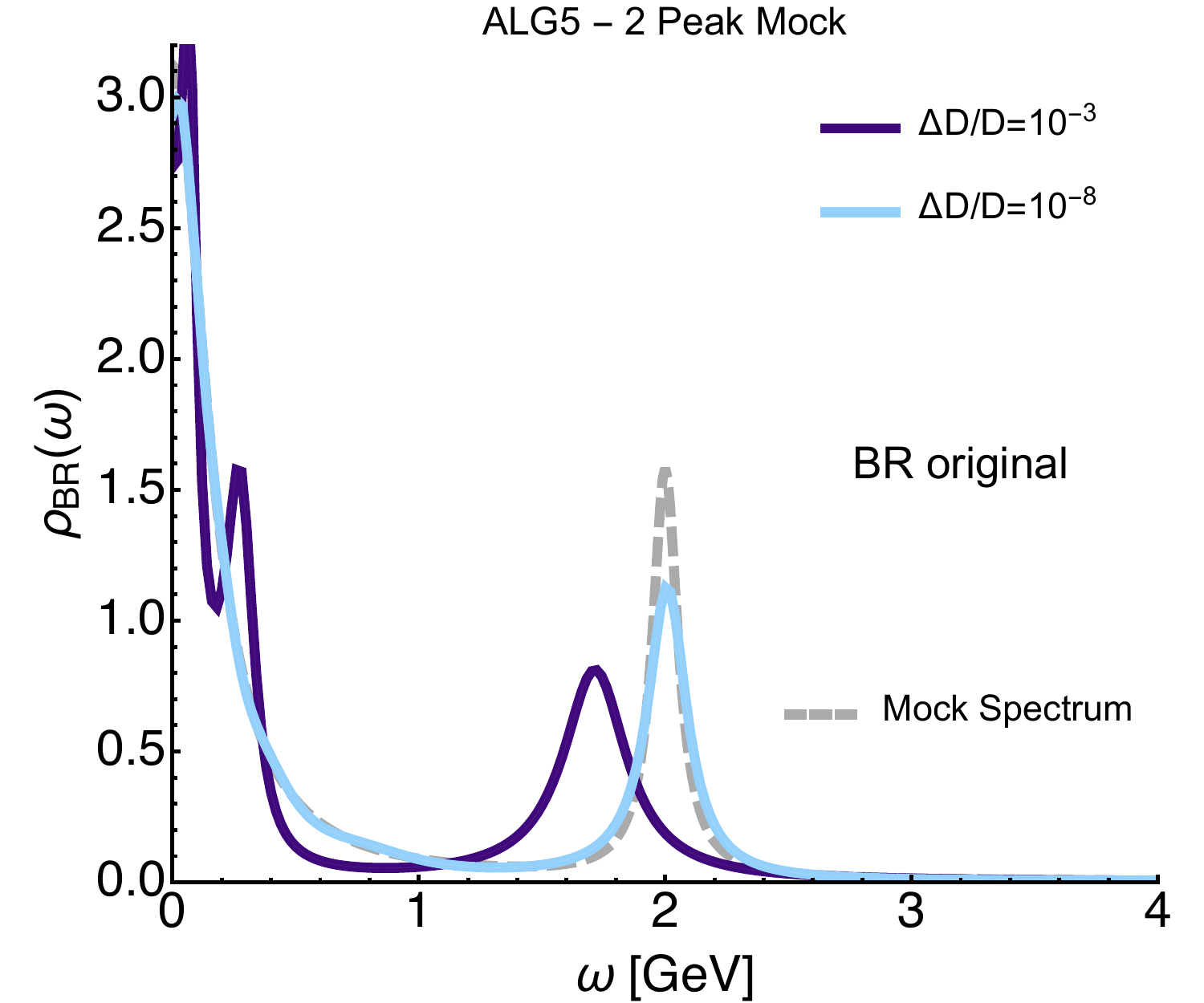}
\caption{Mock data analysis of reconstruction reliability for a
  two-peak scenario (grey dashed). Both the results for the smooth BR
  (top) as well as the original BR (bottom) method are shown. The dark
  curve denotes realistic errors $\Delta D/D=10^{-3}$, while the light
  curve the close to optimal result for $\Delta D/D=10^{-8}$. The
  smooth BR method unambiguously shows only two features and is devoid
  of ringing but it approaches the Bayesian continuum limit more
  slowly than the original BR.}\label{Fig:SpecRecMockTwoPeak2Comp}
\end{figure}

\begin{figure}[t]
\includegraphics[scale=0.5,trim= 0 0 0 0.5cm, clip=true]{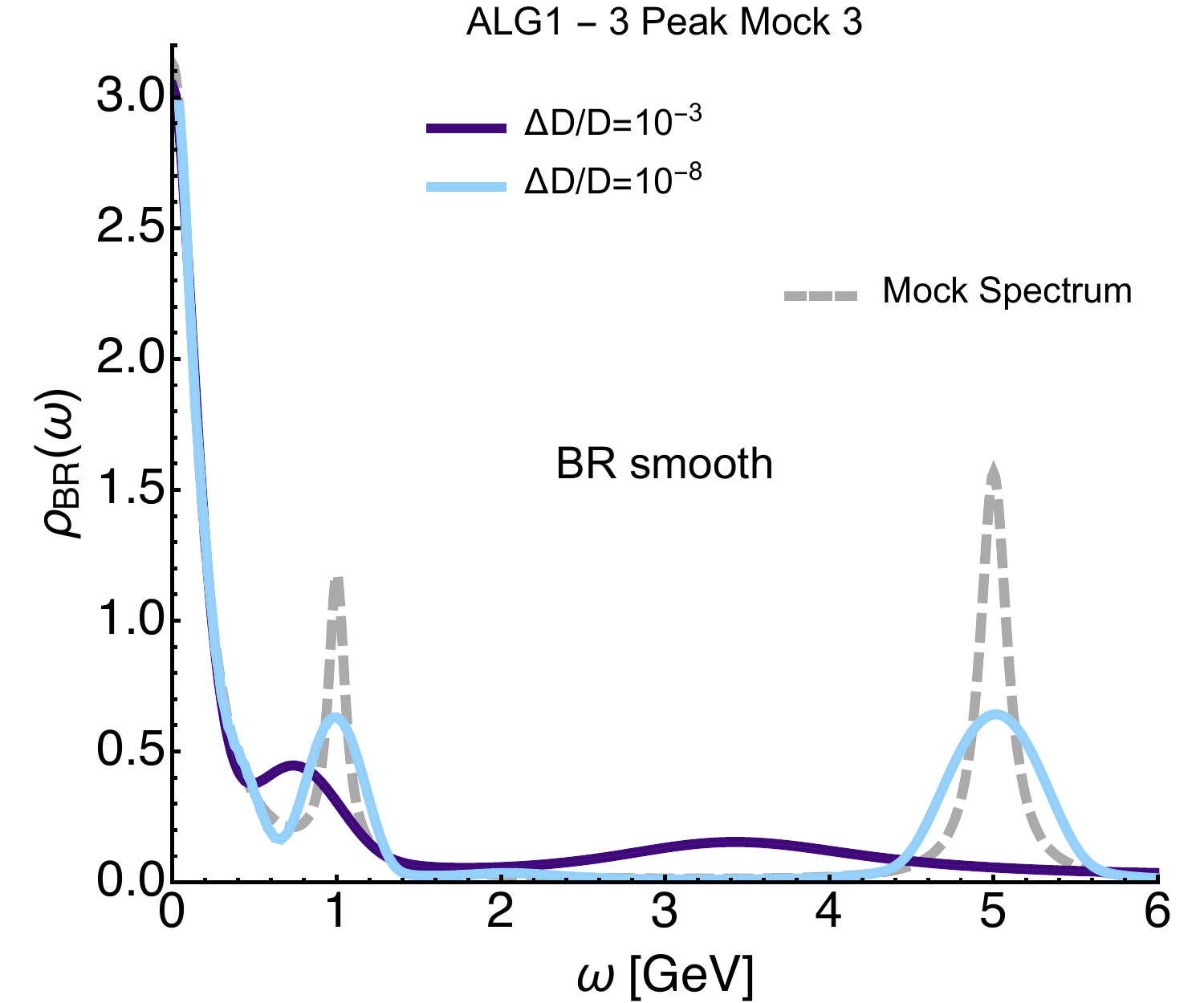} \hspace{1cm}
\includegraphics[scale=0.5,trim= 0 0 0 0.5cm, clip=true]{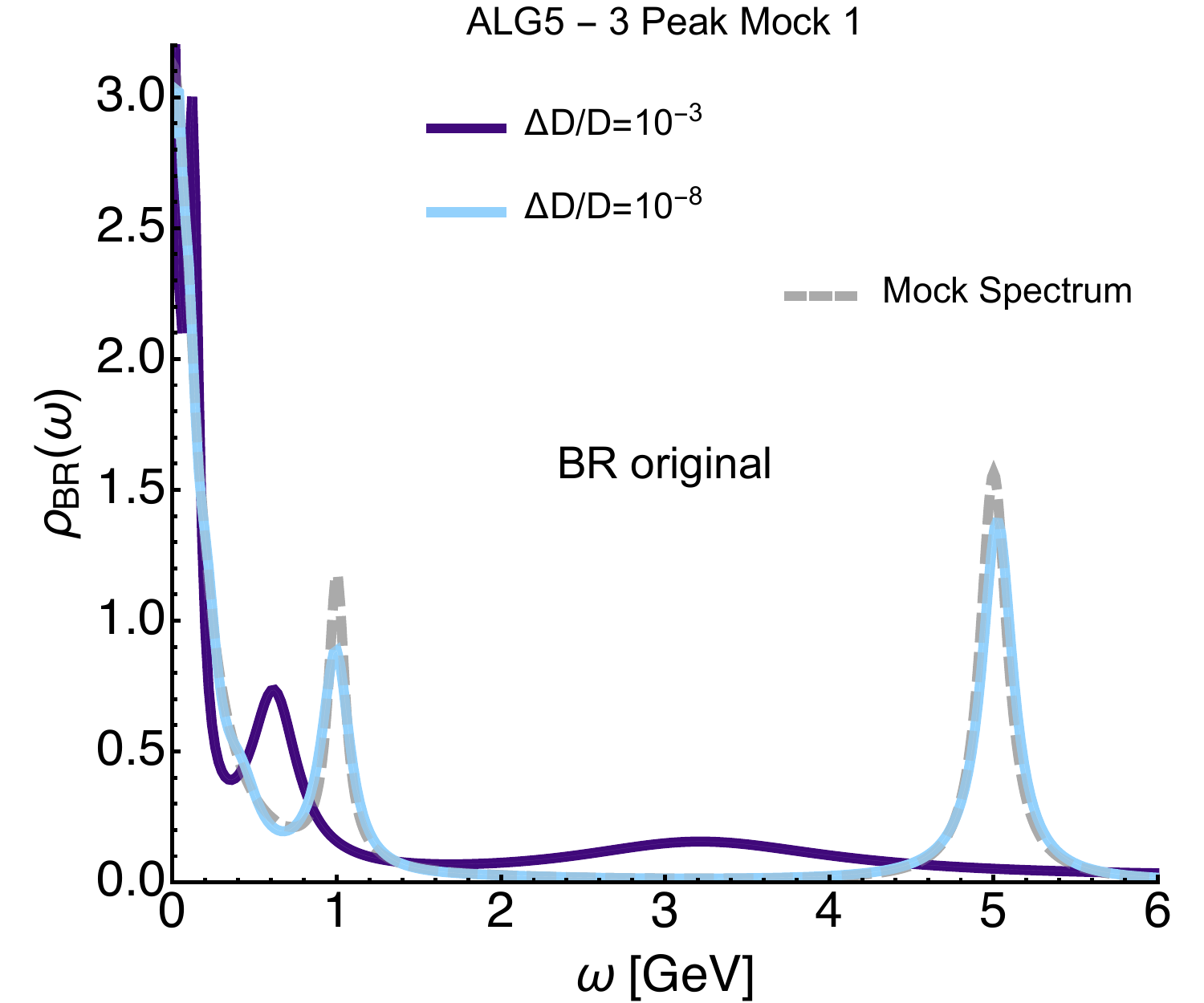}
\caption{Mock data analysis of reconstruction reliability for a
  three-peak scenario (grey dashed) with two peaks closely positioned
  off the origin.Both the results for the smooth BR (top) as well as
  the original BR (bottom) method are shown.The dark curve denotes
  realistic errors $\Delta D/D=10^{-3}$, while the light curve the
  close to optimal result for
  $\Delta D/D=10^{-8}$.}\label{Fig:SpecRecMockThreePeak2Comp}
\end{figure}

In the following we compute the Euclidean frequency correlator $S_4$
according to Eq.~\eqref{spectral} based on mock spectra with different
peak contents. This continuous function is then evaluated at
$N^{\rm mock}_{\rm data}=128$ discretely spaced imaginary frequencies
between $\omega_p\in[0,8]$GeV. This ideal data is fed to the
reconstruction algorithm undistorted but is assigned a constant
relative error $\Delta D/D$. Increasing the number of data points
beyond 128 has not shown significant improvements down to
$\Delta D/D=10^{-8}$.

By successively lowering $\Delta D/D$ we have investigated the approach to the Bayesian continuum limit for fixed $N_{\rm data}$ as explicitly shown in App.\ref{sec:app1}. In the following we will showcase only two of these reconstructions for each mock spectrum, which are relevant for the discussion. One is the best possible outcome using a particular Bayesian implementation, i.e. using a very small $\Delta D/D=10^{-8}$. The other is the realistic case $\Delta D/D=10^{-3}$, which corresponds to the precision easily obtainable in realistic Dyson-Schwinger computations \cite{Welzbacher16}.

As a first step let us have a look at a two-peak scenario with one structure located at the origin and a second at $\omega=2$GeV, as shown by the grey curve in Fig.~\ref{Fig:SpecRecMockTwoPeak2Comp}. The two solid curves of different brightness show the reconstructions either based on the smoothed BR method (top) or its original implementation (bottom). The darker curve corresponds to the error $\Delta D/D=10^{-3}$ and the lightest color to $\Delta D/D=10^{-8}$. The light colored bands around the curves refers to the corresponding dependence on the choice of the prior.

\begin{figure}[t]
\includegraphics[scale=0.5,trim= 0 0 0 0.5cm, clip=true]{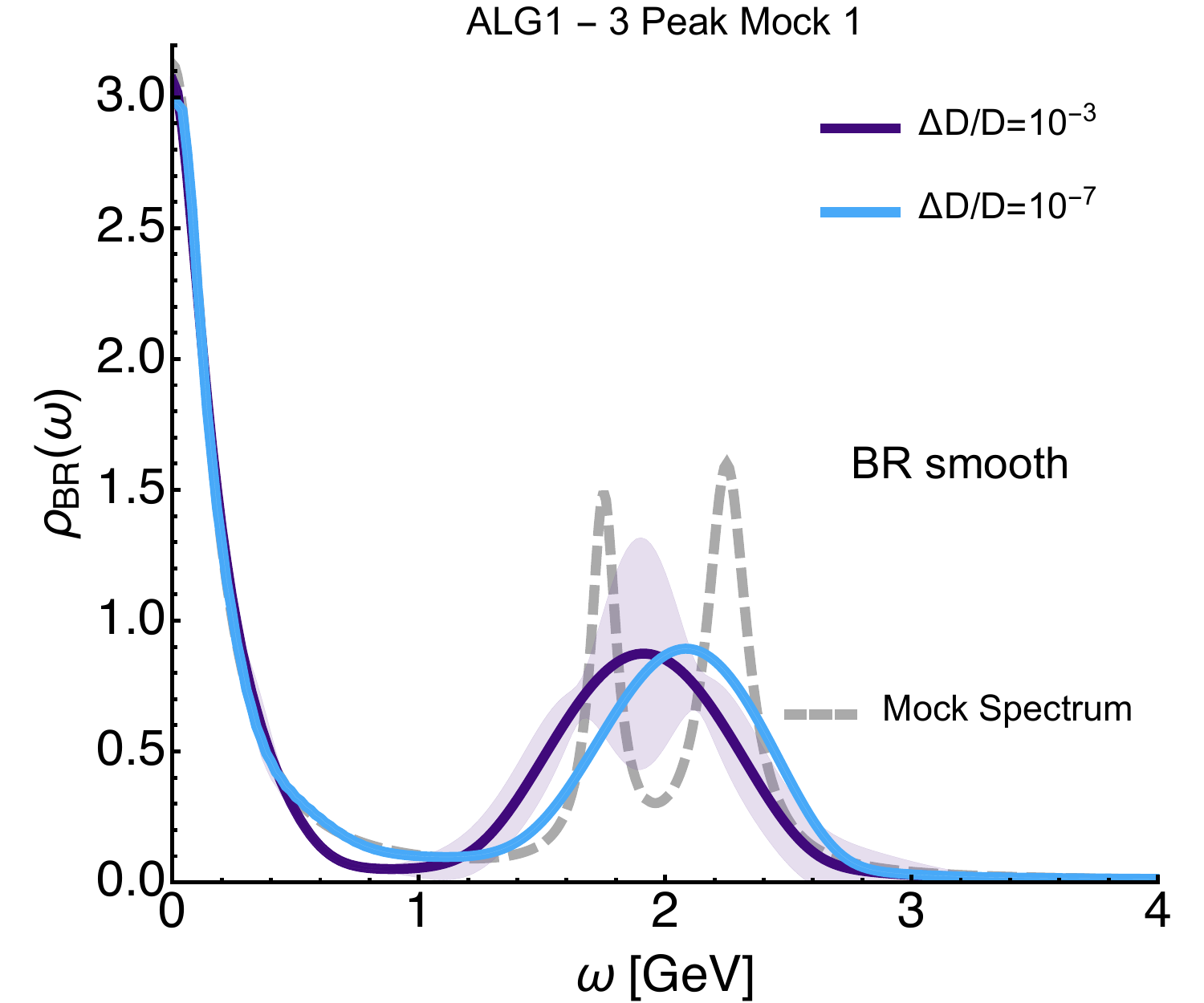} \hspace{1cm}
\includegraphics[scale=0.5,trim= 0 0 0 0.5cm, clip=true]{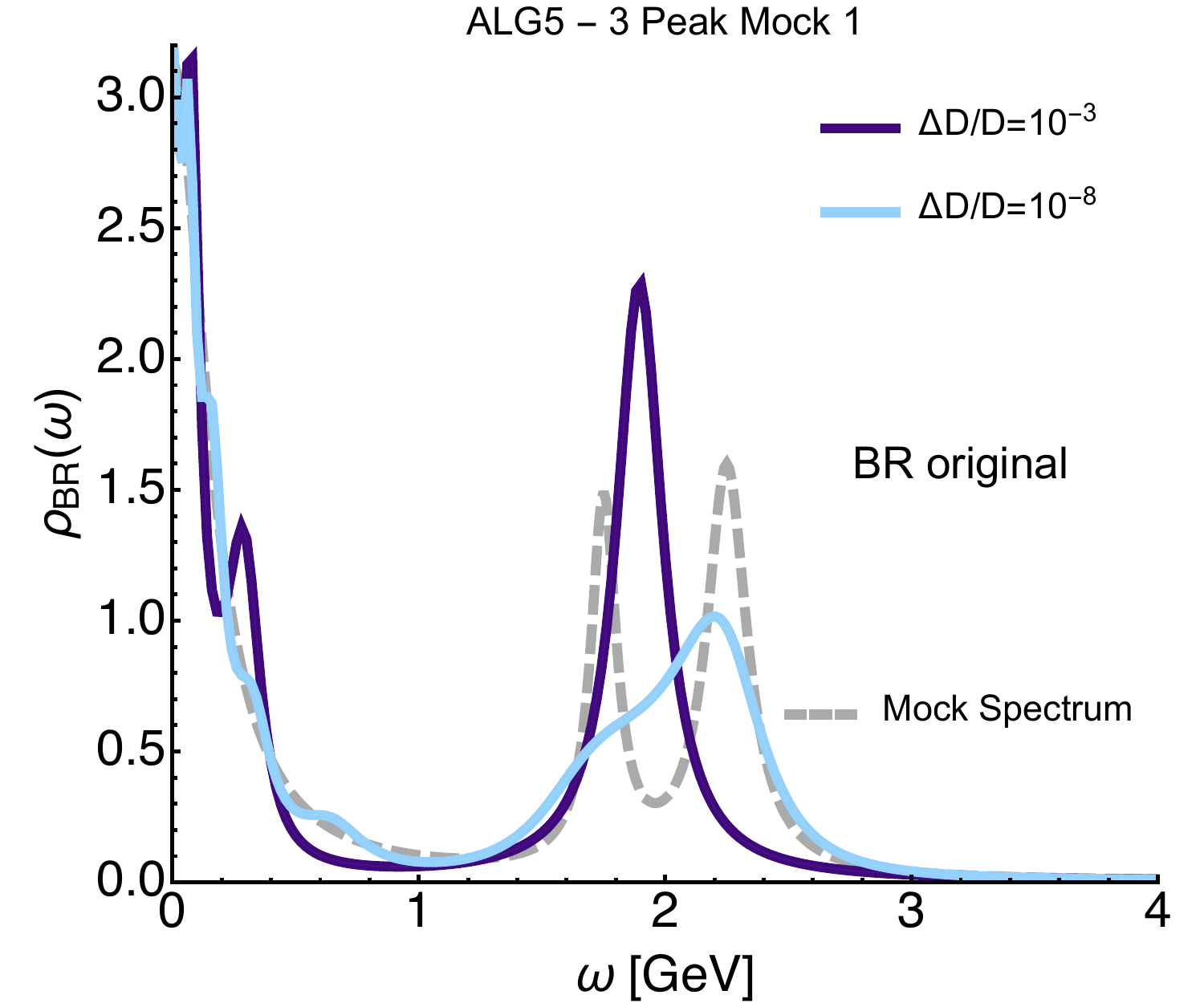}
\caption{Mock data analysis of reconstruction reliability for a three-peak scenario (grey dashed) with two peaks closely positioned off the origin. Both the result for the smooth BR (top) as well as the original BR (bottom) method are shown.The dark curve denotes realistic errors $\Delta D/D=10^{-3}$, while the light curve the close to optimal result for $\Delta D/D=10^{-7}$ or $\Delta D/D=10^{-8}$.}\label{Fig:SpecRecMockThreePeak1Comp}
\end{figure}

Lowering the errors consistently improves the reconstruction of both the peak positions and widths. For $\Delta D/D\leq10^{-3}$ the peak position of the second peak is accurate within $15\%$ for the original BR method, while the  lowest structure is already captured satisfactorily. Note that for the smoothed regulator it is more difficult to reproduce the correct width of the higher lying peak at the same $\Delta D/D$ than for the original BR method. On the other hand the reconstruction based on $S^{\rm smooth}_{\rm BR}$ is free of any of the numerical ringing that appears at intermediate $\Delta D/D$ in the original BR method. Our strategy is proven valid, the smoothed BR reconstruction unambiguously tells us about the presence of two peaked feature and we can use the standard BR method to more accurately extract their properties.

Another possible scenario is the split of one of the peaks into two structures, in particular the emergence of an additional structure close to the origin. If the peak at large frequencies is well separated from these, then again neither the conventional nor the smooth BR method are challenged in identifying the three structures as seen by the results of Fig.~\ref{Fig:SpecRecMockThreePeak2Comp}.

A more difficult scenario for any Bayesian reconstruction arises in the presence of an additional peak positioned closely to the one off the origin. Here close is understood as a distance which is comparable to the width of the peaks, as shown in Fig. \ref{Fig:SpecRecMockThreePeak1Comp} (grey dashed).

Both methods struggle to identify the split between the two peaks even at optimal conditions $\Delta D/D=10^{-7}$ or $\Delta D/D=10^{-8}$. On the other hand it is important to note that the reconstructions of the two different methods show qualitatively different behavior in contrast to all previous scenarios. While before the shape of the peaks in both the conventional and smooth BR method agreed, here we see that at $\Delta D/D=10^{-8}$ the original BR method shows a distorted peak.

We conclude that the combination of the original and smooth BR method is promising for the investigation of quark spectral functions. The latter promises, given small enough errors, to determine the number of peaked features present in the data. The former on the other hand, while susceptible to ringing at realistic $\Delta D/D\sim10^{-3}$ is capable of reproducing the actual peak properties more accurately based on the same quality of data.

\section{Results} \label{sec:results}

In this section we apply the previously tested Bayesian approaches to actual correlator data $S_4$ obtained from Dyson-Schwinger 
computations. We first re-analyse the rainbow-ladder model approach described at the end of section \ref{sec:truncation}. In this 
simple truncation the determination of the correlator is computationally cheap and numerical errors are easily reduced to 
$\Delta D/D<10^{-3}$. Subsequently we discuss our main results based on the truncation scheme with $N_f=2+1$ unquenched light 
flavors, back-coupled to the Yang-Mills sector.

\subsection{Rainbow ladder truncation}

\subsubsection{Chiral limit}

\begin{figure}[t]
\includegraphics[scale=0.5,trim= 0 0 0 0.55cm, clip=true]{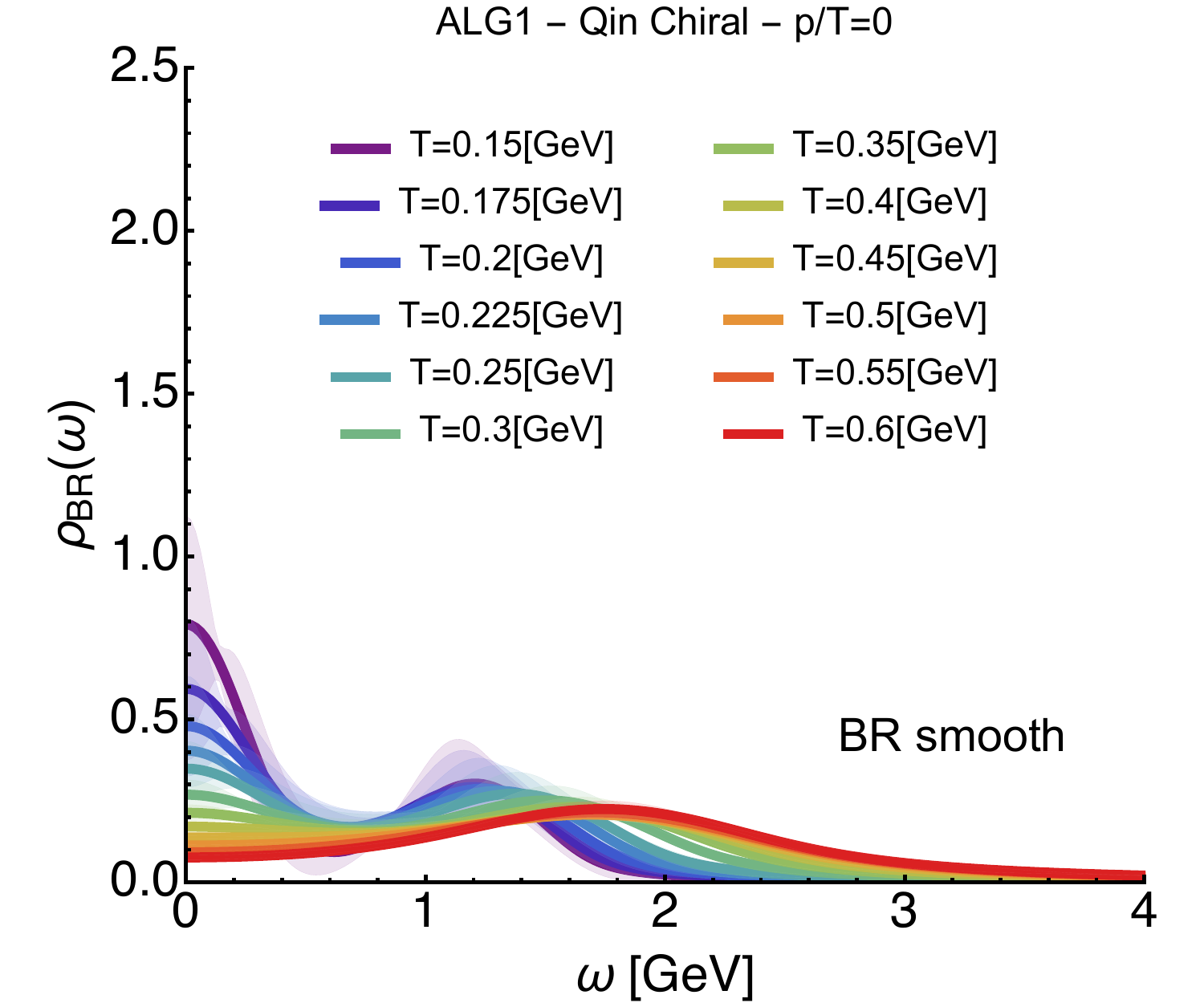}
\includegraphics[scale=0.5,trim= 0 0 0 0.55cm, clip=true]{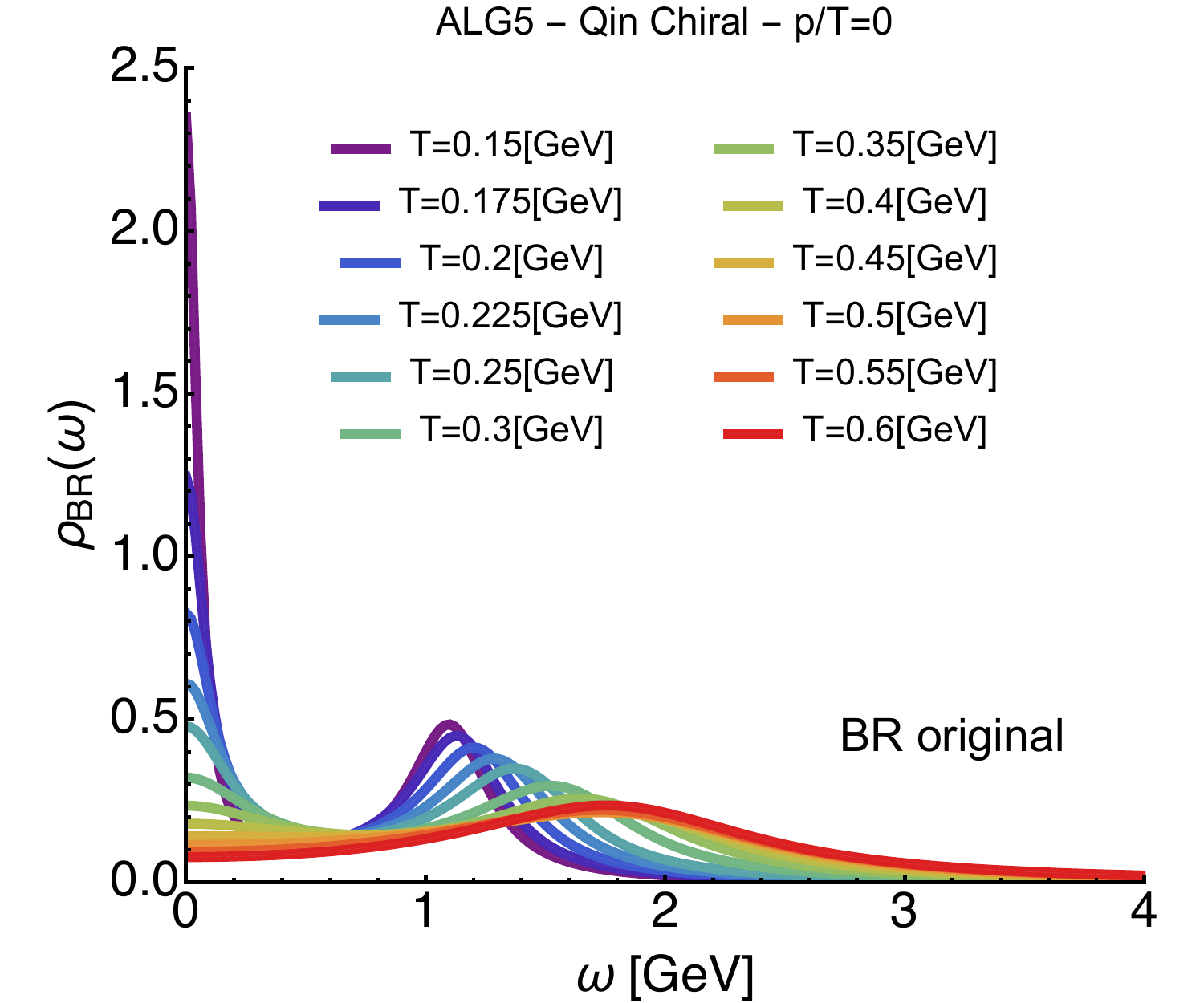}
\caption{Zero momentum reconstruction of the quark spectral function in the chiral limit at finite temperature using the smooth BR (top) and the original BR (bottom) method. Both methods show a clear indication for the presence of two peak structures at low temperature. The structure at the origin weakens with increasing $T$ and as does the higher lying peak, while also shifting.}\label{Fig:SpecRecQinChiralDiffTZerop}
\end{figure}

Let us first discuss the case of the chiral limit, which generates a second order phase transition
at a critical temperature of $T_c = 0.142\,$GeV. We have evaluated the corresponding $S_4$ along Matsubara frequencies at twelve temperatures above $T_c$ in the range of $T\in[0.15,0.60]\,$GeV. Due to the used cutoff $\Lambda = 100$GeV and the discrete nature of the Matsubara frequencies this corresponds to $N_{\rm data}\in[106,27]$. The correlator computations have been checked to carry a numerical error of less than $\Delta D/D\leq10^{-3}$, so that we can assign a corresponding diagonal correlation matrix to it. We deploy the conventional and smooth BR method with a frequency discretization of $N_\omega=1000$ in the interval $\omega\in[0,20]$. Note that even at low temperatures the reconstructions converged swiftly and do not show any signs of spiky defects indicating the presence of negative spectral contributions. The default model is set to to a constant $m(\omega)=m_0$ and we carry out the reconstruction with the different choices $m_0=\{0.1,0.5,1.0,5.0,10\}$. The variance in the outcome is taken as the basis for the error bands depicted in the subsequently shown plots. 

\begin{figure}[b]
\includegraphics[scale=0.5,trim= 0 0 0 0.55cm, clip=true]{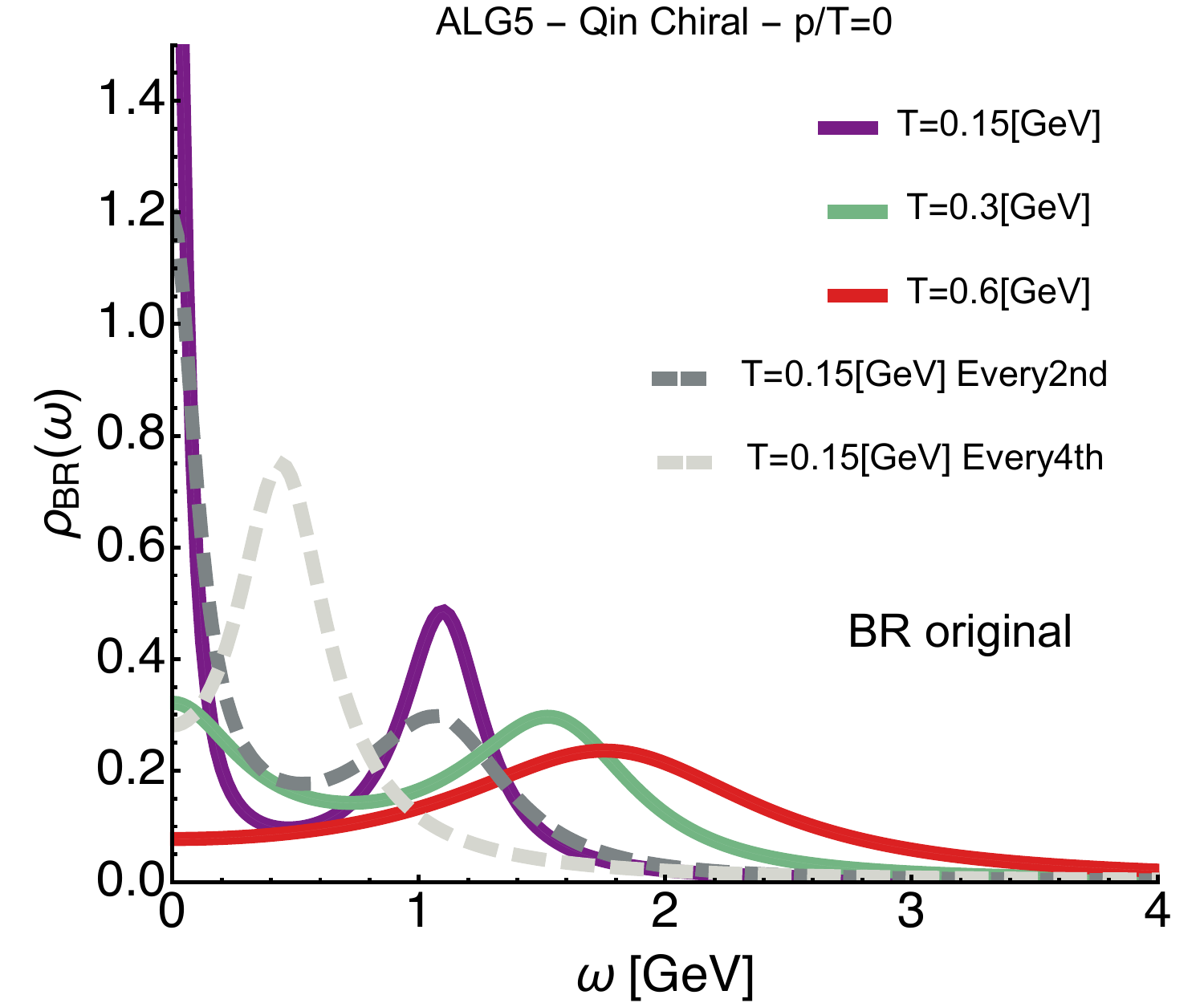}
\caption{Test of the reliability of the reconstruction at different temperatures based on the manually sparsened correlator at $T=0.15\,$GeV. Using only every second data point of the original $N_{\rm data}=106$ lead to diminished peak heights (dark grey) but does not influence peak positions. When sparsened to every fourth the Bayesian method does not recover the peaks satisfactorily any more (light grey). Nonetheless clear in-medium modification at $T>0.15$ GeV is observed.}\label{Fig:SpecRecQinChiralReconstructedCorr}
\end{figure} 

In Fig.~\ref{Fig:SpecRecQinChiralDiffTZerop} we present the zero momentum reconstruction of the quark spectral function for different temperatures.
Note that since we explicitly use the symmetry of the spectral function in Eq.~\eqref{eq:spectralsym} the position of the peak at positive frequency $\omega_+$ is mirrored exactly in the negative frequency domain. In the following it is therefore sufficient to always discuss the positive frequency band only.
In the top panel the outcome of the smooth BR method is shown, while the bottom panel corresponds to the conventional one.  We learn that at low temperatures at least up $T\lesssim0.25\,$GeV two well defined peak structures exist. One is located at the origin and one above $1\,$GeV. With increasing temperature the height of the low lying peak decreases continuously and seems to asymptote to a finite value. The higher lying peak shows a clear tendency to move to higher frequencies, while broadening at the same time.

The reconstructions at different temperatures are based on dataset with different $N_{\rm data}$. Thus before we continue to a more quantitative inspection of the spectra we need to make sure that we can disentangle the actual in-medium modification from the effects of a reduction of data points. To this end we perform the following test: we take the lowest temperature dataset with $N_{\rm data}=106$ in imaginary frequencies and sparsen it by hand. Due to the discrete nature of the Matsubara frequencies this corresponds to a situation, where the $T=0.15\,$GeV spectrum would be encoded in a correlator evaluated at $T=0.3\,$GeV or $T=0.6\,$GeV respectively. In Euclidean time this corresponds to constructing the reconstructed correlator \cite{Ding:2012sp}. From similar tests performed in previous Bayesian studies we expect that with decreasing number of data points the resolution of peaks diminishes, eventually inducing changes in the position and width of the reconstructed features. 
  
\begin{figure}[t]
\includegraphics[scale=0.5,trim= 0 0 0 0.55cm, clip=true]{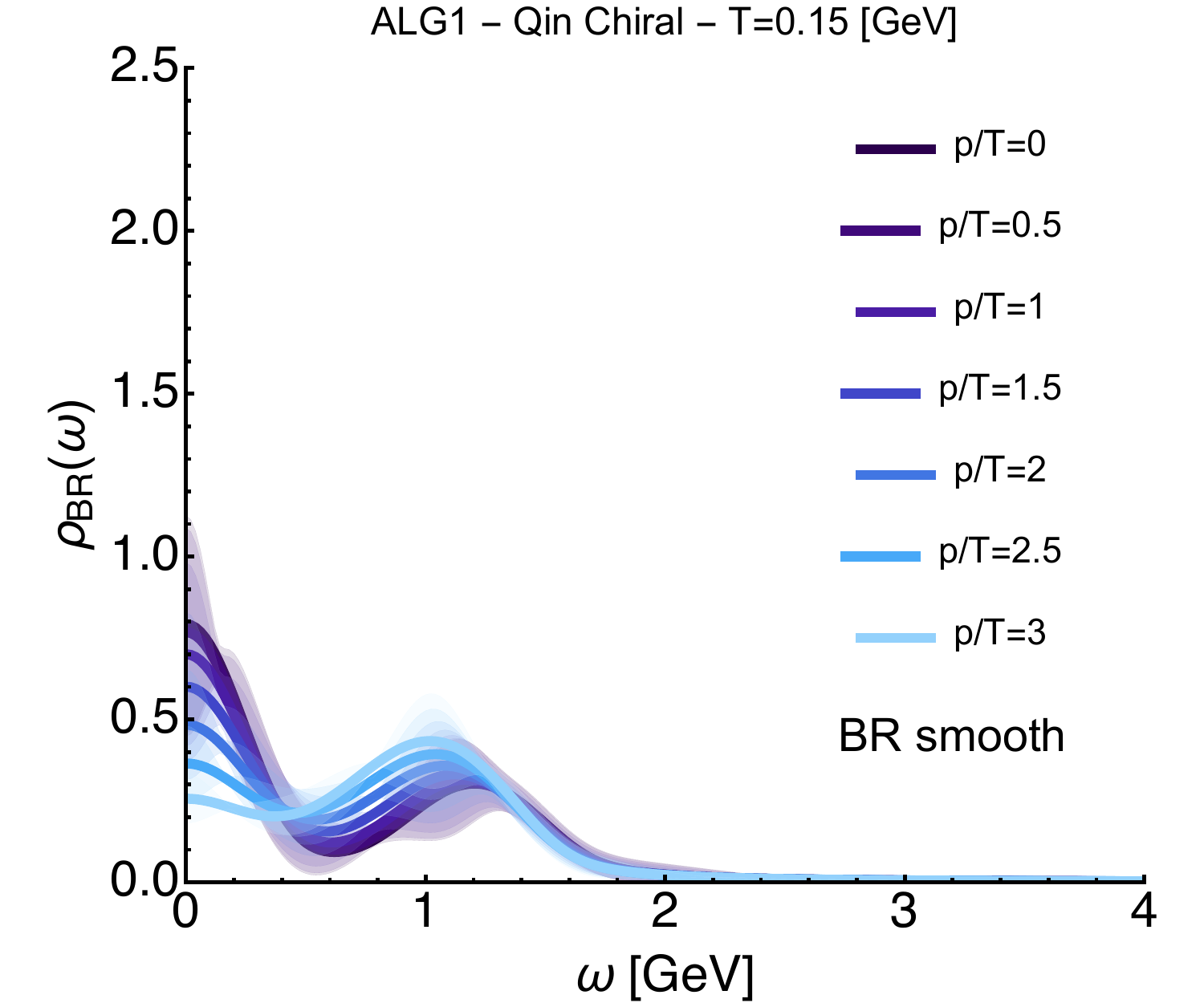} \hspace{1cm}
\includegraphics[scale=0.5,trim= 0 0 0 0.55cm, clip=true]{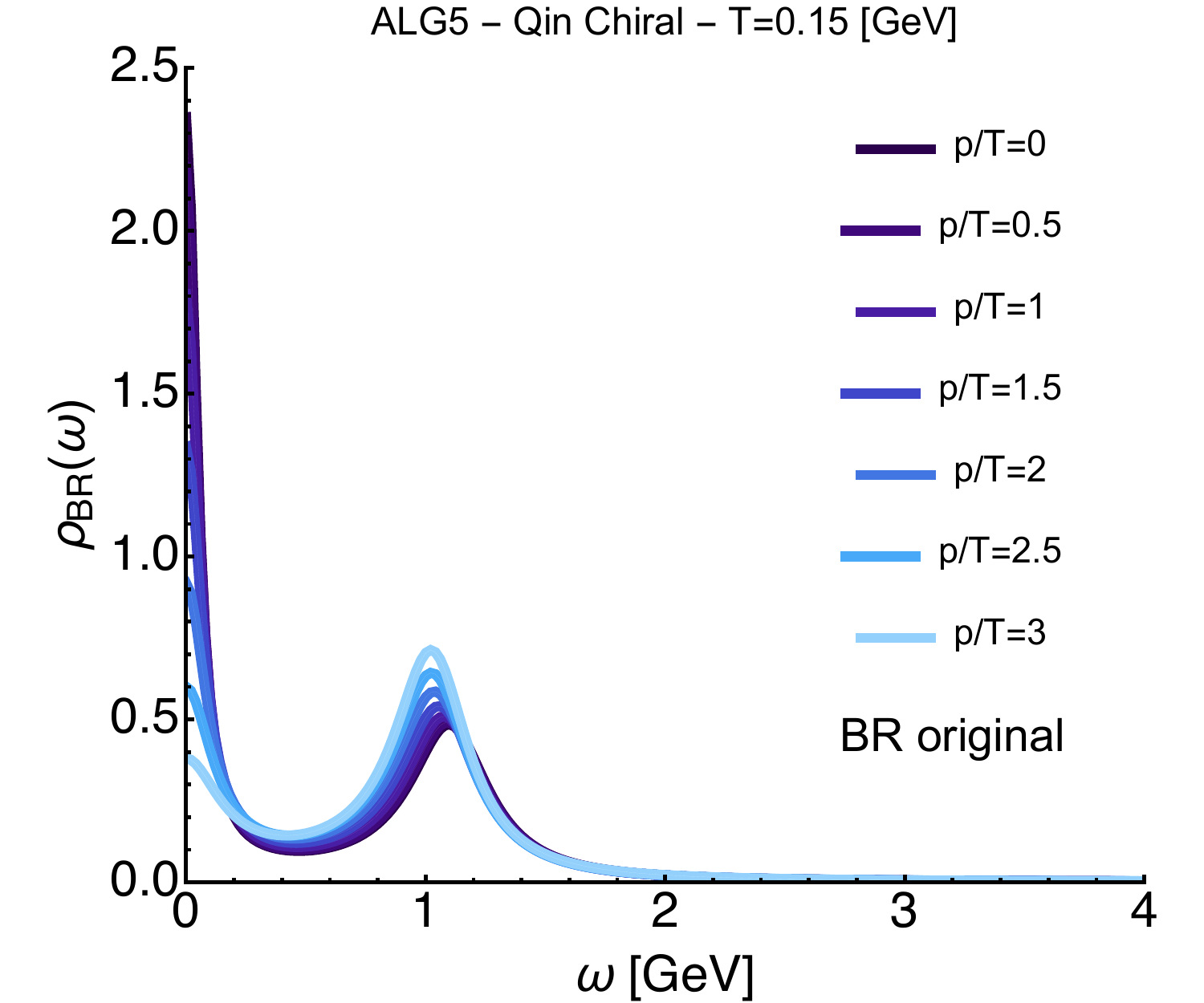}
\caption{Low temperature $T=0.15\,$GeV reconstructions of the quark spectral function at different momenta using the smooth BR (top) and the original BR (bottom) method . We find clear indications of a two peak structure. Both peak heights decrease for larger momenta, the one close to the origin decreases however much more rapidly. A shift of the second peak to higher frequencies is observed.}\label{Fig:SpecRecQinChiralTestDiffpAtLowT}
\end{figure}

And indeed, as shown in Fig.~\ref{Fig:SpecRecQinChiralReconstructedCorr} we find that going from $N_{\rm data}=106$ to $N_{\rm data}=53$  (dark grey dashed) leads to a visible weakening of the peak structures, while their position remains unaffected. Comparing to the reconstruction from actual $T=0.3\,$GeV with the same lower number of data points (green solid) however shows clear differences. Both the diminishing of the lowest lying peak height, as well as the shifting of the second peak to higher frequencies can thus be attributed to genuine in-medium effects. Interestingly the direction of change in the peak position is opposite to that sketched in \cite{Qin:2013ufa,Gao:2014rqa}. At $T=0.6\,$GeV, i.e. $N_{\rm data}=26$ the sparsened data do not allow the reconstruction of the two peaks at all and we must assume that the reconstruction is not trustworthy at this point.
 
 \begin{figure}[t]
\includegraphics[scale=0.5,trim= 0 0 0 0.55cm, clip=true]{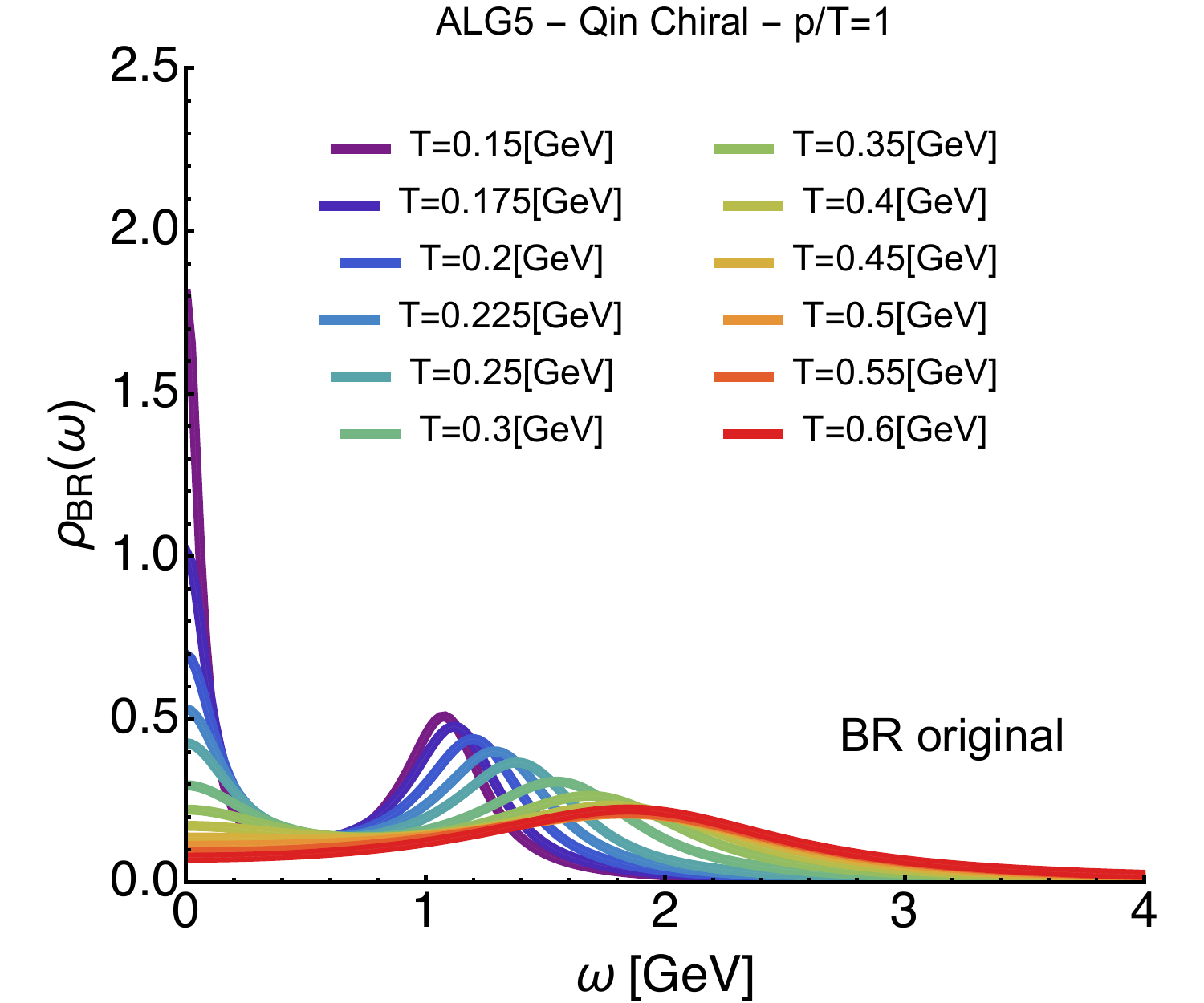}
\includegraphics[scale=0.5,trim= 0 0 0 0.55cm, clip=true]{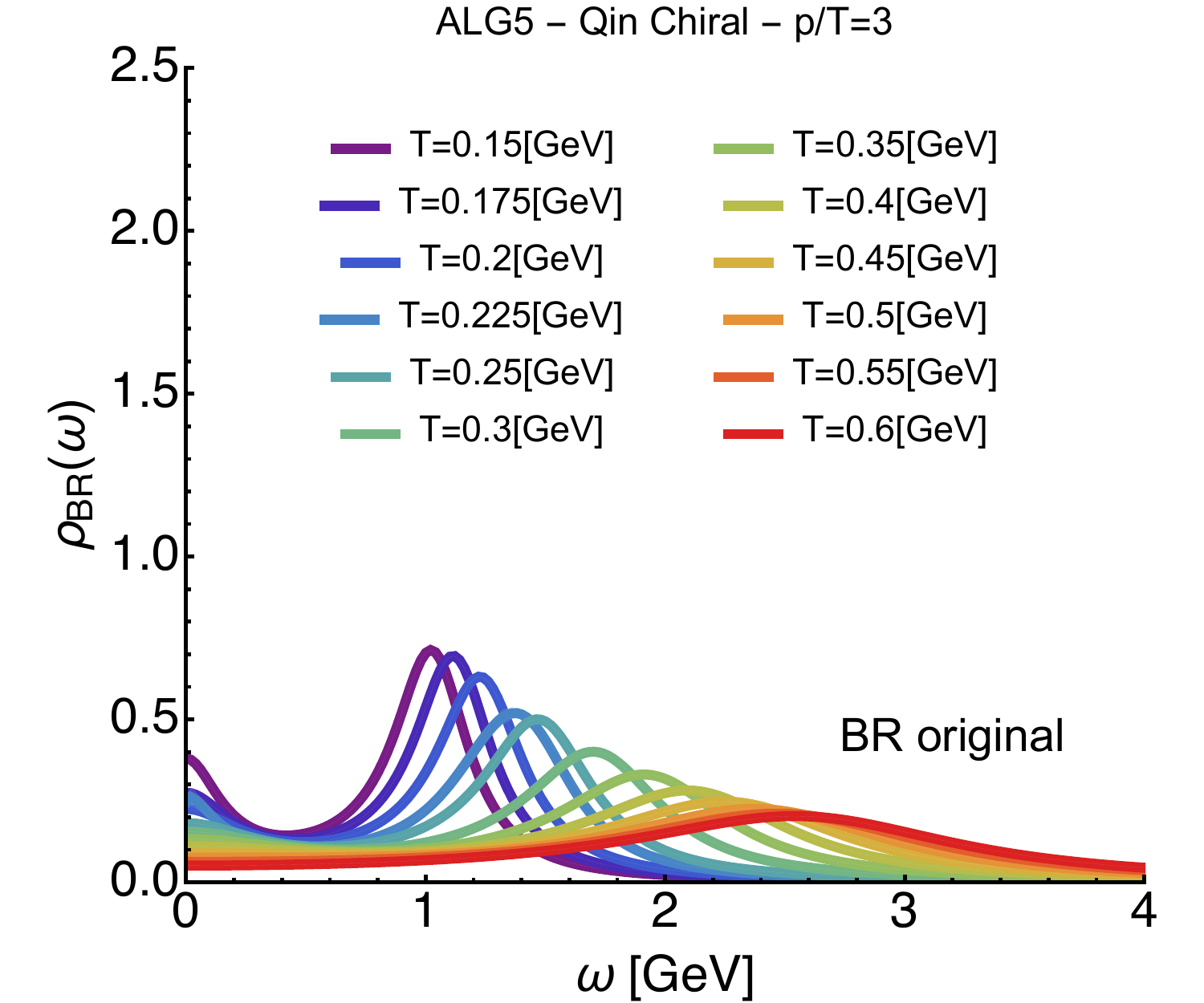}
\caption{Intermediate momentum $\vert\pp\vert/T=1$ (top) and $\vert\pp\vert/T=3$ (bottom) reconstructions of the quark spectral function at different temperatures using the original BR method. The characteristic decrease in the peak height around the origin as well as the shift and broadening of the higher lying peak structure is clearly visible.}\label{Fig:SpecRecQinChiralDiffTDiffp}
\end{figure}
 
The investigation of the effects of finite momentum on the quarks does not suffer from a similar ambiguity, since for fixed $T$ neither the number of data points nor the relative errors change. In Fig.~\ref{Fig:SpecRecQinChiralTestDiffpAtLowT} we show reconstructed spectra at $T=150\,$MeV respectively over a range of momenta $\vert\pp\vert/T=\{0,\frac{1}{2},1,\frac{3}{2},2,\frac{5}{2},3\}$.

As seen before at low temperature both reconstruction approaches unambiguously show the presence of two peaks. One is located around the origin, another one positioned close to $\omega\approx1\,$GeV. Increasing the momentum to $\vert\pp\vert/T=3$ induces changes in the spectrum that are of the same qualitative nature as those from increasing temperature. The peak at the origin decreases significantly in area, while not extending further towards higher $\omega$. The second peak diminishes much more weakly and is seen to shift to higher frequencies, as expected from the naive momentum dependence of the dispersion relation.

\begin{figure}[t]
\includegraphics[scale=0.5,trim= 0 0 0 0.55cm, clip=true]{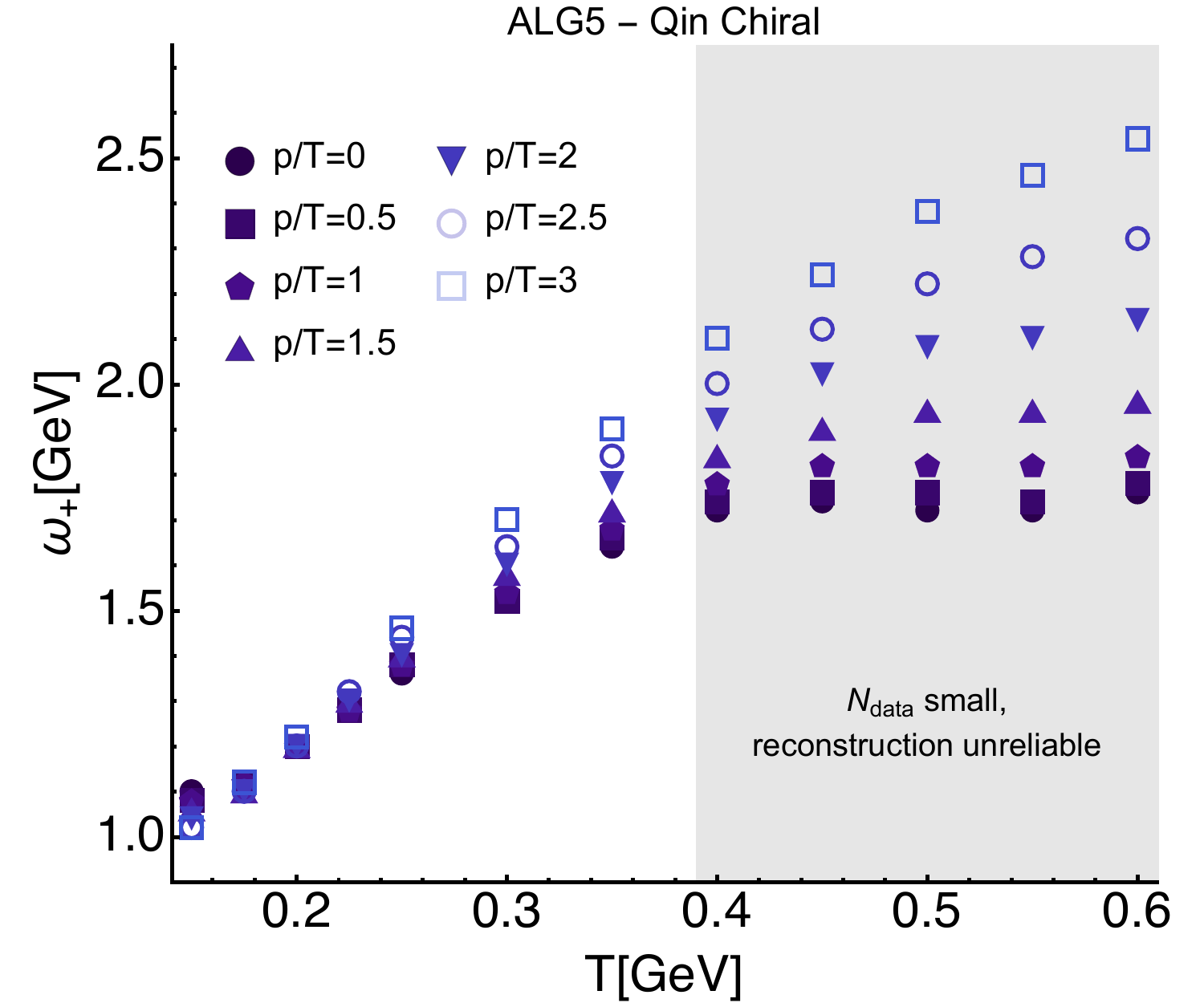}
\includegraphics[scale=0.5,trim= 0 0 0 0.55cm, clip=true]{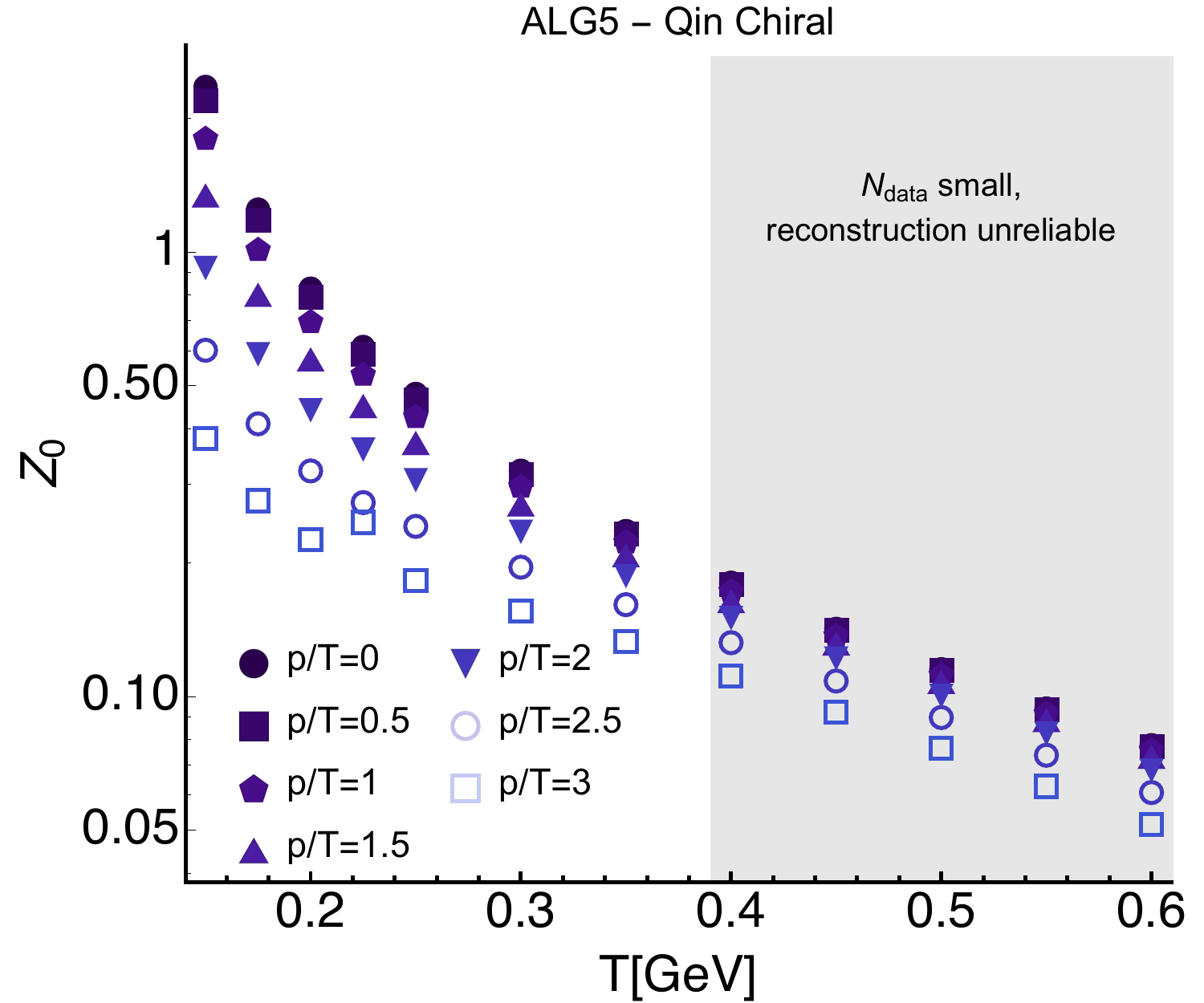}
\caption{(Top) position of the higher lying peak for different temperatures and momenta. At intermediate temperatures, where the reconstruction is reliable, we find a linear rise, which flattens off, once the number of data points becomes too small. (Bottom) Amplitude of the lowest lying peak, which we find to decrease monotonously.}\label{Fig:SpecRecQinChiralOmegaPlusZZero}
\end{figure}

For visualization purposes we present in Fig.~\ref{Fig:SpecRecQinChiralDiffTDiffp} the reconstructions at a fixed intermediate momentum $\vert\pp\vert/T=1$ (top) and $\vert\pp\vert/T=3$ (bottom) for all different temperatures investigated. All the effects on the peak around the origin, as well as the second peak at finite frequencies as discussed above are clearly visible here.

We continue with a quantitative analysis of the in-medium modification of the quark spectral features. Two quantities of interest here are the position of the higher lying peak denotes with $\omega_+$ and the height of the peak around the origin referred to as $Z_0$. Both are shown in Fig.\ref{Fig:SpecRecQinChiralOmegaPlusZZero}. For $\omega_+$ the expectation from resummed hard-thermal loop perturbation theory at small $\vert\pp\vert/T\ll 1$ is a linear dependence on the temperature $\omega_\pm^{\rm HTL}=m_T\pm \frac{\vert \pp \vert}{3}$ with thermal quark mass $m_T\propto T$. While at low temperatures and small momenta we see a rise stronger than linear, at intermediate $T$ our $\omega_+$ indeed shows a behavior compatible with a linear increase. Consistent with our conclusions from the sparsening test, for temperatures much higher than $T=0.3\,$GeV, the reconstruction becomes unreliable and at the same time we see that the linear rise abates and goes over to a constant.

\begin{figure}[b]
\includegraphics[scale=0.5,trim= 0 0 0 0.55cm, clip=true]{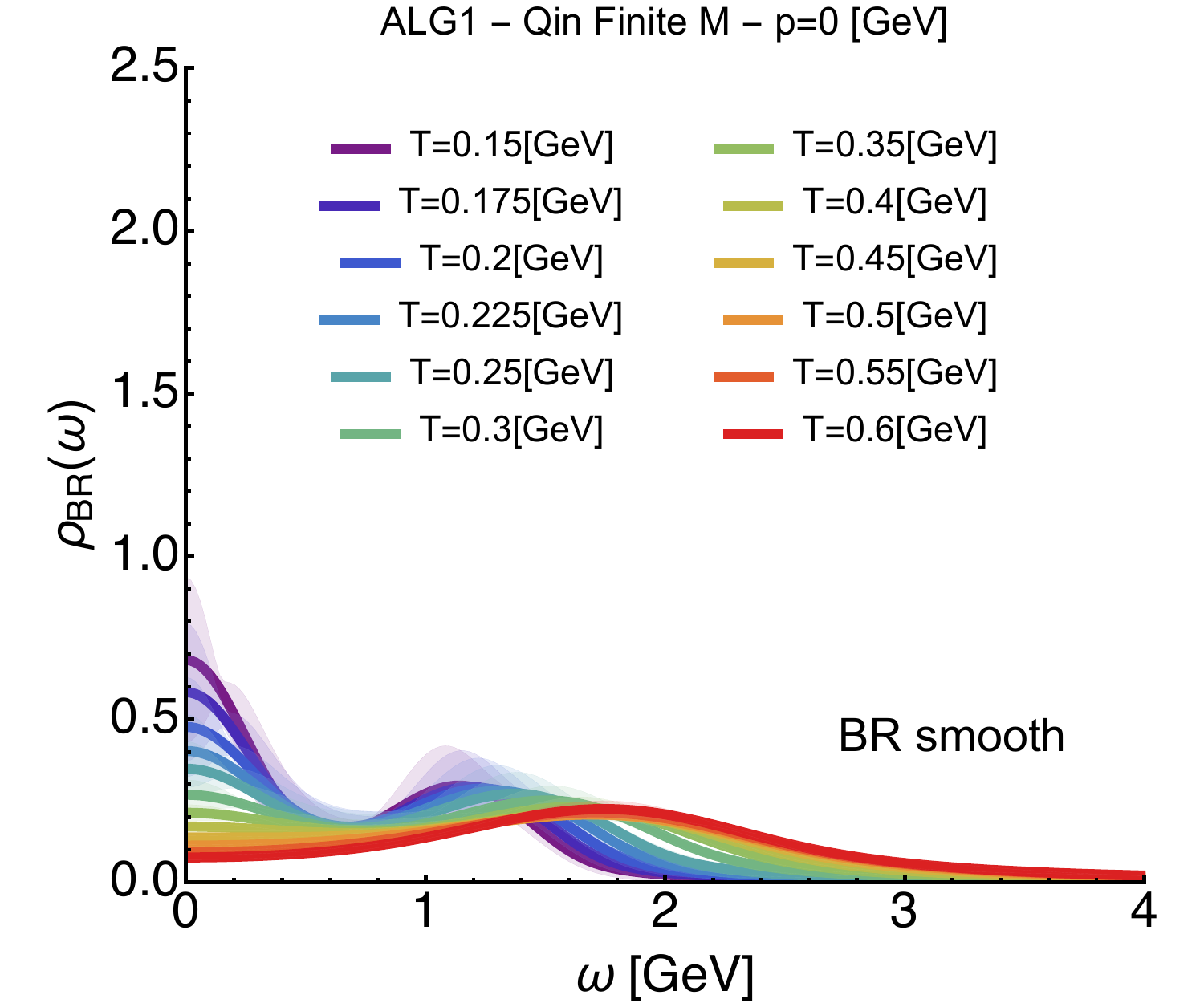} \hspace{1cm}
\includegraphics[scale=0.5,trim= 0 0 0 0.55cm, clip=true]{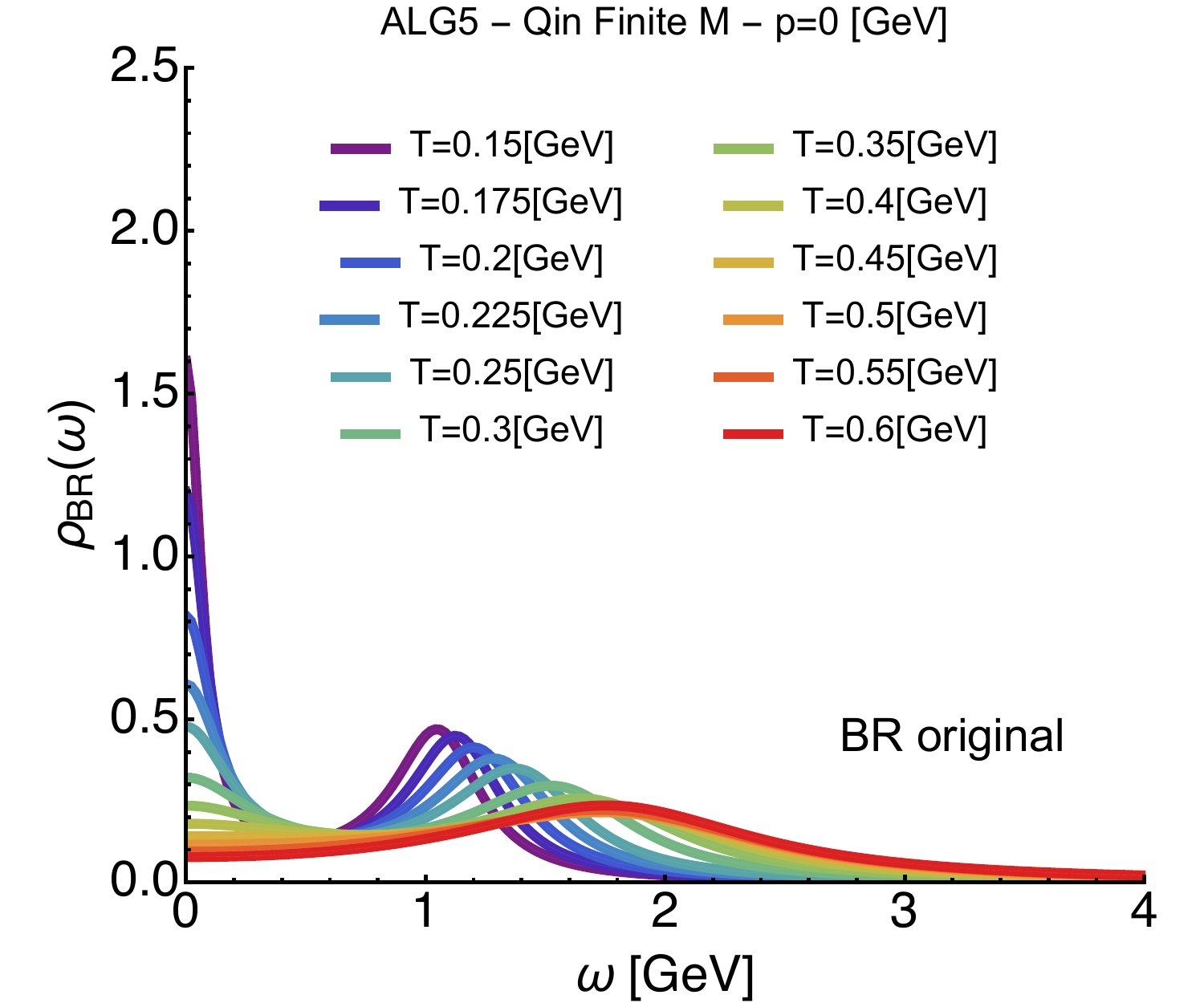}
\caption{Zero momentum reconstruction of the quark spectral function at finite mass and finite temperature using the smooth BR (top) and the original BR (bottom) method. We find qualitatively the same behavior as in the chiral case, the main difference being the height of the lowest lying peak at low temperature.}\label{Fig:SpecRecQinFiniteMass}
\end{figure}

The behavior found here is again different from that presented in \cite{Gao:2014rqa}. Firstly, we find that the peak at 
non-zero frequency $\omega_+$ monotonously moves to larger frequencies with increasing temperature in contrast with the previously reported behavior, where the function $\omega_+(T)$ shows a minimum shortly above $T_c$. The second difference is related to the height of the zero frequency peak, which in the temperature range investigated does not go to zero but stays finite. Since the high temperature regime is not reliably captured due to the sparseness of the Matsubara frequencies we cannot yet conclude whether it eventually vanishes after all.

In all quantitative statements we need to keep in mind that our
current numerical precision for the correlator data is limited to
$\Delta D/D=10^{-3}$. While we are confident, judging from the mock
data tests, that the number of peaks and the direction of changes with
temperature are correctly captured, it is fathomable that the peak
position may still change by up to $20\%$ if the errors are further
reduced. The width of the peaks carries at least the same uncertainty
at this point along the "Bayesian continuum limit".

\subsubsection{Finite mass case}

We proceed with a second set of correlators $S_4$, for which the current quark mass has been set to $m_q=3.7\,$MeV at a renormalization
point of $19\,$ GeV. Here we restrict ourselves to the $p=0$ case. The question to answer is how the spectral structures differ in contrast to the chiral case. We use the same temperature regime and discretization of the correlator data and leave the errors unchanged.

The results for the zero momentum spectral reconstructions at different temperature are given in Fig.~\ref{Fig:SpecRecQinFiniteMass}, with the smooth BR method in the top panel and the conventional one in the bottom one. Qualitatively the figures are very similar to the results in the chiral case. There exist two peaks, one at the origin and one above $1\,$GeV. The position of the second peak moves to higher frequencies as temperature increases, while the height of the lowest lying peak decreases continuously. The only visible difference is that at low temperature the height of the peak around the origin is discernibly smaller than in the chiral case.

We again checked that the changes between the outcome at different temperature is indeed attributable to in-medium effects by manually coarsening the $T=0.15\,$GeV correlator data and repeating the reconstruction with it. The results of this procedure are similar to the ones in the quenched case: the deletion of every second data point, i.e. $N_{\rm data}=56$, weakens the peak strength while leaving the peak position unaffected. 
The reconstruction based on only every fourth data point corresponding to the situation at $T=0.6\,$GeV fails to identify the encoded two peak structure and is therefore deemed not trustworthy.

\begin{figure}[t]
\includegraphics[scale=0.5,trim= 0 0 0 0.55cm, clip=true]{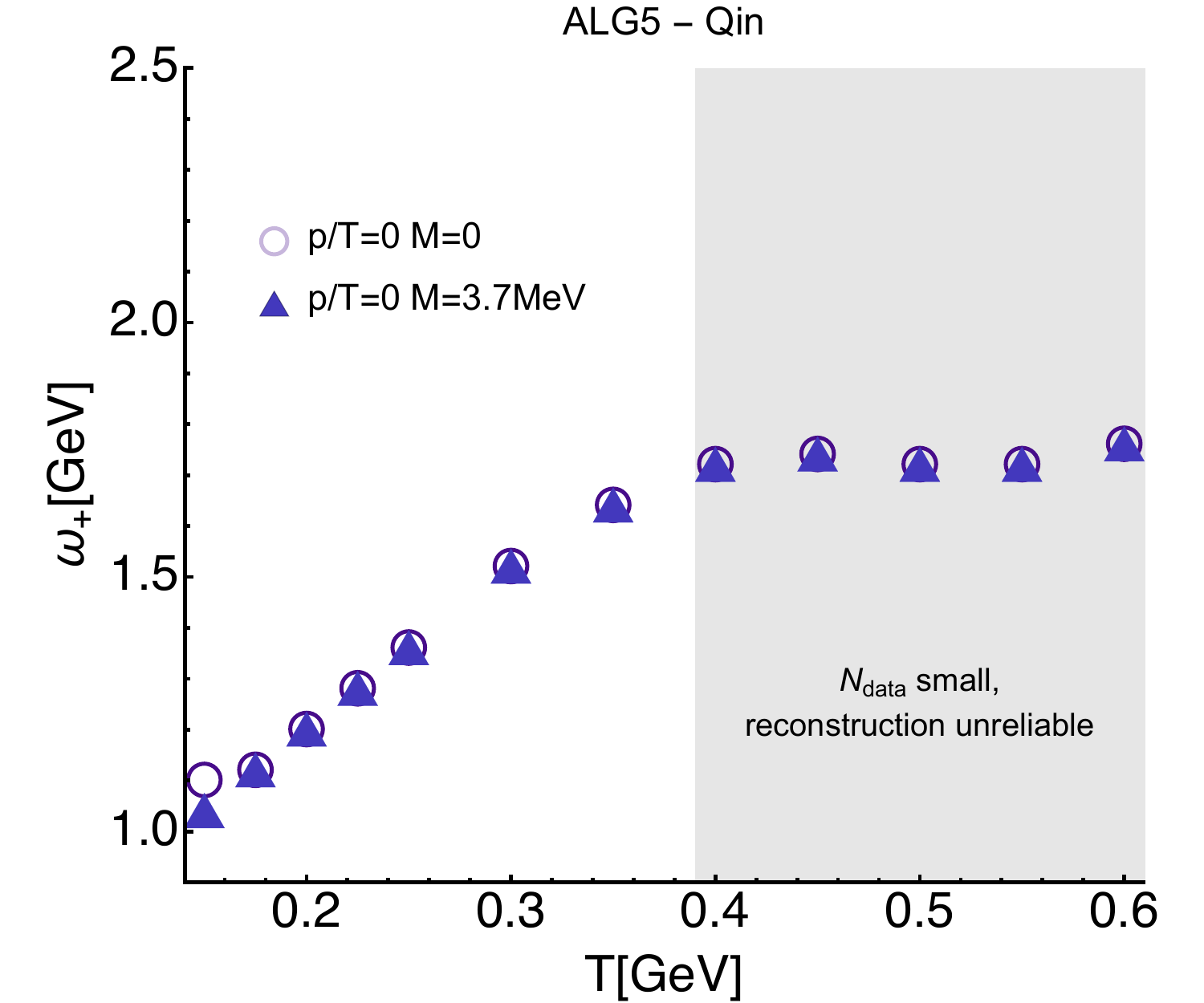}
\includegraphics[scale=0.5,trim= 0 0 0 0.55cm, clip=true]{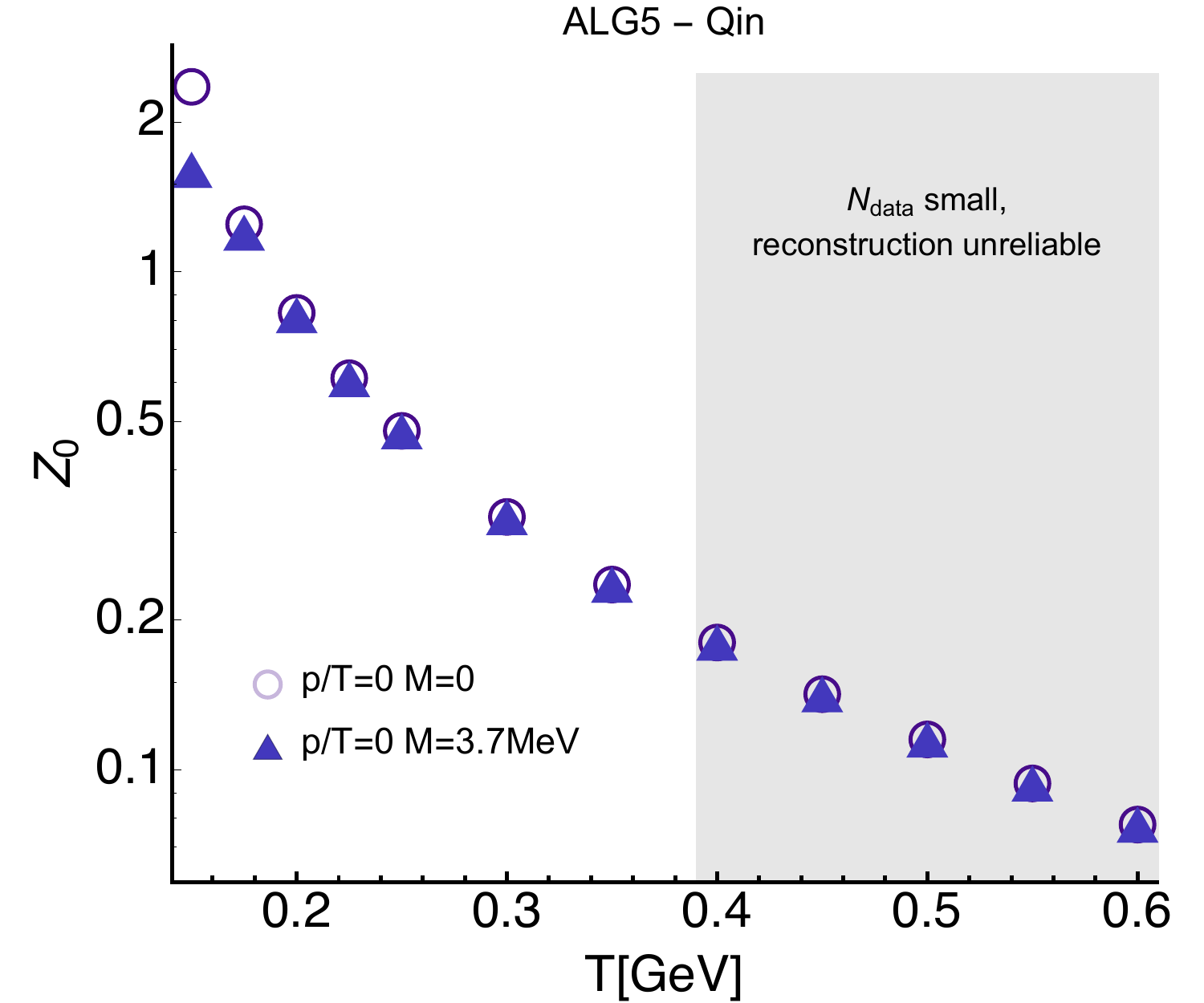}
\caption{(Top) position of the higher lying peak for different temperatures at vanishing momentum for the finite mass case (triangle) and the chiral case (circle). Except for the lowest $T$ no significant difference is observed. (Bottom) Amplitude of the lowest lying peak, which except for very low temperatures show the same behavior between the chiral and finite mass case.}\label{Fig:SpecRecQinFiniteMassOmegaPlusZZero}
\end{figure}

Just as in the chiral case, we determine the position of the second peak $\omega_+$ and the height of the peak around the origin $Z_0$, shown in top and bottom panel of Fig.~\ref{Fig:SpecRecQinFiniteMassOmegaPlusZZero} respectively\footnote{ Note that the quantity $\omega_+$ shown here does not fully coincide with the conventional thermal mass at $\vert\pp\vert=0$, as it is reconstructed  from $S_4$, instead of $S_4\pm S_s$. Since the latter makes the spectral function non-symmetric we postpone its analysis to future work.}. As is to be expected from the close resemblance of the reconstructed spectra, the values obtained do not differ markedly between the finite mass (triangle) and the chiral (circle) case. The only differences are found at low temperatures. The values of $\omega_+$ for finite mass actually show a linear behavior down to $T=0.15\,$GeV, whereas in the chiral case they deviate from a straight line at that point. $Z_0$ is smaller at low temperatures in the presence of a finite quark mass but already at $T=0.2\,$GeV no significant difference remains.

\subsection{Unquenched truncation with back-coupled $N_f=2+1$ quark flavors}
\begin{figure}[b]
\includegraphics[scale=0.5,trim= 0 0 0 0.55cm, clip=true]{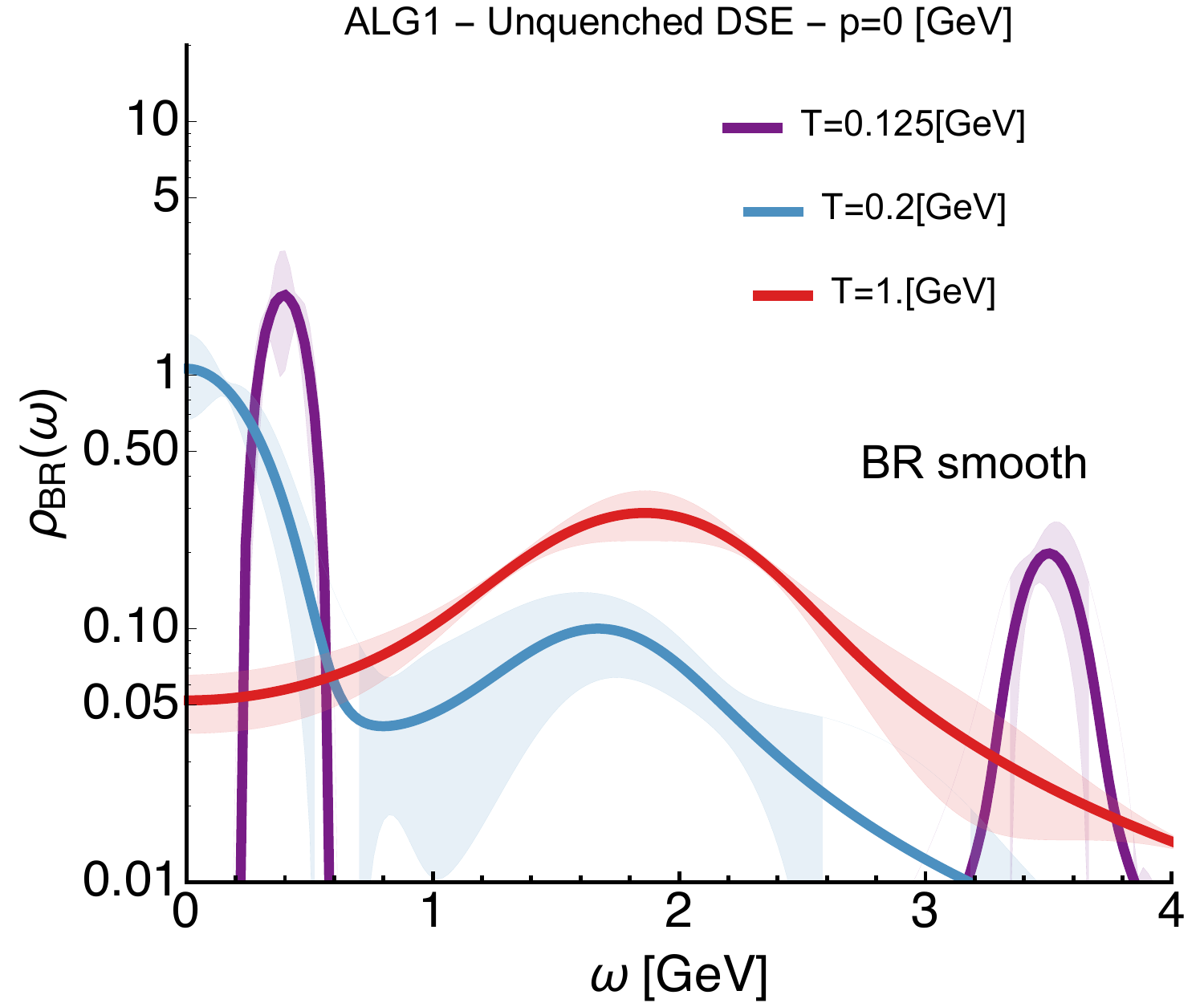} \hspace{1cm}
\includegraphics[scale=0.5,trim= 0 0 0 0.55cm, clip=true]{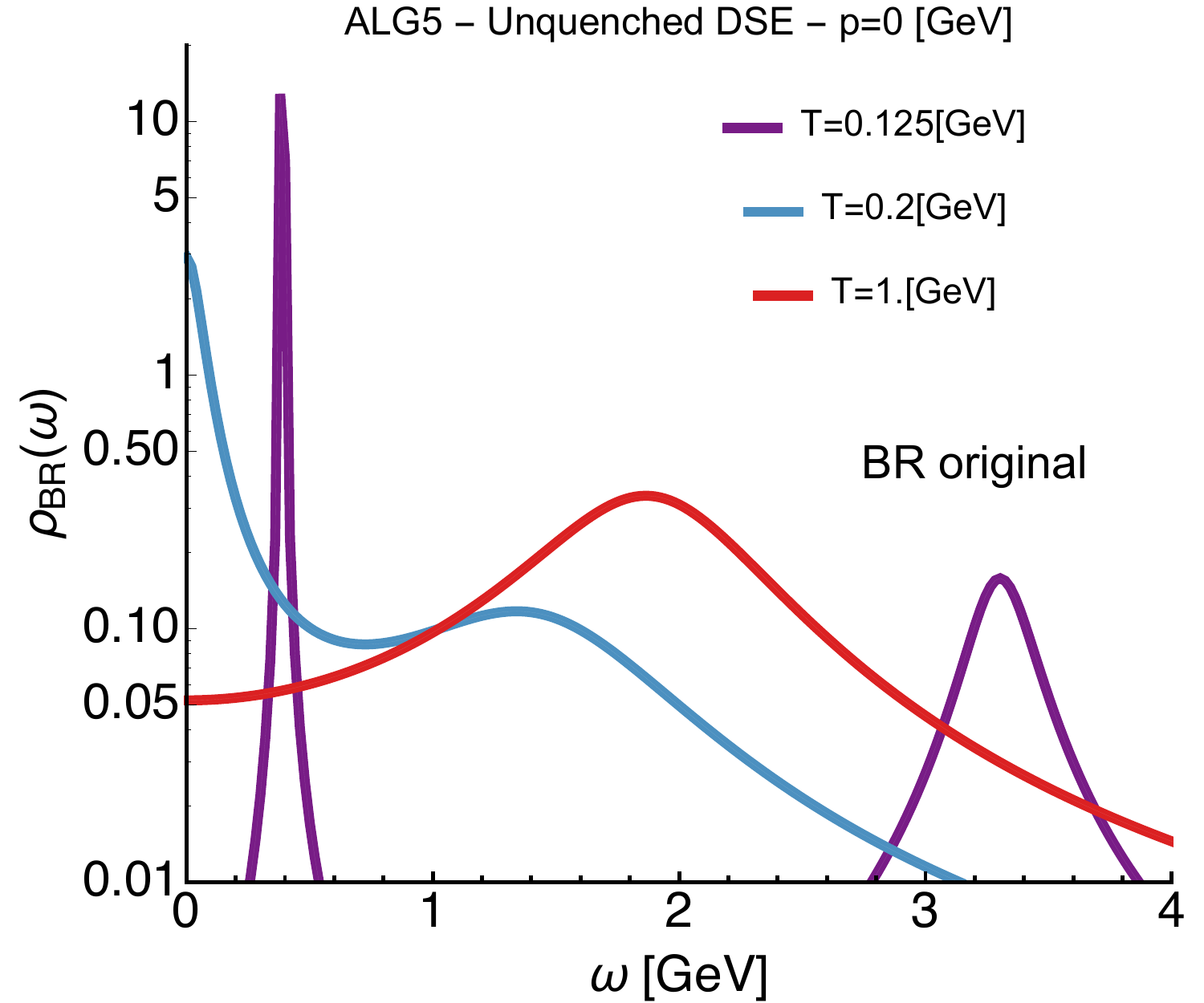}
\caption{Zero momentum reconstruction of the $T>0$ quark spectral function in the modern truncation with unquenched $N_f=2+1$ light quark flavors using the smooth BR (top) and the original BR (bottom) method. For better readability the plot contains only a subset of the reconstructed temperatures and is given in logarithmic scale.}\label{Fig:SpecRecUnquenched}
\end{figure}

Our main result concerns the quark spectrum computed from the quark
and gluon Dyson-Schwinger equations in a modern truncation
incorporating $N_f=2+1$ light quark flavors with physical masses. In
this case we encounter a chiral crossover at a pseudo-critical
temperature of $T = 0.155\,$GeV via the inflection point of the quark
condensate and $T = 0.160\,$GeV via the chiral susceptibility in
agreement with corresponding lattice results. We have computed $S_4$
along Matsubara frequencies at vanishing spatial momentum for eleven
temperatures in a larger temperature range of $T\in[0.125,1.0]\,$GeV,
i.e. also for temperatures below the pseudo-critical one.  At the same
cutoff of $\Lambda=100$GeV as before this now corresponds to
$N_{\rm data}\in[127,16]$. The correlator computations have been
checked to carry a numerical error of less than
$\Delta D/D\leq10^{-3}$, so that we can assign a corresponding
diagonal correlation matrix to it. Both the smooth and original BR
method are carried out with a frequency discretization of
$N_\omega=1000$ in the interval $\omega\in[0,20]$. The default model
is set to to a constant $m(\omega)=m_0$ and we carry out the
reconstruction with the different choices
$m_0=\{0.1,0.5,1.0,5.0,10\}$. The variance in the outcome is taken as
the basis for the error bands depicted in the subsequently shown
plots. In order to keep the presentation of the reconstructed spectra
clear, we show in Fig.\ref{Fig:SpecRecUnquenched} only a subset of the
reconstruction in a temperature range pertinent to the discussion
below. For completeness the full results are plotted in the appendix
\ref{sec:CompleteUnquenchedSpec} in
Fig. \ref{Fig:SpecRecUnquenchedFull}.

The first interesting result is that the reconstructions at low
temperatures show unstable behavior that hints at a failure of the
reconstruction limited to positive definite spectra. We see in
Fig.~\ref{Fig:SpecRecUnquenched} that at $T=0.125\,$GeV two sharp
peaks appear at positions very different from those at higher $T$. The
results at and above the transition region, $T=0.150\,$GeV and
$T=0.175\,$GeV appear to be in better shape at first glance. However,
truncating these datasets (e.g.\ from $N_{\rm data}=106$ to
$N_{\rm data}=94$ for the lowest temperature) actually changes the
behavior of the spectral functions around $\omega=0$. The same tests
for the two analyses in the previous subsections had shown virtually
no effect on the reconstruction, which is what is expected for a well
converged result. We thus conclude, that in the truncation scheme with
back-coupled quarks positivity violations characteristic for the low
temperature spectral functions of the quarks persist for much larger
temperatures than in the simple model case and prohibit
convergence. Only the spectra at and above $T=0.2\,$GeV do not show
such artificial behavior and are therefore deemed trustworthy.

\begin{figure}[b]
\includegraphics[scale=0.5,trim= 0 0 0 0.55cm, clip=true]{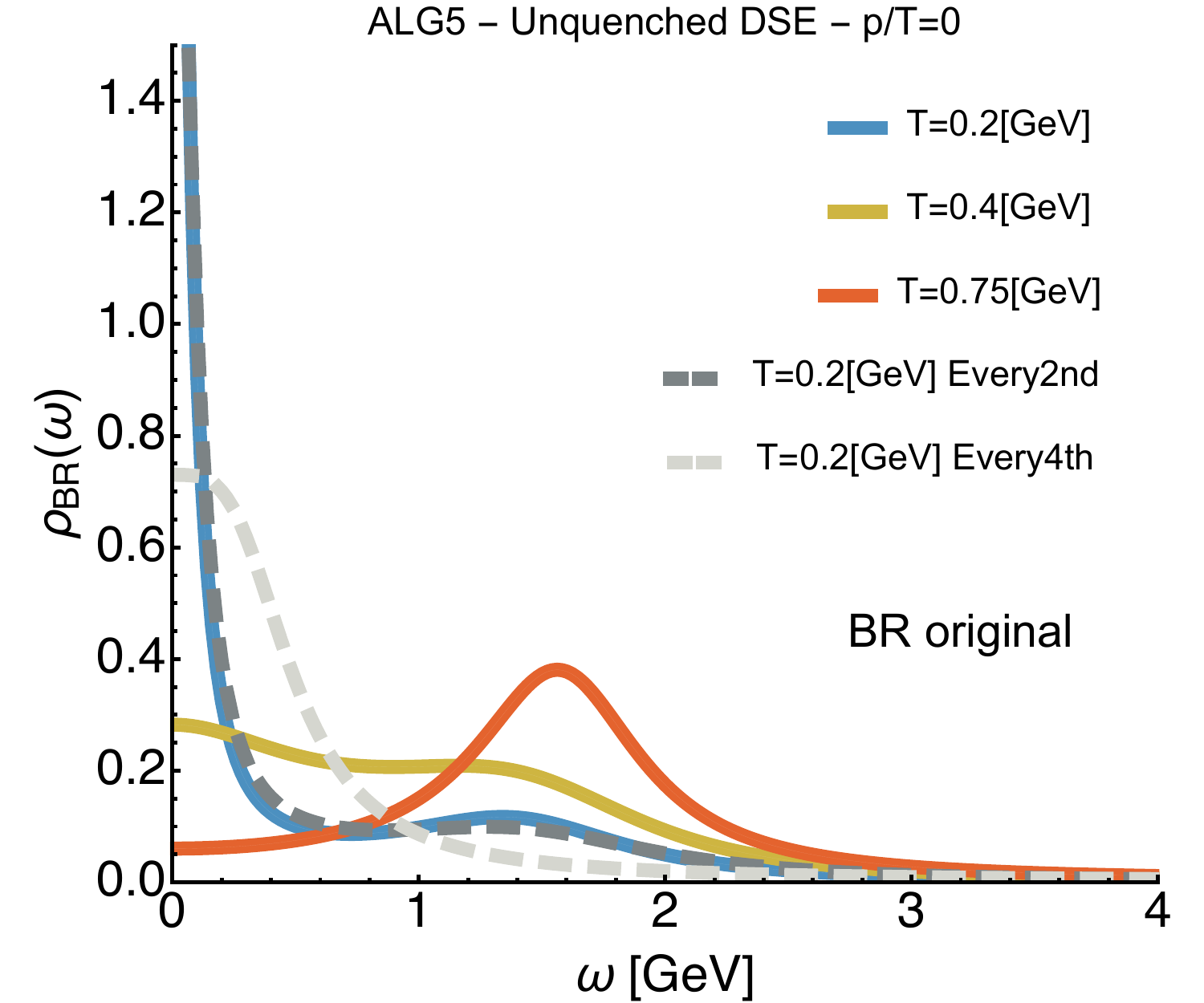}
\caption{Test of the reliability of the reconstruction at different temperatures based on the manually sparsened correlator at $T=0.2$GeV. Using only every second data point of the original $N_{\rm data}=80$ (dark grey) leaves the result virtually unchanged. When sparsened to every fourth point, the Bayesian method only recovers the single peak at the origin (light grey). Nonetheless clear in-medium modification at $T>0.2$GeV is observed.}\label{Fig:SpecRecUnquenchedReconstructedCorr}
\end{figure} 

This is also indicated in Fig.~\ref{Fig:SpecRecUnquenchedReconstructedCorr}, where we show the reconstruction outcome of taking the correlator at $T=0.2\,$GeV and $N_{\rm data}=80$ and sparsening it by factor two (dark grey dashed) or factor four (light grey dashed). For $N_{\rm data}=40$ the reconstruction is only very weakly affected, while for $N_{\rm data}=20$ a sole peak at the origin remains. We conclude as before that the reconstruction eventually becomes unreliable at high temperature but that at intermediate $T$ we are able to observe genuine in-medium modification.

We now analyze the position of the peaks. Although we find the same number of peaks present 
as in the model calculation, their behavior under variations of temperature seems to be different. Whereas the amplitude of the lowest lying 
peak still decreases with increasing temperature, the position of the second peak does not show a clear pattern. In particular it appears 
to not move monotonously to higher values of frequency with increasing temperature.

\begin{figure}[t]
\includegraphics[scale=0.5,trim= 0 0 0 0.55cm, clip=true]{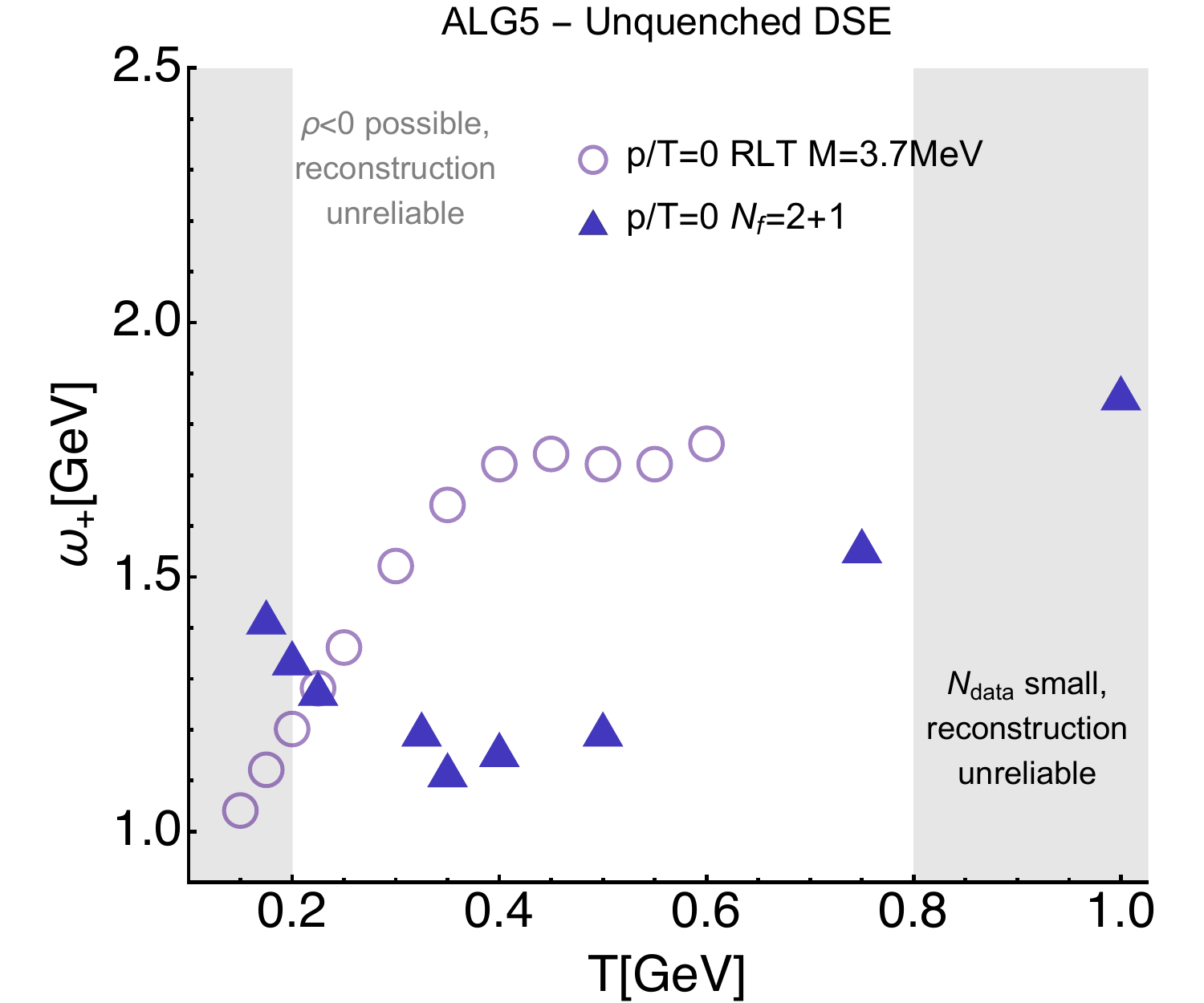}
\includegraphics[scale=0.5,trim= 0 0 0 0.55cm, clip=true]{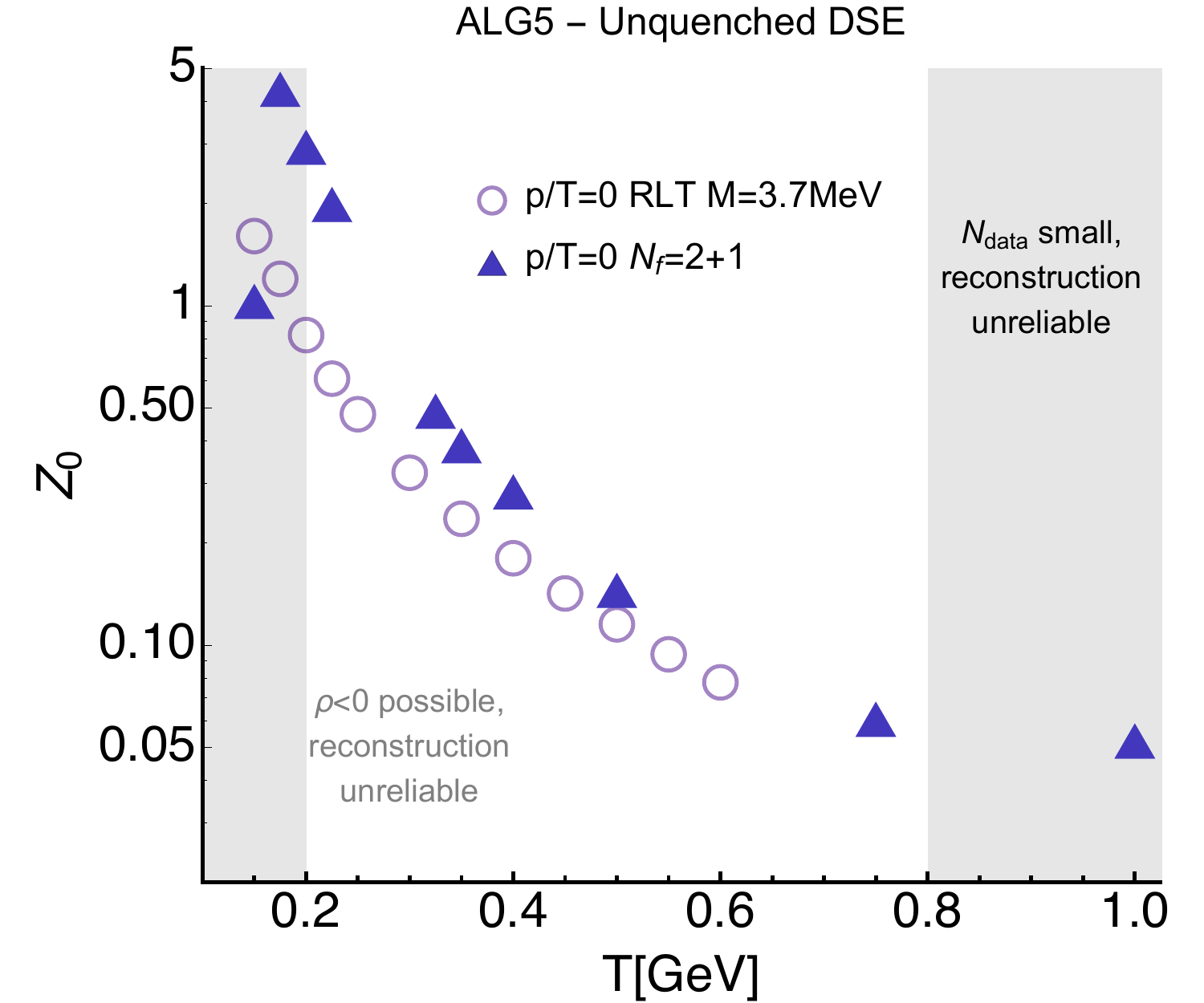}
\caption{(Top) position of the higher lying peak for different temperatures at vanishing momentum for the finite mass case (triangle) and the chiral case (circle). Except for the lowest $T$ no significant difference is observed. (Bottom) Amplitude of the lowest lying peak, which except for very low temperatures show the same behavior between the chiral and finite mass case.}\label{Fig:SpecRecUnquenchedOmegaPlusZZero}
\end{figure}

This is obvious from Fig.~\ref{Fig:SpecRecUnquenchedOmegaPlusZZero}, where we plot again the location $\omega_+$ of the second peak (upper panel)
and the amplitude $Z_0$ of the zero mode (lower panel). For purpose of comparison we also depict the results from the finite mass rainbow ladder truncation as circles. Clear qualitative and quantitative differences are visible. Instead of monotonously rising in value, $\omega_+$ appears to decrease first up to around $T=0.4\,$GeV before then increasing again in an almost linear fashion towards higher temperatures. Interestingly the initial downward trend starts in the low temperature regime, where we did not deem the reconstruction reliable due to the possible presence of positivity violation. Then we must further clarify whether the behavior of $\omega_+$ up to $T=0.4\,$GeV might still suffer from the influence of residual positivity violation. This will require the application of a reconstruction algorithm for general spectral functions, which is foreseen as next step in this line of study.

$Z_0$ on the other hand behaves at least qualitatively similar in the region where we trust the reconstruction. Below $T=0.15\,$GeV it also shows a clear dip, which is related to the appearance of the artificial spiky structures at intermediate frequencies there. Above $T=0.2\,$GeV it decreases monotonously.

Compared to the values reported in \cite{Gao:2014rqa} the behavior of $\omega_+$ here is quite similar. If the reconstructions between $T=0.2\,$GeV and $T=0.4\,$GeV are reliable, in particular as they do not show any obvious pathologies, then we also observe a dip in $\omega_+$ at intermediate temperatures. The height of the central peak on the other hand never fully vanishes in our case. 

\section{Conclusions} \label{sec:conclusion}

We have investigated the spectral properties of quarks in the Landau
gauge, based on Dyson-Schwinger equations according to two different
truncation schemes. In the rainbow-ladder approximation model both the
chiral and finite current quark mass case have been considered, while
our main result concerns quark spectra in a modern truncation with
$N_f=2+1$ unquenched flavors of light medium quarks.

The reconstruction of the spectral functions is based on a recently
developed Bayesian approach, the so called BR method, formulated in
imaginary frequencies. We further developed in this study a low-gain
variant of the BR method, which successfully suppresses numerical
ringing, which can affect the original BR method and in turn helps us
to unambiguously determine the number of physical peaks in the
spectrum. The accuracy of the reconstruction further benefits from the
use of the K\"allen-Lehmann kernel instead of the Euclidean one.

In mock data tests we have shown the capabilities and limitations of
our Bayesian reconstruction approach for either a best-case scenario
with correlator precision $\Delta D/D=10^{-8}$ and a real-world
setting with $\Delta D/D=10^{-3}$. In cases with two or three peak
features, which are expected to be relevant for our study the
combination of the conventional and smooth BR method allowed us to
unambiguously identify the number of encoded peaks and estimate their
properties. In the most challenging but least likely case that two
rather broad peaks at high frequency are located close to each other
it required the best case scenario to infer the presence of all
features.

The reconstructed spectra for the rainbow ladder truncation model with
vanishing and finite current quark mass showed very similar
behavior. At low temperatures two peaks are present, one at the
frequency origin, another one above $\omega=1$GeV. Changing the
temperature or changing the spatial momenta induced qualitatively
similar changes. The lowest lying peak height diminishes but does not
vanish up to the highest parameter values investigated. The second
peak both broadens and moves to higher frequencies.

A quantitative analysis of the height of the low lying peak $Z_0$ and
position of the second peak $\omega_+$ revealed a different behavior
as reported in previous studies. We did not find any indication of a
non-monotonicity in $\omega_+$ with respect to temperature and our
value for $Z_0$ always took on finite values in contrast to a
vanishing $\omega=0$ peak in \cite{Gao:2014rqa}.

We have made sure that the observed changes in $Z_0$ and $\omega_+$
with temperature can be attributed to the thermal medium. To
disentangle the effects from a degrading of the reconstruction due to
less available data points at high temperature, we repeated the
reconstructions with manually sparsened correlator data sets and
identified the regime, where the Bayesian method is reliable. And
indeed, in the region where the reconstruction can be trusted we find
that $\omega_+$ shows a linear rise with $T$ qualitatively compatible
with hard-thermal loop predictions. At the same point where the
reconstruction becomes unreliable we also see that the linear rise
begins to artificially flatten off.

In the unquenched truncation scheme with $N_f=2+1$ flavors of light
medium quarks the positive definite Bayesian approach is challenged at
low and high temperature. For $T<0.2\,$GeV we find indications that
non-positive spectral contributions are present, which lead to
artificially spiky structures, while a sparsening test shows that for
$N_{\rm data}<20$ the reconstruction also becomes unreliable. In the
intermediate temperature window we observe again two peak structures
with the lower one decreasing monotonously in height. The second peak
however behaves very differently than before as it now appears to
exhibit a dip in $\omega_+$, similar to the behavior reported in
\cite{Gao:2014rqa}.

In order to unambiguously determine, whether the non-monotonous
behavior of $\omega_+$ can be attributed to physics encoded in the
correlator, we will have to extend the analysis of this study in the
future to non-positive spectral functions. In the context of gluon
spectral functions in lattice QCD, a generalization of the BR method
has been proven a useful tool \cite{Ilgenfritz:2017kkp}. Implementing
a smooth version of this generalized BR method will constitute an
important step towards a robust and quantitative picture of the low
temperature regime of quark spectra, which is work in progress.

\section*{Acknowledgement}
C.F. and C.W. were supported by the German Federal Ministry of
Education and Research (BMBF) under Contract No. 05P15RGFCA and the
Helmholtz International Center for FAIR within the LOEWE program of
the State of Hesse. J.P. and A.R. acknowledge support by EMMI, the
grants ERC-AdG-290623, BMBF 05P12VHCTG as well as that this work is
part of and supported by the DFG Collaborative Research Centre "SFB
1225 (ISOQUANT)".

\FloatBarrier

\appendix
\section{Mock test Bayesian continuum limit}
\label{sec:app1}
\begin{figure*}[t]
\includegraphics[scale=0.5,trim= 0 0 0 0.5cm, clip=true]{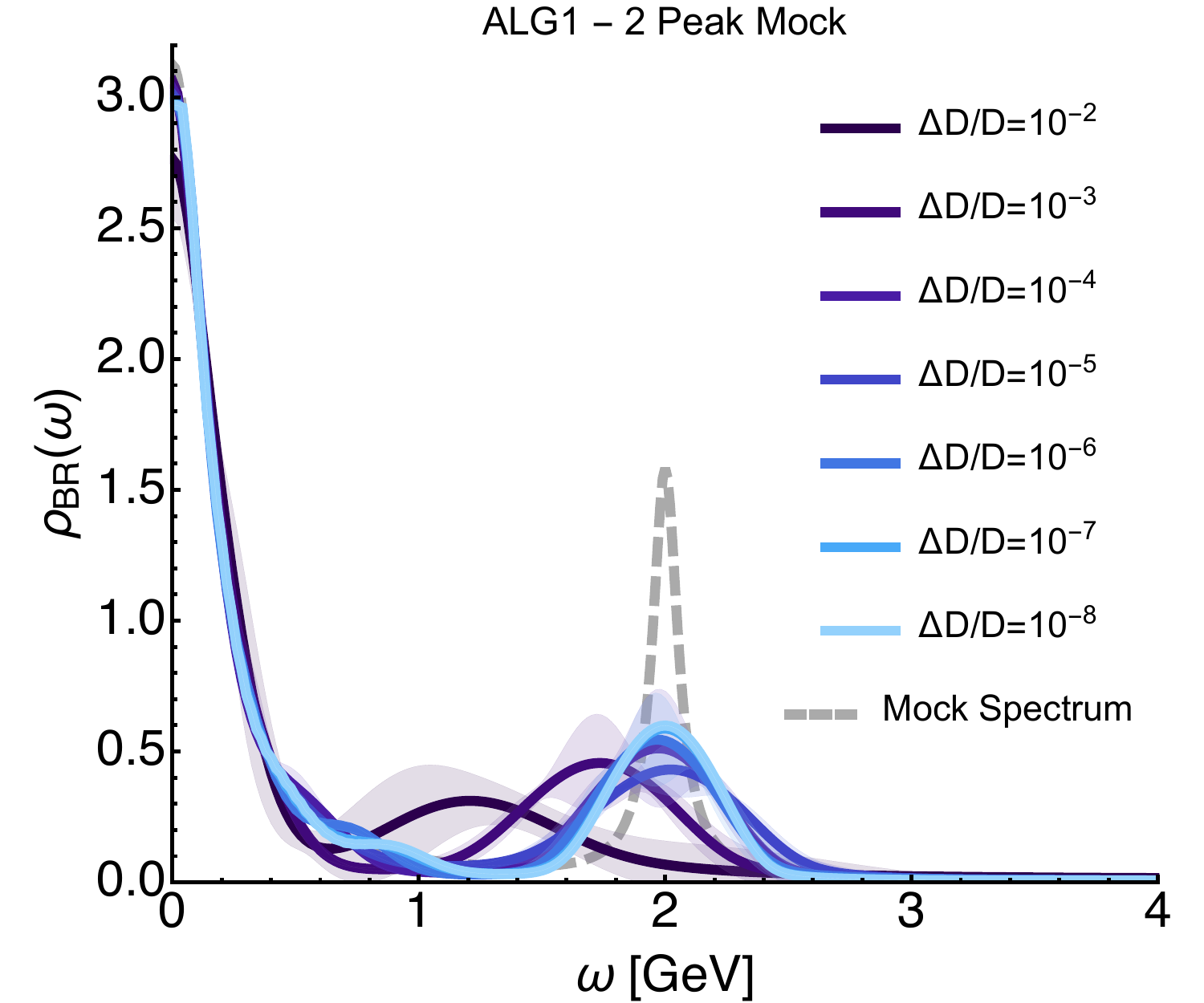}
\includegraphics[scale=0.5,trim= 0 0 0 0.5cm, clip=true]{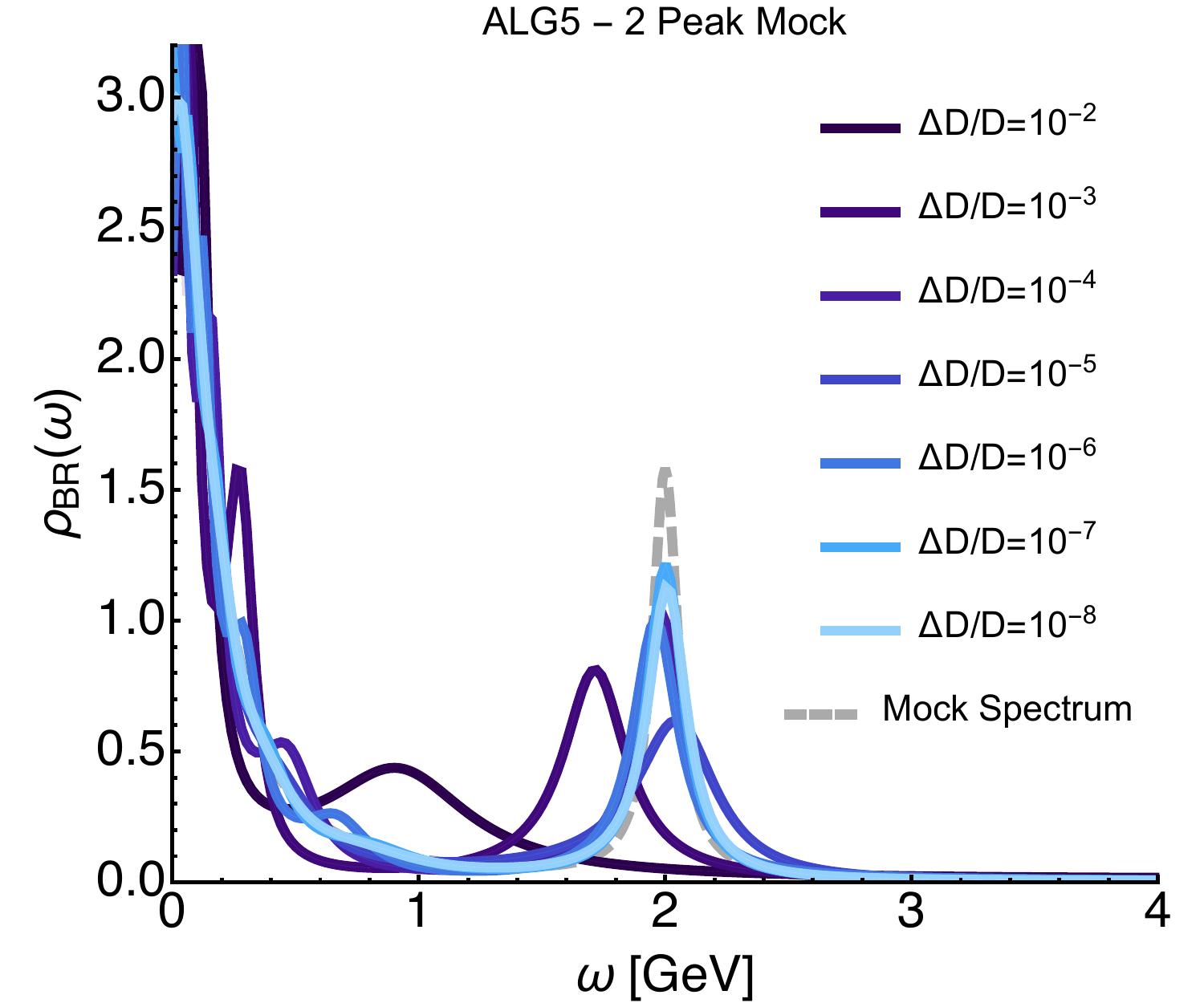}
\caption{Mock data analysis of reconstruction reliability for a two-peak scenario (grey dashed). Both the smooth BR (left) as well as the original BR (right) method are shown. While the latter unambiguously shows only two features and is devoid of ringing it approaches the Bayesian continuum limit more slowly than the original BR.}\label{Fig:SpecRecMockTwoPeak2Comp2}
\end{figure*}

\begin{figure*}[t]
\includegraphics[scale=0.5,trim= 0 0 0 0.5cm, clip=true]{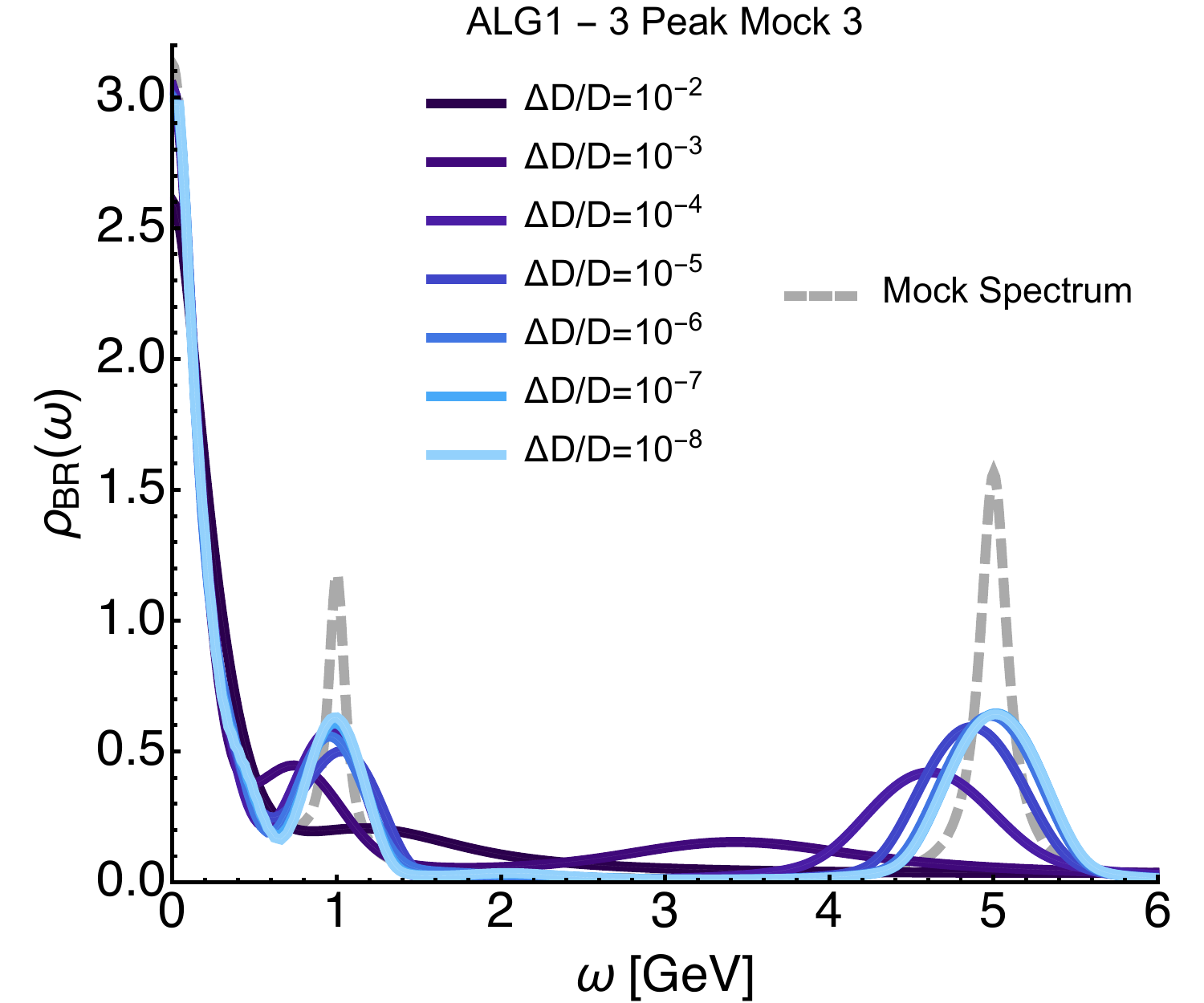} \hspace{1cm}
\includegraphics[scale=0.5,trim= 0 0 0 0.5cm, clip=true]{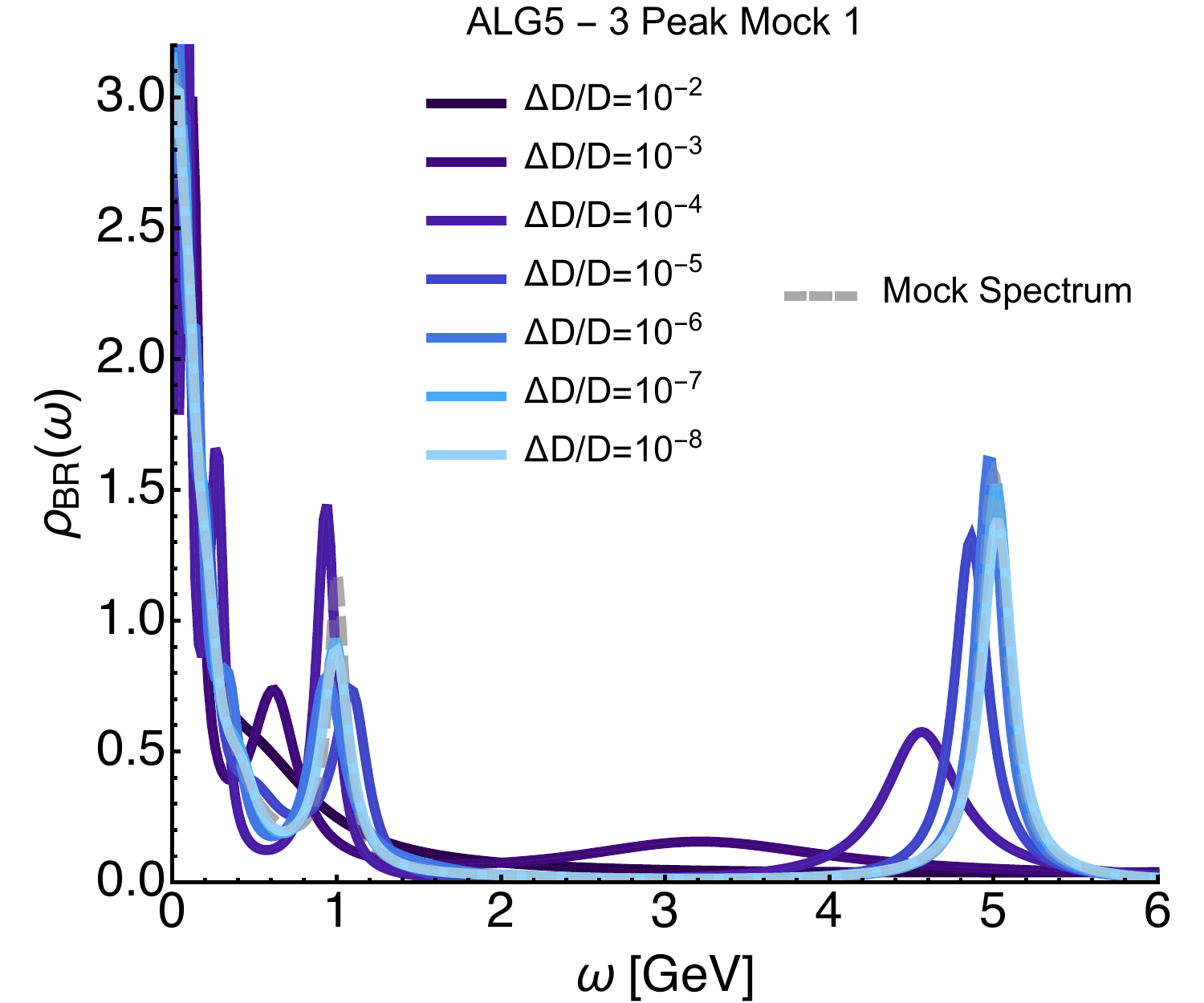}
\caption{Mock data analysis of reconstruction reliability for a three-peak scenario (grey dashed) with two peaks closely positioned off the origin.Both the smooth BR (left) as well as the original BR (right) method are shown.}\label{Fig:SpecRecMockThreePeak2Comp2}
\end{figure*}

\begin{figure*}[t]
\includegraphics[scale=0.5,trim= 0 0 0 0.5cm, clip=true]{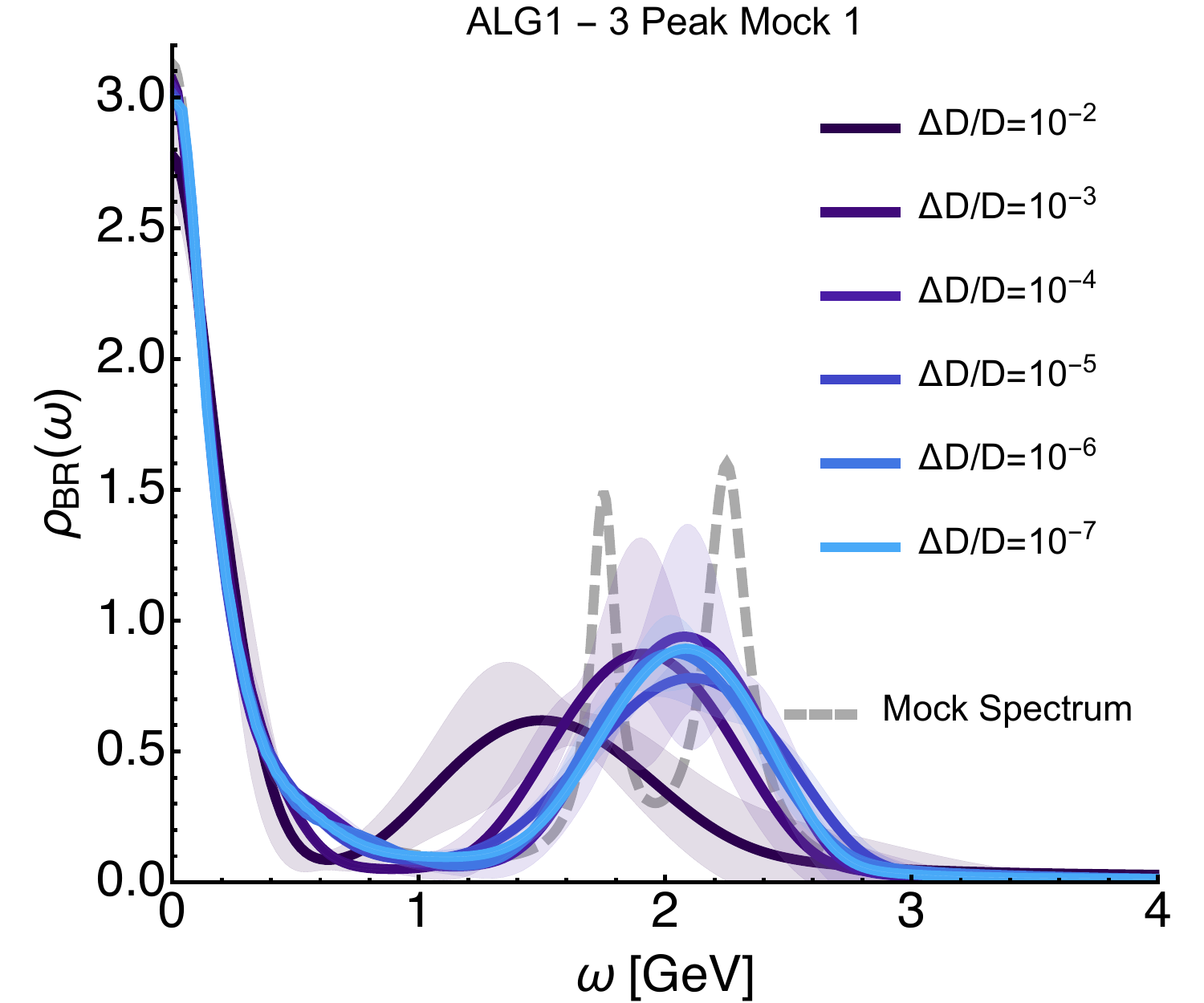} \hspace{1cm}
\includegraphics[scale=0.5,trim= 0 0 0 0.5cm, clip=true]{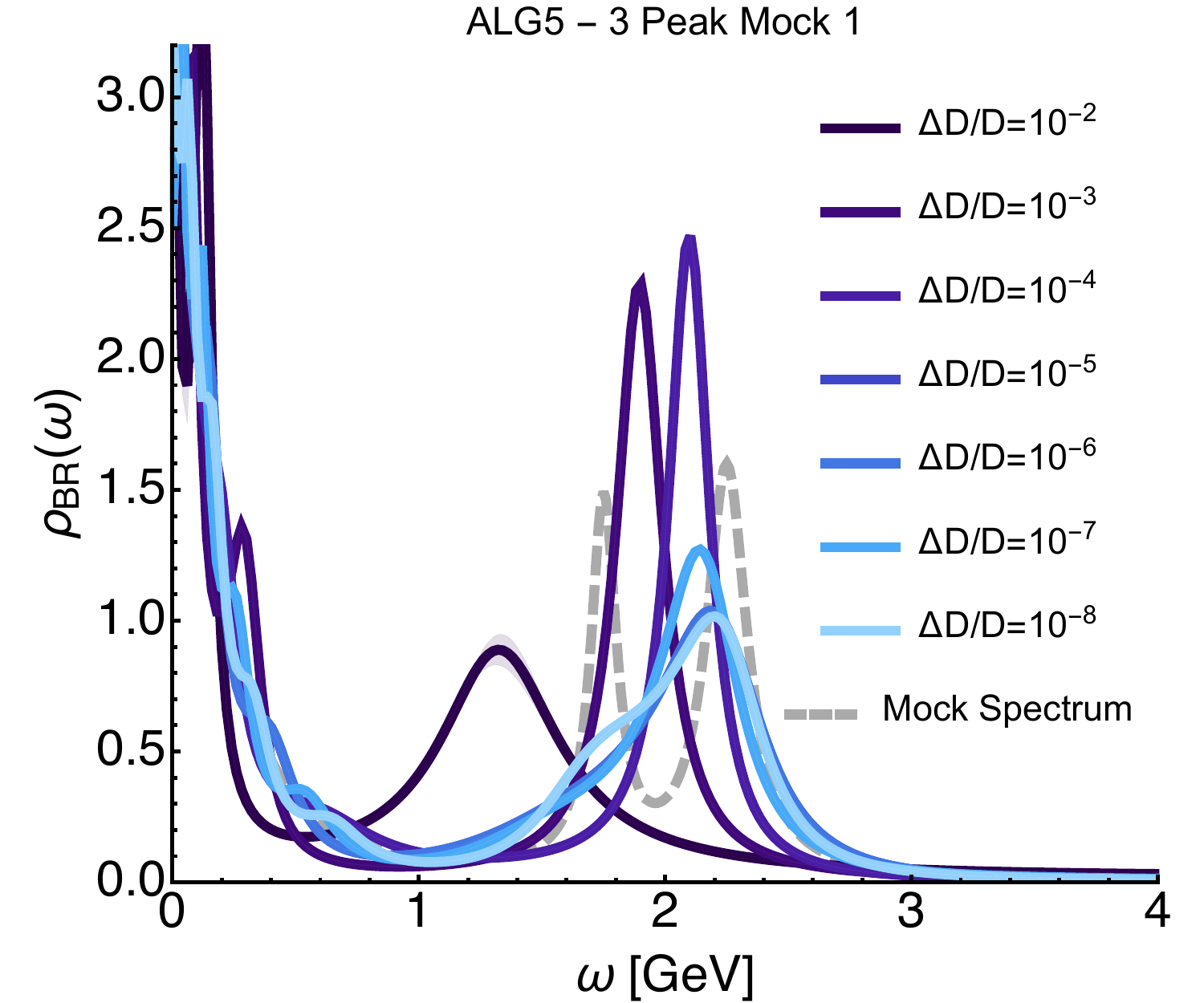}
\caption{Mock data analysis of reconstruction reliability for a three-peak scenario (grey dashed) with two peaks closely positioned off the origin.Both the smooth BR (left) as well as the original BR (right) method are shown.}\label{Fig:SpecRecMockThreePeak1Comp2}
\end{figure*}

In this appendix we present figures for the explicit approach of the
mock spectral reconstructions of Sec.\ref{sec:mocktest} towards the
Bayesian continuum limit at fixed $N_{\rm data}=128$. Each of the
figures contains seven solid curves, denoting the reconstruction
according to an assigned relative error
$\Delta D/D=[10^{-8},10^{-2}]$. In the upper panel the smooth BR is
deployed, while in the lower one it is the original BR method.The
darkest curve corresponds to the largest error. In
Fig.~\ref{Fig:SpecRecMockTwoPeak2Comp2} the two-peak scenario is
shown, while Fig.~\ref{Fig:SpecRecMockThreePeak2Comp2} contains the
results for three peak scenario with the second peak lying close to
that at the origin. The last Fig.~\ref{Fig:SpecRecMockThreePeak1Comp2}
shows the outcome for a three peak scenario with two closely places
structure at finite $\omega$.

\section{Complete spectral reconstructions for the unquenched case}
\label{sec:CompleteUnquenchedSpec}

\begin{figure*}[t]
\includegraphics[scale=0.5,trim= 0 0 0 0.55cm, clip=true]{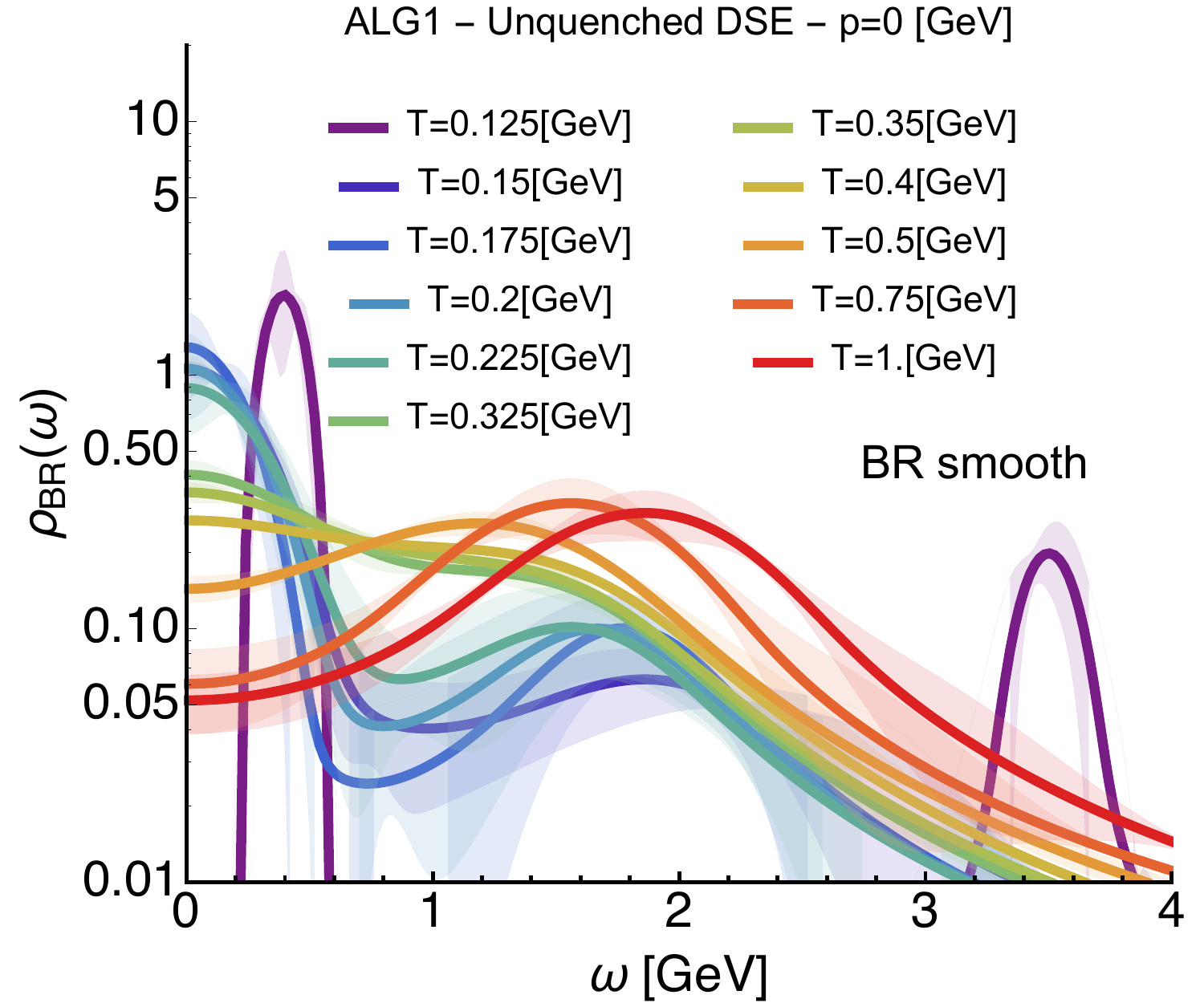} \hspace{1cm}
\includegraphics[scale=0.5,trim= 0 0 0 0.55cm, clip=true]{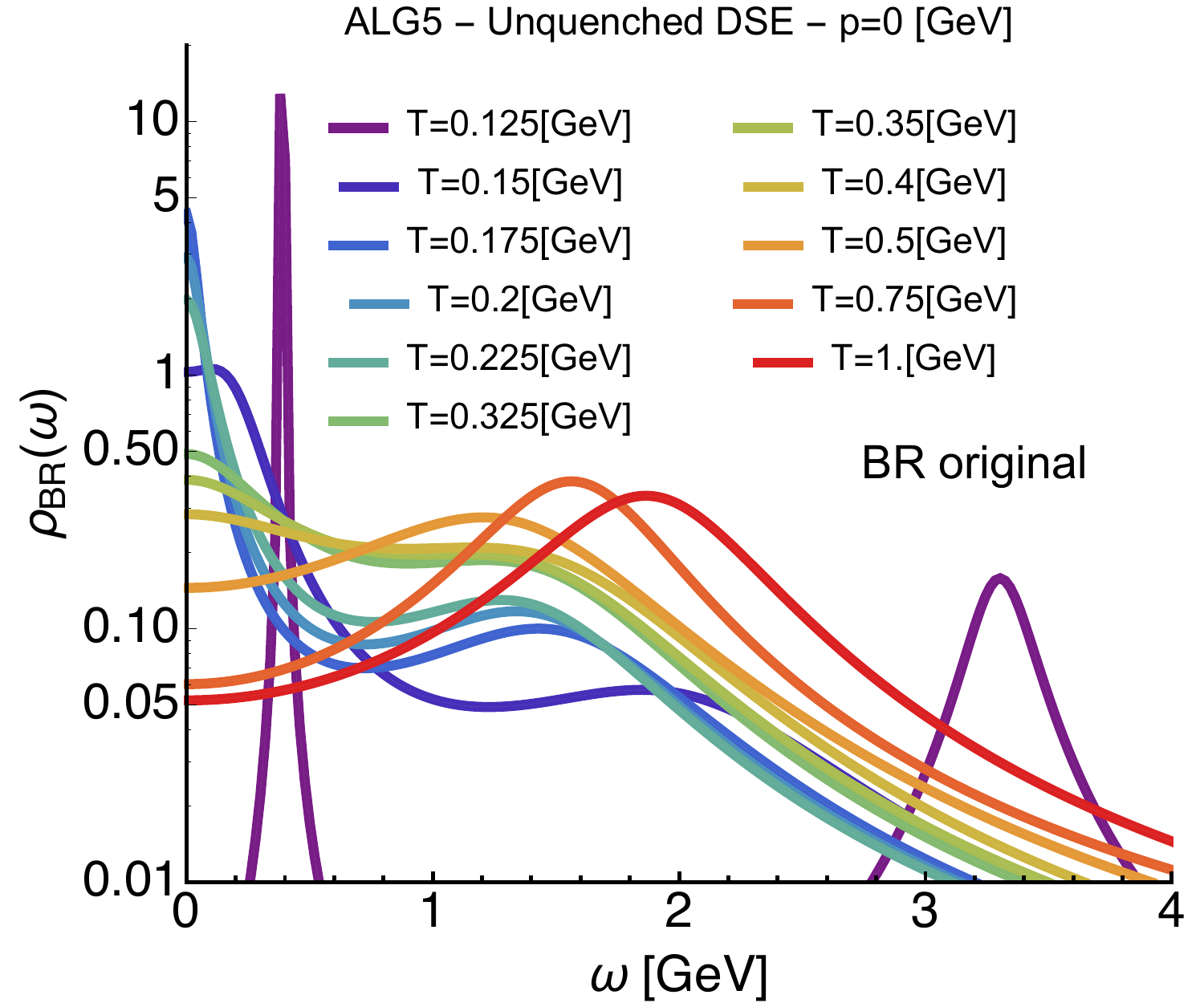}
\caption{Zero momentum reconstruction of the $T>0$ quark spectral function in the modern truncation with unquenched $N_f=2+1$ light quark flavors using the smooth BR (top) and the original BR (bottom) method. For better readability the plot is given in logarithmic scale.}\label{Fig:SpecRecUnquenchedFull}
\end{figure*}

Here we plot in Fig.\ref{Fig:SpecRecUnquenchedFull} for completeness the spectral reconstructions for the unquenched truncation with back-coupled $N_f=2+1$ quark flavors. One observes the appearance of artificial peaked structures at the lowest teperature $T=0.125$GeV, while at $T=0.2$GeV and above we obtain the same number of peaks as in the model computations. As was discussed in the context of Fig.\ref{Fig:SpecRecUnquenchedOmegaPlusZZero} the peak position of the peak located at finite frequencies however displays a qualitatively different behavior here than in the model truncation.

\section{Test of the evidence probability computation}
\label{sec:MEMcheck}

In contrast to the BR method, where the hyperparameter $\alpha$ is self-consistently integrated out apriori, MEM-like approaches marginalize $\alpha$ at the end of the reconstruction procedure \cite{Jarrell:1996,Asakawa:2000tr}. I.e in MEM one conventionally computes the corresponding spectrum $\rho_\alpha$ for many different values of $\alpha$ and then determines the probability distribution $P[\alpha|\rho,D,I]$. The individual $\rho_\alpha$ are subsequently averaged, weighted by $P[\alpha|\rho,D,I]$. In order to compute $P[\alpha|\rho,D,I]$ one however relies both on the assumption that the posterior probability is highly peaked and that it allows for a Gaussian approximation. Both are not tested in practice.

Common lore states that $P[\alpha|\rho,D,I]$ will have a peak at a finite $\alpha$ for which one particular spectrum contributes most strongly. If the maximum were at $\alpha=0$ the method reverts to an under-determined $\chi^2$ fit and no unique extremum exists. Here we give numerical evidence that the existence of a peak in the approximated $P[\alpha|\rho,D,I]$ depends on the choice of search space used. Furthermore if the search space is extended to the full size of the problem (in which there still exists a unique Bayesian answer) we find that only a maximum at $\alpha=0$ remains.

\begin{figure}[t]
\includegraphics[scale=0.35]{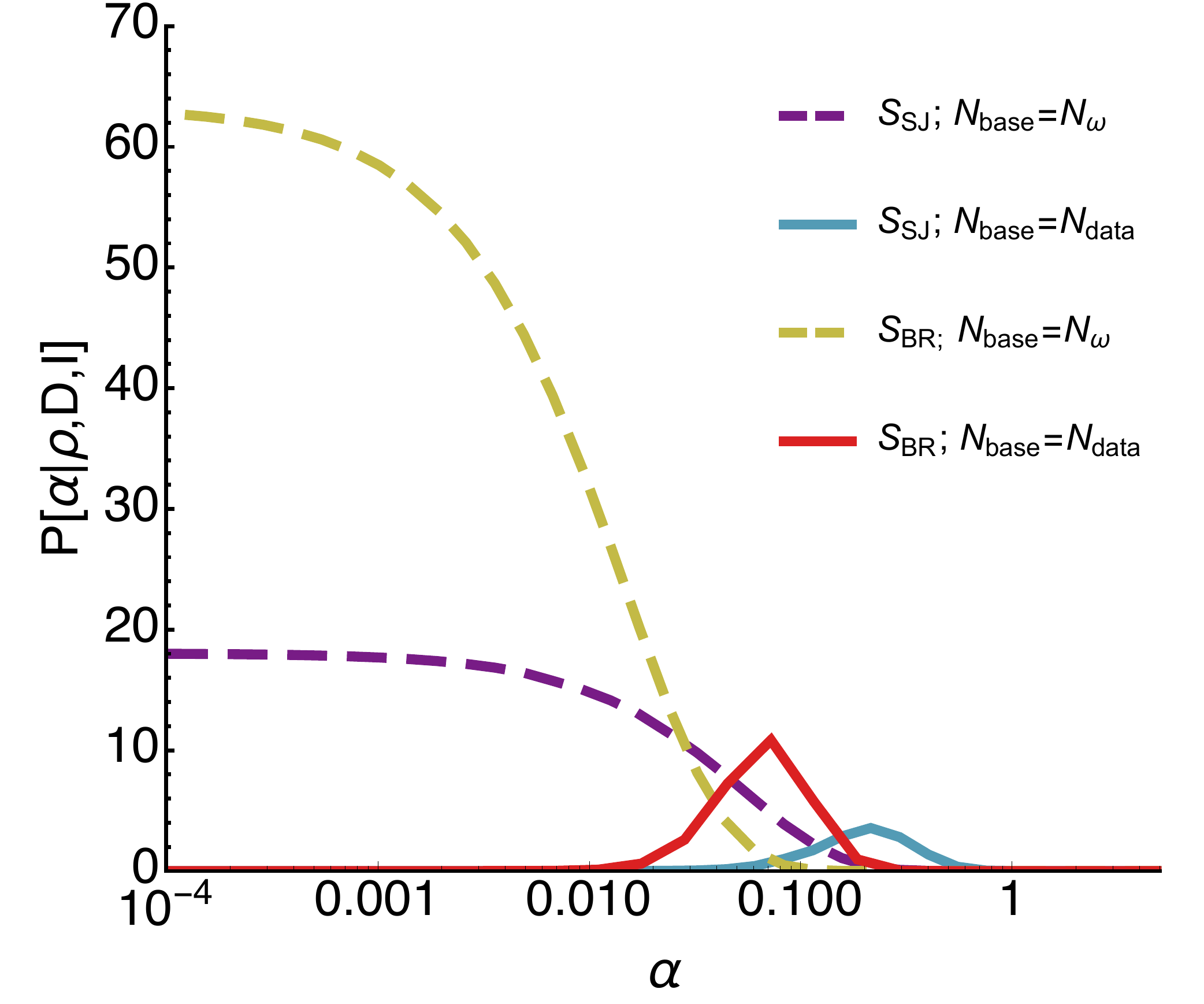}
\caption{Approximate probability distribution for the hyperparameter $P[\alpha|\rho,D,I]$ computed according to four different scenarios. Shannon-Jaynes Entropy $S_{\rm BR}$ with Bryan $N_{\rm base}=N_{\rm data}$ or full search space $N_{\rm base}=N_\omega$. The other two are BR prior $S_{\rm BR}$ with Bryan $N_{\rm base}=N_{\rm data}$ or full search space $N_{\rm base}=N_\omega$. Only by artificially restricting the search space do we find a maximum at finite $\alpha$.}\label{Fig:alphaTest}
\end{figure}

We use the same mock data as in the two-peak scenario in sec.\ref{sec:mocktest} and compare four different scenarios. We deploy the MEM with limited search space $S_{\rm BR}$ and $N_{\rm base}=N_{\rm data}$ according to Bryan and compare with (solid line) an implementation without restriction, where $N_{\rm base}=N_{\omega}$ (dashed line). The Bayesian result is selected with a step tolerance in the minimizer of $\Delta=5\times 10^{-8}$. In addition we replace the Shannon-Jaynes Entropy by the BR prior and repeat the reconstruction with $S_{\rm BR}$ and restricted search space $N_{\rm base}=N_{\rm data}$ (solid line) or without, i.e. using $S_{\rm BR}$ and $N_{\rm base}=N_{\omega}$ (dashed line). We have of course adapted the computation of $P[\alpha|\rho,D,I]$ to this new prior. The results for the probabilities are shown in Fig.~\ref{Fig:alphaTest}. 

We find indeed that only for the restricted search space a peak at finite values of $\alpha$ remains. This issue is independent of the actual regulator used, both $S_{\rm SJ}$ and $S_{\rm BR}$ show the same trend. We believe that the underlying reason is that in the presence of a restricted search space the minimizer is at some point not able to lower the value of $L$, while in the full search space it can be brought very close to zero. This finite minimal value of $L$ then prohibits the probability to rise further. Since the Bayesian answer is unique if it exists \cite{Asakawa:2000tr}, we conclude that it is the approximations made to determine $P[\alpha|\rho,D,I]$, which prevent us from obtaining that unique result in the full search space.

As a consequence we revert to the historic MEM choice of setting $\alpha$ such that $L=N_{\rm data}$ in case of the smooth BR method, where an apriori marginalization of the hyperparameter is not analytically feasible.

\bibliography{finite_T_mu.bib}

\begin{thebibliography}{52}
\expandafter\ifx\csname natexlab\endcsname\relax\def\natexlab#1{#1}\fi
\expandafter\ifx\csname bibnamefont\endcsname\relax
  \def\bibnamefont#1{#1}\fi
\expandafter\ifx\csname bibfnamefont\endcsname\relax
  \def\bibfnamefont#1{#1}\fi
\expandafter\ifx\csname citenamefont\endcsname\relax
  \def\citenamefont#1{#1}\fi
\expandafter\ifx\csname url\endcsname\relax
  \def\url#1{\texttt{#1}}\fi
\expandafter\ifx\csname urlprefix\endcsname\relax\def\urlprefix{URL }\fi
\providecommand{\bibinfo}[2]{#2}
\providecommand{\eprint}[2][]{\url{#2}}

\bibitem[{\citenamefont{Mueller et~al.}(2010)\citenamefont{Mueller, Fischer,
  and Nickel}}]{Mueller:2010ah}
\bibinfo{author}{\bibfnamefont{J.~A.} \bibnamefont{Mueller}},
  \bibinfo{author}{\bibfnamefont{C.~S.} \bibnamefont{Fischer}},
  \bibnamefont{and} \bibinfo{author}{\bibfnamefont{D.}~\bibnamefont{Nickel}},
  \bibinfo{journal}{Eur. Phys. J.} \textbf{\bibinfo{volume}{C70}},
  \bibinfo{pages}{1037} (\bibinfo{year}{2010}), \eprint{1009.3762}.

\bibitem[{\citenamefont{Burnier and Rothkopf}(2013)}]{Burnier:2013nla}
\bibinfo{author}{\bibfnamefont{Y.}~\bibnamefont{Burnier}} \bibnamefont{and}
  \bibinfo{author}{\bibfnamefont{A.}~\bibnamefont{Rothkopf}},
  \bibinfo{journal}{Phys. Rev. Lett.} \textbf{\bibinfo{volume}{111}},
  \bibinfo{pages}{182003} (\bibinfo{year}{2013}), \eprint{1307.6106}.

\bibitem[{\citenamefont{Muller and Nagle}(2006)}]{Muller:2006ee}
\bibinfo{author}{\bibfnamefont{B.}~\bibnamefont{Muller}} \bibnamefont{and}
  \bibinfo{author}{\bibfnamefont{J.~L.} \bibnamefont{Nagle}},
  \bibinfo{journal}{Ann. Rev. Nucl. Part. Sci.} \textbf{\bibinfo{volume}{56}},
  \bibinfo{pages}{93} (\bibinfo{year}{2006}), \eprint{nucl-th/0602029}.

\bibitem[{\citenamefont{Shuryak}(2009)}]{Shuryak:2008eq}
\bibinfo{author}{\bibfnamefont{E.}~\bibnamefont{Shuryak}},
  \bibinfo{journal}{Prog. Part. Nucl. Phys.} \textbf{\bibinfo{volume}{62}},
  \bibinfo{pages}{48} (\bibinfo{year}{2009}), \eprint{0807.3033}.

\bibitem[{\citenamefont{Braun-Munzinger and
  Wambach}(2009)}]{BraunMunzinger:2008tz}
\bibinfo{author}{\bibfnamefont{P.}~\bibnamefont{Braun-Munzinger}}
  \bibnamefont{and} \bibinfo{author}{\bibfnamefont{J.}~\bibnamefont{Wambach}},
  \bibinfo{journal}{Rev. Mod. Phys.} \textbf{\bibinfo{volume}{81}},
  \bibinfo{pages}{1031} (\bibinfo{year}{2009}), \eprint{0801.4256}.

\bibitem[{\citenamefont{Andronic}(2014)}]{Andronic:2014zha}
\bibinfo{author}{\bibfnamefont{A.}~\bibnamefont{Andronic}},
  \bibinfo{journal}{Int. J. Mod. Phys.} \textbf{\bibinfo{volume}{A29}},
  \bibinfo{pages}{1430047} (\bibinfo{year}{2014}), \eprint{1407.5003}.

\bibitem[{\citenamefont{Foka and Janik}(2016)}]{Foka:2016zdb}
\bibinfo{author}{\bibfnamefont{P.}~\bibnamefont{Foka}} \bibnamefont{and}
  \bibinfo{author}{\bibfnamefont{M.~A.} \bibnamefont{Janik}},
  \bibinfo{journal}{Rev. Phys.} \textbf{\bibinfo{volume}{1}},
  \bibinfo{pages}{172} (\bibinfo{year}{2016}), \eprint{1702.07231}.

\bibitem[{\citenamefont{Nickel}(2007)}]{Nickel:2006mm}
\bibinfo{author}{\bibfnamefont{D.}~\bibnamefont{Nickel}},
  \bibinfo{journal}{Annals Phys.} \textbf{\bibinfo{volume}{322}},
  \bibinfo{pages}{1949} (\bibinfo{year}{2007}), \eprint{hep-ph/0607224}.

\bibitem[{\citenamefont{Harada et~al.}(2008)\citenamefont{Harada, Nemoto, and
  Yoshimoto}}]{Harada:2007gg}
\bibinfo{author}{\bibfnamefont{M.}~\bibnamefont{Harada}},
  \bibinfo{author}{\bibfnamefont{Y.}~\bibnamefont{Nemoto}}, \bibnamefont{and}
  \bibinfo{author}{\bibfnamefont{S.}~\bibnamefont{Yoshimoto}},
  \bibinfo{journal}{Prog. Theor. Phys.} \textbf{\bibinfo{volume}{119}},
  \bibinfo{pages}{117} (\bibinfo{year}{2008}), \eprint{0708.3351}.

\bibitem[{\citenamefont{Harada and Yoshimoto}(2009)}]{Harada:2009zq}
\bibinfo{author}{\bibfnamefont{M.}~\bibnamefont{Harada}} \bibnamefont{and}
  \bibinfo{author}{\bibfnamefont{S.}~\bibnamefont{Yoshimoto}}
  (\bibinfo{year}{2009}), \eprint{0903.5495}.

\bibitem[{\citenamefont{Karsch and Kitazawa}(2007)}]{Karsch:2007wc}
\bibinfo{author}{\bibfnamefont{F.}~\bibnamefont{Karsch}} \bibnamefont{and}
  \bibinfo{author}{\bibfnamefont{M.}~\bibnamefont{Kitazawa}},
  \bibinfo{journal}{Phys. Lett.} \textbf{\bibinfo{volume}{B658}},
  \bibinfo{pages}{45} (\bibinfo{year}{2007}), \eprint{0708.0299}.

\bibitem[{\citenamefont{Karsch and Kitazawa}(2009)}]{Karsch:2009tp}
\bibinfo{author}{\bibfnamefont{F.}~\bibnamefont{Karsch}} \bibnamefont{and}
  \bibinfo{author}{\bibfnamefont{M.}~\bibnamefont{Kitazawa}},
  \bibinfo{journal}{Phys. Rev.} \textbf{\bibinfo{volume}{D80}},
  \bibinfo{pages}{056001} (\bibinfo{year}{2009}), \eprint{0906.3941}.

\bibitem[{\citenamefont{Qin et~al.}(2011)\citenamefont{Qin, Chang, Liu, and
  Roberts}}]{Qin:2010pc}
\bibinfo{author}{\bibfnamefont{S.-x.} \bibnamefont{Qin}},
  \bibinfo{author}{\bibfnamefont{L.}~\bibnamefont{Chang}},
  \bibinfo{author}{\bibfnamefont{Y.-x.} \bibnamefont{Liu}}, \bibnamefont{and}
  \bibinfo{author}{\bibfnamefont{C.~D.} \bibnamefont{Roberts}},
  \bibinfo{journal}{Phys. Rev.} \textbf{\bibinfo{volume}{D84}},
  \bibinfo{pages}{014017} (\bibinfo{year}{2011}), \eprint{1010.4231}.

\bibitem[{\citenamefont{Qin and Rischke}(2013)}]{Qin:2013ufa}
\bibinfo{author}{\bibfnamefont{S.-x.} \bibnamefont{Qin}} \bibnamefont{and}
  \bibinfo{author}{\bibfnamefont{D.~H.} \bibnamefont{Rischke}},
  \bibinfo{journal}{Phys. Rev.} \textbf{\bibinfo{volume}{D88}},
  \bibinfo{pages}{056007} (\bibinfo{year}{2013}), \eprint{1304.6547}.

\bibitem[{\citenamefont{Gao et~al.}(2014)\citenamefont{Gao, Qin, Liu, Roberts,
  and Schmidt}}]{Gao:2014rqa}
\bibinfo{author}{\bibfnamefont{F.}~\bibnamefont{Gao}},
  \bibinfo{author}{\bibfnamefont{S.-X.} \bibnamefont{Qin}},
  \bibinfo{author}{\bibfnamefont{Y.-X.} \bibnamefont{Liu}},
  \bibinfo{author}{\bibfnamefont{C.~D.} \bibnamefont{Roberts}},
  \bibnamefont{and} \bibinfo{author}{\bibfnamefont{S.~M.}
  \bibnamefont{Schmidt}}, \bibinfo{journal}{Phys. Rev.}
  \textbf{\bibinfo{volume}{D89}}, \bibinfo{pages}{076009}
  (\bibinfo{year}{2014}), \eprint{1401.2406}.

\bibitem[{\citenamefont{Strauss et~al.}(2012)\citenamefont{Strauss, Fischer,
  and Kellermann}}]{Strauss:2012dg}
\bibinfo{author}{\bibfnamefont{S.}~\bibnamefont{Strauss}},
  \bibinfo{author}{\bibfnamefont{C.~S.} \bibnamefont{Fischer}},
  \bibnamefont{and}
  \bibinfo{author}{\bibfnamefont{C.}~\bibnamefont{Kellermann}},
  \bibinfo{journal}{Phys. Rev. Lett.} \textbf{\bibinfo{volume}{109}},
  \bibinfo{pages}{252001} (\bibinfo{year}{2012}), \eprint{1208.6239}.

\bibitem[{\citenamefont{Haas et~al.}(2014)\citenamefont{Haas, Fister, and
  Pawlowski}}]{Haas:2013hpa}
\bibinfo{author}{\bibfnamefont{M.}~\bibnamefont{Haas}},
  \bibinfo{author}{\bibfnamefont{L.}~\bibnamefont{Fister}}, \bibnamefont{and}
  \bibinfo{author}{\bibfnamefont{J.~M.} \bibnamefont{Pawlowski}},
  \bibinfo{journal}{Phys.Rev.} \textbf{\bibinfo{volume}{D90}},
  \bibinfo{pages}{091501} (\bibinfo{year}{2014}), \eprint{1308.4960}.

\bibitem[{\citenamefont{Christiansen et~al.}(2015)\citenamefont{Christiansen,
  Haas, Pawlowski, and Strodthoff}}]{Christiansen:2014ypa}
\bibinfo{author}{\bibfnamefont{N.}~\bibnamefont{Christiansen}},
  \bibinfo{author}{\bibfnamefont{M.}~\bibnamefont{Haas}},
  \bibinfo{author}{\bibfnamefont{J.~M.} \bibnamefont{Pawlowski}},
  \bibnamefont{and}
  \bibinfo{author}{\bibfnamefont{N.}~\bibnamefont{Strodthoff}},
  \bibinfo{journal}{Phys. Rev. Lett.} \textbf{\bibinfo{volume}{115}},
  \bibinfo{pages}{112002} (\bibinfo{year}{2015}), \eprint{1411.7986}.

\bibitem[{\citenamefont{Dudal et~al.}(2014)\citenamefont{Dudal, Oliveira, and
  Silva}}]{Dudal:2013yva}
\bibinfo{author}{\bibfnamefont{D.}~\bibnamefont{Dudal}},
  \bibinfo{author}{\bibfnamefont{O.}~\bibnamefont{Oliveira}}, \bibnamefont{and}
  \bibinfo{author}{\bibfnamefont{P.~J.} \bibnamefont{Silva}},
  \bibinfo{journal}{Phys. Rev.} \textbf{\bibinfo{volume}{D89}},
  \bibinfo{pages}{014010} (\bibinfo{year}{2014}), \eprint{1310.4069}.

\bibitem[{\citenamefont{Ilgenfritz et~al.}(2017)\citenamefont{Ilgenfritz,
  Pawlowski, Rothkopf, and Trunin}}]{Ilgenfritz:2017kkp}
\bibinfo{author}{\bibfnamefont{E.-M.} \bibnamefont{Ilgenfritz}},
  \bibinfo{author}{\bibfnamefont{J.~M.} \bibnamefont{Pawlowski}},
  \bibinfo{author}{\bibfnamefont{A.}~\bibnamefont{Rothkopf}}, \bibnamefont{and}
  \bibinfo{author}{\bibfnamefont{A.}~\bibnamefont{Trunin}}
  (\bibinfo{year}{2017}), \eprint{1701.08610}.

\bibitem[{\citenamefont{Braaten et~al.}(1990)\citenamefont{Braaten, Pisarski,
  and Yuan}}]{Braaten:1990wp}
\bibinfo{author}{\bibfnamefont{E.}~\bibnamefont{Braaten}},
  \bibinfo{author}{\bibfnamefont{R.~D.} \bibnamefont{Pisarski}},
  \bibnamefont{and} \bibinfo{author}{\bibfnamefont{T.-C.} \bibnamefont{Yuan}},
  \bibinfo{journal}{Phys. Rev. Lett.} \textbf{\bibinfo{volume}{64}},
  \bibinfo{pages}{2242} (\bibinfo{year}{1990}).

\bibitem[{\citenamefont{Peshier and Thoma}(2000)}]{Peshier:1999dt}
\bibinfo{author}{\bibfnamefont{A.}~\bibnamefont{Peshier}} \bibnamefont{and}
  \bibinfo{author}{\bibfnamefont{M.~H.} \bibnamefont{Thoma}},
  \bibinfo{journal}{Phys. Rev. Lett.} \textbf{\bibinfo{volume}{84}},
  \bibinfo{pages}{841} (\bibinfo{year}{2000}), \eprint{hep-ph/9907268}.

\bibitem[{\citenamefont{Arnold et~al.}(2002)\citenamefont{Arnold, Moore, and
  Yaffe}}]{Arnold:2002ja}
\bibinfo{author}{\bibfnamefont{P.~B.} \bibnamefont{Arnold}},
  \bibinfo{author}{\bibfnamefont{G.~D.} \bibnamefont{Moore}}, \bibnamefont{and}
  \bibinfo{author}{\bibfnamefont{L.~G.} \bibnamefont{Yaffe}},
  \bibinfo{journal}{JHEP} \textbf{\bibinfo{volume}{06}}, \bibinfo{pages}{030}
  (\bibinfo{year}{2002}), \eprint{hep-ph/0204343}.

\bibitem[{\citenamefont{Kim et~al.}(2015{\natexlab{a}})\citenamefont{Kim,
  Asakawa, and Kitazawa}}]{Kim:2015poa}
\bibinfo{author}{\bibfnamefont{T.}~\bibnamefont{Kim}},
  \bibinfo{author}{\bibfnamefont{M.}~\bibnamefont{Asakawa}}, \bibnamefont{and}
  \bibinfo{author}{\bibfnamefont{M.}~\bibnamefont{Kitazawa}},
  \bibinfo{journal}{Phys. Rev.} \textbf{\bibinfo{volume}{D92}},
  \bibinfo{pages}{114014} (\bibinfo{year}{2015}{\natexlab{a}}),
  \eprint{1505.07195}.

\bibitem[{\citenamefont{Braaten and Pisarski}(1990)}]{Braaten:1989mz}
\bibinfo{author}{\bibfnamefont{E.}~\bibnamefont{Braaten}} \bibnamefont{and}
  \bibinfo{author}{\bibfnamefont{R.~D.} \bibnamefont{Pisarski}},
  \bibinfo{journal}{Nucl. Phys.} \textbf{\bibinfo{volume}{B337}},
  \bibinfo{pages}{569} (\bibinfo{year}{1990}).

\bibitem[{\citenamefont{Baym et~al.}(1992)\citenamefont{Baym, Blaizot, and
  Svetitsky}}]{Baym:1992eu}
\bibinfo{author}{\bibfnamefont{G.}~\bibnamefont{Baym}},
  \bibinfo{author}{\bibfnamefont{J.-P.} \bibnamefont{Blaizot}},
  \bibnamefont{and}
  \bibinfo{author}{\bibfnamefont{B.}~\bibnamefont{Svetitsky}},
  \bibinfo{journal}{Phys. Rev.} \textbf{\bibinfo{volume}{D46}},
  \bibinfo{pages}{4043} (\bibinfo{year}{1992}).

\bibitem[{\citenamefont{Blaizot and Ollitrault}(1993)}]{Blaizot:1993bb}
\bibinfo{author}{\bibfnamefont{J.-P.} \bibnamefont{Blaizot}} \bibnamefont{and}
  \bibinfo{author}{\bibfnamefont{J.-Y.} \bibnamefont{Ollitrault}},
  \bibinfo{journal}{Phys. Rev.} \textbf{\bibinfo{volume}{D48}},
  \bibinfo{pages}{1390} (\bibinfo{year}{1993}), \eprint{hep-th/9303070}.

\bibitem[{\citenamefont{Schaefer and Thoma}(1999)}]{Schaefer:1998wd}
\bibinfo{author}{\bibfnamefont{A.}~\bibnamefont{Schaefer}} \bibnamefont{and}
  \bibinfo{author}{\bibfnamefont{M.~H.} \bibnamefont{Thoma}},
  \bibinfo{journal}{Phys. Lett.} \textbf{\bibinfo{volume}{B451}},
  \bibinfo{pages}{195} (\bibinfo{year}{1999}), \eprint{hep-ph/9811364}.

\bibitem[{\citenamefont{Kitazawa et~al.}(2006)\citenamefont{Kitazawa, Kunihiro,
  and Nemoto}}]{Kitazawa:2005mp}
\bibinfo{author}{\bibfnamefont{M.}~\bibnamefont{Kitazawa}},
  \bibinfo{author}{\bibfnamefont{T.}~\bibnamefont{Kunihiro}}, \bibnamefont{and}
  \bibinfo{author}{\bibfnamefont{Y.}~\bibnamefont{Nemoto}},
  \bibinfo{journal}{Phys. Lett.} \textbf{\bibinfo{volume}{B633}},
  \bibinfo{pages}{269} (\bibinfo{year}{2006}), \eprint{hep-ph/0510167}.

\bibitem[{\citenamefont{Kitazawa et~al.}(2007)\citenamefont{Kitazawa, Kunihiro,
  and Nemoto}}]{Kitazawa:2006zi}
\bibinfo{author}{\bibfnamefont{M.}~\bibnamefont{Kitazawa}},
  \bibinfo{author}{\bibfnamefont{T.}~\bibnamefont{Kunihiro}}, \bibnamefont{and}
  \bibinfo{author}{\bibfnamefont{Y.}~\bibnamefont{Nemoto}},
  \bibinfo{journal}{Prog. Theor. Phys.} \textbf{\bibinfo{volume}{117}},
  \bibinfo{pages}{103} (\bibinfo{year}{2007}), \eprint{hep-ph/0609164}.

\bibitem[{\citenamefont{Kamikado et~al.}(2014)\citenamefont{Kamikado,
  Strodthoff, von Smekal, and Wambach}}]{Kamikado:2013sia}
\bibinfo{author}{\bibfnamefont{K.}~\bibnamefont{Kamikado}},
  \bibinfo{author}{\bibfnamefont{N.}~\bibnamefont{Strodthoff}},
  \bibinfo{author}{\bibfnamefont{L.}~\bibnamefont{von Smekal}},
  \bibnamefont{and} \bibinfo{author}{\bibfnamefont{J.}~\bibnamefont{Wambach}},
  \bibinfo{journal}{Eur.Phys.J.} \textbf{\bibinfo{volume}{C74}},
  \bibinfo{pages}{2806} (\bibinfo{year}{2014}), \eprint{1302.6199}.

\bibitem[{\citenamefont{Tripolt et~al.}(2014)\citenamefont{Tripolt, von Smekal,
  and Wambach}}]{Tripolt:2014wra}
\bibinfo{author}{\bibfnamefont{R.-A.} \bibnamefont{Tripolt}},
  \bibinfo{author}{\bibfnamefont{L.}~\bibnamefont{von Smekal}},
  \bibnamefont{and} \bibinfo{author}{\bibfnamefont{J.}~\bibnamefont{Wambach}},
  \bibinfo{journal}{Phys. Rev.} \textbf{\bibinfo{volume}{D90}},
  \bibinfo{pages}{074031} (\bibinfo{year}{2014}), \eprint{1408.3512}.

\bibitem[{\citenamefont{Pawlowski and Strodthoff}(2015)}]{Pawlowski:2015mia}
\bibinfo{author}{\bibfnamefont{J.~M.} \bibnamefont{Pawlowski}}
  \bibnamefont{and}
  \bibinfo{author}{\bibfnamefont{N.}~\bibnamefont{Strodthoff}},
  \bibinfo{journal}{Phys. Rev.} \textbf{\bibinfo{volume}{D92}},
  \bibinfo{pages}{094009} (\bibinfo{year}{2015}), \eprint{1508.01160}.

\bibitem[{\citenamefont{Strodthoff}(2017)}]{Strodthoff:2016pxx}
\bibinfo{author}{\bibfnamefont{N.}~\bibnamefont{Strodthoff}},
  \bibinfo{journal}{Phys. Rev.} \textbf{\bibinfo{volume}{D95}},
  \bibinfo{pages}{076002} (\bibinfo{year}{2017}), \eprint{1611.05036}.

\bibitem[{\citenamefont{Jarrell and Gubernatis}(1996)}]{Jarrell:1996}
\bibinfo{author}{\bibfnamefont{M.}~\bibnamefont{Jarrell}} \bibnamefont{and}
  \bibinfo{author}{\bibfnamefont{J.}~\bibnamefont{Gubernatis}},
  \bibinfo{journal}{Phys. Rep. 269, 133}  (\bibinfo{year}{1996}).

\bibitem[{\citenamefont{Press et~al.}(1992)\citenamefont{Press, Teukolsy, W.T.,
  and Flannery}}]{nr:1997}
\bibinfo{author}{\bibfnamefont{W.}~\bibnamefont{Press}},
  \bibinfo{author}{\bibfnamefont{S.}~\bibnamefont{Teukolsy}},
  \bibinfo{author}{\bibfnamefont{V.}~\bibnamefont{W.T.}}, \bibnamefont{and}
  \bibinfo{author}{\bibfnamefont{B.}~\bibnamefont{Flannery}},
  \emph{\bibinfo{title}{Numerical Recipes in C: The Art of Scientific
  Computing, Second Edition}} (\bibinfo{publisher}{Cambridge University Press},
  \bibinfo{year}{1992}).

\bibitem[{\citenamefont{Asakawa et~al.}(2001)\citenamefont{Asakawa, Hatsuda,
  and Nakahara}}]{Asakawa:2000tr}
\bibinfo{author}{\bibfnamefont{M.}~\bibnamefont{Asakawa}},
  \bibinfo{author}{\bibfnamefont{T.}~\bibnamefont{Hatsuda}}, \bibnamefont{and}
  \bibinfo{author}{\bibfnamefont{Y.}~\bibnamefont{Nakahara}},
  \bibinfo{journal}{Prog. Part. Nucl. Phys.} \textbf{\bibinfo{volume}{46}},
  \bibinfo{pages}{459} (\bibinfo{year}{2001}), \eprint{hep-lat/0011040}.

\bibitem[{\citenamefont{Rothkopf}(2017)}]{Rothkopf:2016luz}
\bibinfo{author}{\bibfnamefont{A.}~\bibnamefont{Rothkopf}},
  \bibinfo{journal}{Phys. Rev.} \textbf{\bibinfo{volume}{D95}},
  \bibinfo{pages}{056016} (\bibinfo{year}{2017}), \eprint{1611.00482}.

\bibitem[{\citenamefont{Kim et~al.}(2015{\natexlab{b}})\citenamefont{Kim,
  Petreczky, and Rothkopf}}]{Kim:2014iga}
\bibinfo{author}{\bibfnamefont{S.}~\bibnamefont{Kim}},
  \bibinfo{author}{\bibfnamefont{P.}~\bibnamefont{Petreczky}},
  \bibnamefont{and} \bibinfo{author}{\bibfnamefont{A.}~\bibnamefont{Rothkopf}},
  \bibinfo{journal}{Phys. Rev.} \textbf{\bibinfo{volume}{D91}},
  \bibinfo{pages}{054511} (\bibinfo{year}{2015}{\natexlab{b}}),
  \eprint{1409.3630}.

\bibitem[{\citenamefont{Fischer and Luecker}(2013)}]{Fischer:2012vc}
\bibinfo{author}{\bibfnamefont{C.~S.} \bibnamefont{Fischer}} \bibnamefont{and}
  \bibinfo{author}{\bibfnamefont{J.}~\bibnamefont{Luecker}},
  \bibinfo{journal}{Phys. Lett.} \textbf{\bibinfo{volume}{B718}},
  \bibinfo{pages}{1036} (\bibinfo{year}{2013}), \eprint{1206.5191}.

\bibitem[{\citenamefont{Aouane et~al.}(2013)\citenamefont{Aouane, Burger,
  Ilgenfritz, Müller-Preussker, and Sternbeck}}]{Aouane:2012bk}
\bibinfo{author}{\bibfnamefont{R.}~\bibnamefont{Aouane}},
  \bibinfo{author}{\bibfnamefont{F.}~\bibnamefont{Burger}},
  \bibinfo{author}{\bibfnamefont{E.~M.} \bibnamefont{Ilgenfritz}},
  \bibinfo{author}{\bibfnamefont{M.}~\bibnamefont{Müller-Preussker}},
  \bibnamefont{and}
  \bibinfo{author}{\bibfnamefont{A.}~\bibnamefont{Sternbeck}},
  \bibinfo{journal}{Phys. Rev.} \textbf{\bibinfo{volume}{D87}},
  \bibinfo{pages}{114502} (\bibinfo{year}{2013}), \eprint{1212.1102}.

\bibitem[{\citenamefont{Borsanyi et~al.}(2010)\citenamefont{Borsanyi, Fodor,
  Hoelbling, Katz, Krieg, Ratti, and Szabo}}]{Borsanyi:2010bp}
\bibinfo{author}{\bibfnamefont{S.}~\bibnamefont{Borsanyi}},
  \bibinfo{author}{\bibfnamefont{Z.}~\bibnamefont{Fodor}},
  \bibinfo{author}{\bibfnamefont{C.}~\bibnamefont{Hoelbling}},
  \bibinfo{author}{\bibfnamefont{S.~D.} \bibnamefont{Katz}},
  \bibinfo{author}{\bibfnamefont{S.}~\bibnamefont{Krieg}},
  \bibinfo{author}{\bibfnamefont{C.}~\bibnamefont{Ratti}}, \bibnamefont{and}
  \bibinfo{author}{\bibfnamefont{K.~K.} \bibnamefont{Szabo}}
  (\bibinfo{collaboration}{Wuppertal-Budapest}), \bibinfo{journal}{JHEP}
  \textbf{\bibinfo{volume}{09}}, \bibinfo{pages}{073} (\bibinfo{year}{2010}),
  \eprint{1005.3508}.

\bibitem[{\citenamefont{Fischer
  et~al.}(2014{\natexlab{a}})\citenamefont{Fischer, Fister, Luecker, and
  Pawlowski}}]{Fischer:2013eca}
\bibinfo{author}{\bibfnamefont{C.~S.} \bibnamefont{Fischer}},
  \bibinfo{author}{\bibfnamefont{L.}~\bibnamefont{Fister}},
  \bibinfo{author}{\bibfnamefont{J.}~\bibnamefont{Luecker}}, \bibnamefont{and}
  \bibinfo{author}{\bibfnamefont{J.~M.} \bibnamefont{Pawlowski}},
  \bibinfo{journal}{Phys.Lett.} \textbf{\bibinfo{volume}{B732}},
  \bibinfo{pages}{273} (\bibinfo{year}{2014}{\natexlab{a}}),
  \eprint{1306.6022}.

\bibitem[{\citenamefont{Fischer
  et~al.}(2014{\natexlab{b}})\citenamefont{Fischer, Luecker, and
  Welzbacher}}]{Fischer:2014ata}
\bibinfo{author}{\bibfnamefont{C.~S.} \bibnamefont{Fischer}},
  \bibinfo{author}{\bibfnamefont{J.}~\bibnamefont{Luecker}}, \bibnamefont{and}
  \bibinfo{author}{\bibfnamefont{C.~A.} \bibnamefont{Welzbacher}},
  \bibinfo{journal}{Phys. Rev.} \textbf{\bibinfo{volume}{D90}},
  \bibinfo{pages}{034022} (\bibinfo{year}{2014}{\natexlab{b}}),
  \eprint{1405.4762}.

\bibitem[{\citenamefont{Eichmann et~al.}(2016)\citenamefont{Eichmann, Fischer,
  and Welzbacher}}]{Eichmann:2015kfa}
\bibinfo{author}{\bibfnamefont{G.}~\bibnamefont{Eichmann}},
  \bibinfo{author}{\bibfnamefont{C.~S.} \bibnamefont{Fischer}},
  \bibnamefont{and} \bibinfo{author}{\bibfnamefont{C.~A.}
  \bibnamefont{Welzbacher}}, \bibinfo{journal}{Phys. Rev.}
  \textbf{\bibinfo{volume}{D93}}, \bibinfo{pages}{034013}
  (\bibinfo{year}{2016}), \eprint{1509.02082}.

\bibitem[{\citenamefont{Fischer}(2009)}]{Fischer:2009wc}
\bibinfo{author}{\bibfnamefont{C.~S.} \bibnamefont{Fischer}},
  \bibinfo{journal}{Phys. Rev. Lett.} \textbf{\bibinfo{volume}{103}},
  \bibinfo{pages}{052003} (\bibinfo{year}{2009}), \eprint{0904.2700}.

\bibitem[{\citenamefont{Fischer et~al.}(2010)\citenamefont{Fischer, Maas, and
  Muller}}]{Fischer:2010fx}
\bibinfo{author}{\bibfnamefont{C.~S.} \bibnamefont{Fischer}},
  \bibinfo{author}{\bibfnamefont{A.}~\bibnamefont{Maas}}, \bibnamefont{and}
  \bibinfo{author}{\bibfnamefont{J.~A.} \bibnamefont{Muller}},
  \bibinfo{journal}{Eur. Phys. J.} \textbf{\bibinfo{volume}{C68}},
  \bibinfo{pages}{165} (\bibinfo{year}{2010}), \eprint{1003.1960}.

\bibitem[{\citenamefont{Maas et~al.}(2012)\citenamefont{Maas, Pawlowski, von
  Smekal, and Spielmann}}]{Maas:2011ez}
\bibinfo{author}{\bibfnamefont{A.}~\bibnamefont{Maas}},
  \bibinfo{author}{\bibfnamefont{J.~M.} \bibnamefont{Pawlowski}},
  \bibinfo{author}{\bibfnamefont{L.}~\bibnamefont{von Smekal}},
  \bibnamefont{and}
  \bibinfo{author}{\bibfnamefont{D.}~\bibnamefont{Spielmann}},
  \bibinfo{journal}{Phys. Rev.} \textbf{\bibinfo{volume}{D85}},
  \bibinfo{pages}{034037} (\bibinfo{year}{2012}), \eprint{1110.6340}.

\bibitem[{\citenamefont{Ball and Chiu}(1980)}]{Ball:1980ay}
\bibinfo{author}{\bibfnamefont{J.~S.} \bibnamefont{Ball}} \bibnamefont{and}
  \bibinfo{author}{\bibfnamefont{T.-W.} \bibnamefont{Chiu}},
  \bibinfo{journal}{Phys. Rev.} \textbf{\bibinfo{volume}{D22}},
  \bibinfo{pages}{2542} (\bibinfo{year}{1980}).

\bibitem[{\citenamefont{Pawlowski and Rothkopf}(2016)}]{Pawlowski:2016eck}
\bibinfo{author}{\bibfnamefont{J.}~\bibnamefont{Pawlowski}} \bibnamefont{and}
  \bibinfo{author}{\bibfnamefont{A.}~\bibnamefont{Rothkopf}}
  (\bibinfo{year}{2016}), \eprint{1610.09531}.

\bibitem[{\citenamefont{Welzbacher}(2016)}]{Welzbacher16}
\bibinfo{author}{\bibfnamefont{C.~A.} \bibnamefont{Welzbacher}}, Ph.D. thesis,
  \bibinfo{school}{Justus-Liebig-Universit\"at} (\bibinfo{year}{2016}).

\bibitem[{\citenamefont{Ding et~al.}(2012)\citenamefont{Ding, Francis,
  Kaczmarek, Karsch, Satz, and Soeldner}}]{Ding:2012sp}
\bibinfo{author}{\bibfnamefont{H.~T.} \bibnamefont{Ding}},
  \bibinfo{author}{\bibfnamefont{A.}~\bibnamefont{Francis}},
  \bibinfo{author}{\bibfnamefont{O.}~\bibnamefont{Kaczmarek}},
  \bibinfo{author}{\bibfnamefont{F.}~\bibnamefont{Karsch}},
  \bibinfo{author}{\bibfnamefont{H.}~\bibnamefont{Satz}}, \bibnamefont{and}
  \bibinfo{author}{\bibfnamefont{W.}~\bibnamefont{Soeldner}},
  \bibinfo{journal}{Phys. Rev.} \textbf{\bibinfo{volume}{D86}},
  \bibinfo{pages}{014509} (\bibinfo{year}{2012}), \eprint{1204.4945}.

\end{thebibliography}

\end{document}